\documentclass[a4paper,12pt,oneside]{article}  % document class
\usepackage{geometry}
\usepackage[english]{babel}
\usepackage{microtype}
\usepackage{natbib} %bibliography

\usepackage{import, multicol,lipsum} 
\usepackage[utf8]{inputenc} % easy accents
\usepackage[T1]{fontenc}
\usepackage{subcaption}
\usepackage{pifont}
\usepackage{xcolor, blindtext}
\usepackage{url}
\usepackage{hyperref} 
\hypersetup{
colorlinks=true, %   colorlinks=false,
linkcolor=red,
citecolor=blue,     
urlcolor=blue
}

\usepackage{emptypage} % prevents page numbers and headings from appearing on empty pages
\usepackage[export]{adjustbox} %label next image
\usepackage{amsthm, amsmath, amssymb, amsfonts, mathtools} % math
\usepackage{nicefrac}
\usepackage{enumitem}
\usepackage{graphicx} %images and tables
\usepackage{multirow, array} %tables
\usepackage{nicematrix} %tables
\usepackage{makecell}
\graphicspath{{img/}} %graphics path of 
\usepackage{wrapfig}
\usepackage{float} %force image position with [H]
\usepackage{tikz} %graph

\usepackage{verbatim}
\usepackage{bbm}
\RequirePackage{fix-cm}

\newtheorem{theorem}{Theorem}%[section]
\newtheorem{definition}[theorem]{Definition} 
\newtheorem{proposition}{Proposition} 
\usepackage[ruled,vlined]{algorithm2e}

\newcommand{\set}[1]{\left\{#1\right\}}

\newcommand{\expfun}[1]{\exp\set{#1}}
\newcommand{\indicator}[1]{\mathbbm{1}\set{#1}}

\def \R {\mathbb{R}}
\def \Nat {\mathbb{N}}
\def \Z {\mathbb{Z}}

\def \E {\mathcal{E}}
\def \G {\mathcal{G}}
\def \I {\mathcal{I}}
\def \K {\mathcal{K}}

\def \N {\mathcal{N}}

\def \dd {\boldsymbol{d}}
\def \ss {\boldsymbol{s}}

\def \aalpha {\boldsymbol{\alpha}}
\def \bbeta {\boldsymbol{\beta}}
\def \ttheta {\boldsymbol{\theta}}
\def \ssigma {\boldsymbol{\sigma}}
\def \vveps {\boldsymbol{\varepsilon}}

\def \XX {\mathbf{X}}
\def \xx {\mathbf{x}}
\def \YY {\mathbf{Y}}
\def \ZZ {\mathbf{Z}}
\def \zz {\mathbf{z}}

\def \gA {\tilde A}

\title{\bf Long memory network time series}
\author{
Chiara Boetti\\University of Bath \and Matthew A. Nunes \\University of Bath \and Marina I. Knight\\ University of York}
\date{\today}

\begin{document}
\maketitle
\thispagestyle{empty}

\begin{abstract}
%\footnotesize
Many scientific areas, from computer science to the environmental sciences and finance, give rise to multivariate time series which exhibit long memory, or loosely put, a slow decay in their autocorrelation structure. Efficient modelling and estimation in such settings is key for a number of analysis tasks, such as accurate prediction. However, traditional approaches for modelling such data, for example long memory vector autoregressive processes, are challenging even in modest dimensions, as the number of parameters grows quadratically with the number of modelled variables.  Additionally, in many practical data settings, the observed series is accompanied by a (possibly inferred) network that provides information about the presence or absence of between-component associations via the graph edge topology. This article proposes two new models for capturing the dynamics of long memory time series where a network is accounted for. Our approach not only facilitates the analysis of graph-structured long memory time series, but also improves computational efficiency over traditional multivariate long memory models by leveraging the inherent low-dimensional parameter space by adapting likelihood-based estimation algorithms to the network setting. Simulation studies show that our proposed estimation is more stable than traditional models, and is able to tackle data scenarios where current models fail due to computational challenges. While widely applicable, here we demonstrate the efficacy of our proposed models on datasets arising in environmental science and finance.
\end{abstract}


%-------------------------------------------------------------------
%  INTRODUCTION 
%-------------------------------------------------------------------
\section{Introduction}
Long memory time series models are commonly used to describe data characterised by a slow, power-law decaying autocovariance function. Due to their strong dependence on past observations, these series are also commonly referred to as long-range dependent processes. The phenomenon of long-range dependence (or persistence) has been observed and modelled in a number of scientific areas, such as climatology \citep{fraedrich03:scaling, rehman09:wavelet, knight16:a}, the geophysical and environmental sciences \citep{ventosa14:long, knight2019long}, financial applications \citep{cont2005long, martinez2018impact} and internet data modelling \citep{karagiannis2004long} to name but a few.

While univariate long memory models are now widely applied and well-understood theoretically \citep{beran2013long, pipiras2017longrange}, their multivariate counterparts present both significant theoretical and computational challenges. In particular, for parametric multivariate models a primary difficulty in {\em exact} estimation lies in the evaluation of the autocovariance function. This was first tackled numerically by \cite{sowell1989maximum}, but such an implementation is computationally inefficient as it relies on the evaluation of hypergeometric functions. Building on this work, \cite{sela2009computationally} proposed faster {\em approximate} algorithms for single lag autoregressive models, while \cite{tsay2010maximum} developed a conditional likelihood approach based on the Durbin-Levinson algorithm. Other notable contributions include the work of \cite{ravishanker1997bayesian} (with a later corrigendum by \cite{doppelt2019posterior}), who were among the first to apply Bayesian methods for estimating multivariate fractionally integrated models, as well as \cite{pai2009maximum} and \cite{pai2009multivariate}, who utilised the expectation-maximisation (EM) algorithm and the multivariate preconditioned conjugate gradient (PCG) algorithm, respectively.

These methods are worthy advancements to modelling multivariate long memory processes, but they become rapidly computationally infeasible as the dimensionality of the process increases: nearly all examples in the literature on their practical use for exact maximum likelihood estimation have been limited to two- or three-dimensional processes. In addition, they are currently not well-suited for handling time-dependent data defined on a graph structure, since they will fail to capture the sparsity in the connections among variables.

On the other hand, there is a growing body of work dedicated to {\em network time series}, originally proposed by \cite{Knight2016}, that allows for a modelling framework in which a known or inferred network structure is incorporated in order to capture inter-variable relationships in multivariate time series data, see also \cite{zhu2017network}. The  Generalised Network AutoRegressive (GNAR) processes of \cite{knight2020generalized} have been shown to yield benefits for analysis tasks such as forecasting despite, or precisely because, its use of very few model parameters compared to their non-network counterparts \citep{knight2020generalized, nason2025forecasting}. Recently, this framework has been extended to other settings that include non-Gaussian time series and count processes \citep{armillotta2022generalized, armillotta2024count}, nonlinear network relationships \citep{armillotta2023nonlinear} and  
ARCH-type models \citep{zhou2020network, pan2024threshold}. Network time series approaches have now found to provide scientific insight in a number of applications, including the geosciences \citep{Knight2016}, economics \citep{nason2022quantifying}, epidemiology \citep{nason2023new} and political science \citep{nason2024modelling}.  These existing network time series models, however, are not specifically designed to handle data which exhibit long-range dependence.  %In the network vector autoregressive (GNAR) setting, this approach 

In this article, we introduce two new statistical models for network-structured long memory data, motivated by reducing computational complexity in multivariate long memory models and settings such as computer traffic monitoring, financial networks and environmental processes, in which a network structure is often observed alongside the time series.  The article is structured as follows. Section \ref{sec:background} reviews long memory processes and provides an overview of the autoregressive network time series model of \cite{knight2020generalized}. Our proposed long memory network models are introduced in Section \ref{sec:proposal}, including parameter estimation methods specifically designed for the network setting in Section \ref{sec:parest}. A detailed simulation study validating our estimation techniques can be found in Section \ref{sec:simstudy}. Section \ref{sec:forecasting} considers the task of forecasting time series. Section \ref{sec:application} presents an application of our models to two real-world datasets arising in meteorology and finance, and we provide concluding remarks in Section \ref{sec:concs}.

%-------------------------------------------------------------------
%  MULTIVARIATE TIME SERIES
%-------------------------------------------------------------------
\section{Background}\label{sec:background}
In this section, we review long memory behaviour in the multivariate time series setting, as well as give a brief overview of the notation and network time series model of \cite{knight2020generalized}. These two aspects provide the necessary background to formulate our proposed modelling framework for network-structured long memory processes in Section \ref{sec:proposal}.

\subsection{Multivariate Long Memory Time Series}\label{sec:longmemo}

In the multivariate case, time series can exhibit long-range dependence in one or more of their components. Several authors have proposed models which account for long memory behaviour in multivariate data, see for example \cite{chambers1995simulation} and \cite{didierpipiras}.  In this article we focus on the stationary time series setting, in which persistent behaviour is captured by the autocovariance matrix function $\Omega(h)$, which decays with a power-law rate as the lag $h$ increases; this phenomenon can be similarly characterised by the asymptotic behaviour of the spectral density matrix as the frequency approaches zero.  

Within the broader class of multivariate long memory processes, long-range dependence is incorporated into parametric models using the long memory filter, also known as $\dd$-th differencing at lag 1. More precisely, given an $N$-dimensional time series, the $N$-dimensional vector of long memory parameters $\dd = \left(d_{1}, \ldots, d_{N}\right)$ encodes the magnitude of long-range dependence in the process. The long memory filter is expressed as $(1 - L)^{\dd} = \operatorname{diag}\left((1 - L)^{d_{1}}, \ldots, (1 - L)^{d_{N}}\right)$, where $L$ is the backshift operator and $L^{0}=I$ is the identity operator. Thus, the parameter $d_{i} \in (0, \frac{1}{2})$ represents the long memory coefficient associated with the $i$-th component of the $N$-dimensional series; $d_{i} < 0$ indicates anti-persistence. %\newline

The simplest parametric model for multivariate long memory processes is the so-called fractionally integrated white noise (FIWN), a natural extension of the univariate model obtained by fractionally integrating the white noise at each component. Specifically, given an $N$-dimensional Gaussian process $\vveps_t = \left(\varepsilon_{1,t}, \ldots, \varepsilon_{N,t} \right)^{\top}$  for $t\in\Z$, the process $\ZZ_t = \left(Z_{1,t}, \ldots, Z_{N,t} \right)^{\top}$ follows a FIWN($\dd$) model if $(1 - L)^{\dd} \ZZ_t = \vveps_t$. \cite{lobato1997consistency} generalised this framework by additionally including a short-memory component; specifically, the FIVAR and the VARFI models with long memory and autoregressive orders $\dd$ and $p$, respectively, are %defined as 
\begin{equation}\label{fivar-varfi}
    (1-L)^{\dd} A(L) \XX_t = \vveps_t \quad \text{ and } \quad A(L) (1-L)^{\dd} \XX_t = \vveps_t,
\end{equation}
where $A(L) = I + \sum_{j=1}^{p} A_{j} L^{j} $ is the usual vector autoregressive (VAR) filter of order $p$. Due to the non-commutative nature of multivariate filter composition, these models are not equivalent, unlike in the univariate case where such a distinction does not exist. 
Further extensions of \eqref{fivar-varfi} include allowing for a moving average component, and incorporating a general phase in the cross-spectrum, as considered in the bivariate case by \cite{kechagias2015definitions} and \cite{kechagias2020modeling}. 

%-------------------------------------------------------------------
\subsection{Network Time Series}\label{sec:GNAR}
%In many settings data naturally arise as connections between individual nodes in a graph. Some examples include collaboration networks in social media, internet traffic data, pollution records. Alternatively, data originating from other domains, like images, can be transformed into graph-like structures to facilitate analysis. 

Network time series have been developed to naturally account for the connections among time-dependent variables. Specifically, a network time series is a multivariate process defined over a network or {\em graph}, in which each node corresponds to a univariate time series, and the dependence between nodes are modelled via the edge connections of the given network.  In what follows, we adopt the notation of \cite{knight2020generalized}.

In our framework, a graph $\G = (\K, \E)$ is defined by a fixed set of nodes $\K = \set{1,\ldots,N}$ and a set of edges $\E$. In this article, we restrict our attention to graphs with undirected edges, thus the edge set is defined as $\E = \set{(i,j)\in \K : i\leftrightsquigarrow j}$, where $i \leftrightsquigarrow j$ indicates that an edge connection exists between node $i$ and node $j$. %\newline

For a nodal subset $\I \subset \K$, the \textit{neighbour set of} $\I$ is $\N(\I) = \set{j\in\K\setminus \I : i \leftrightsquigarrow j, i\in \I}$. \cite{knight2020generalized} iteratively define, and use in modelling, the $r$-th stage neighbours of node $i$ %as 
\begin{equation*}
	\begin{dcases}
        & \N^{(1)}(i) = \N(\set{i}),\\
		& \N^{(r)}(i) = \N\set{\N^{(r-1)}(i)}\setminus \left[ \set{ \bigcup_{q=1}^{r-1} \N^{(q)}(i)} \cup \set{i} \right] \quad \text{for } r=2,3,\ldots .
	\end{dcases}
\end{equation*}
Notice that $\N(\I)$ is the set of all first stage neighbours of $\I$. %\newline

Edges in the network $\G$ can be associated to a connection weight $\omega\in[0,1]$ carrying information which may be used to model the strength of association between two nodal time series.  For example, for networks in which the nodes relate to geographical locations, the distance between two nodes can be used to form {\em inverse distance weights} so that a nodal time series is more serially correlated to one of its closer neighbouring nodes, whilst the process exhibits weaker dependence between nodes which are further away in the network, see e.g. \cite{knight2020generalized}. These weights may depend on the size of the neighbour set and can also be time-varying to describe a time-dependent graph structure. Finally, we can encode other edge effects into a graph via covariates. As an example, if there are $C \in \mathbb{N}$ types of edges in a network, a covariate may assign an index $c =1,\ldots, C$ to each effect. %\newline

%-------------------------------------------------------------------
% \subsection{GNAR Time Series}
The generalised network autoregressive (GNAR) model proposed by \cite{knight2020generalized} set out to describe multivariate time series processes observed at nodes of time-varying network structures. Given a zero-mean Gaussian process $\vveps$ with covariance matrix $\Sigma_{\vveps} = \operatorname{diag}\left(\sigma_{1}^{2}, \ldots, \sigma_{N}^{2} \right)$, the process $\{\XX_t = \left( X_{1,t}, \dots, X_{N,t}\right)^\top\}_t$ is said to follow a GNAR($p,[\mathbf{s}]= \left(s_{1}, \ldots, s_{p} \right)$) model with fixed network structure if
\begin{equation}\label{eq:gnar}
    \gA(L)\XX_{t} = \vveps_t,
\end{equation} 
where $\gA(L)= I + \sum_{j=1}^{p} \tilde{A}_{j} L^{j}$ is the GNAR filter defined in terms of the individual time-lag $j$, $(N \times N)$-dimensional matrices
\begin{equation}\label{eq:gnarfilter}
    \gA_j = \operatorname{diag}(\alpha_{1,j},\ldots,\alpha_{N,j}) + \sum_{c=1}^{C} \sum_{r=1}^{s_j} \beta_{j,r,c} W^{(r,c)}.
\end{equation}
Here, $p \in \Nat$ refers to the largest autoregressive (AR) lag, $[\ss] = \left(s_{1}, \ldots, s_{p} \right)$ and $s_{j} \in \Nat$ is the maximal stage of neighbour dependence for a time-lag $j$. Then $\alpha_{i, j} \in \R$ is the usual AR parameter at lag $j$ for node $i$, and $\beta_{j, r, c} \in \R$ is the {\em network-spatial} AR parameter, which encodes the effect of the $r$-th stage neighbours at lag $j$ according to covariate $c$. We denote with $\aalpha=(\alpha_{i, j})_{i,j}$ and $\boldsymbol{\beta}=(\beta_{j, r, c})_{j, r, c}$ the vectors of usual and network AR parameters, respectively. The weight matrices $ W^{(r,c)}$ encode the network structure and their entries are $\left( W^{(r,c)} \right)_{i,l} = \omega_{i,l,c} \indicator{l\in\N^{(r)}(i)}$.  Note that the GNAR filter $\gA$ with components as in \eqref{eq:gnarfilter} can be seen as a special (constrained) form of the VAR filter in \eqref{fivar-varfi} which crucially  incorporates the network structure: absence of edges in the network $\G$ corresponds to zero entries in the lagged coefficient matrices $\{\gA_j\}_j$.  

The model is flexible enough to capture cross-nodal dependence whilst also being parsimonious, since the network parameters $\bbeta$ are not specific to individual nodes, $i$. We refer to specification~\eqref{eq:gnarfilter} as the individual-$\alpha$ GNAR($p,[\mathbf{s}]$). An even simpler variant of Model \eqref{eq:gnar} is the global-$\alpha$ GNAR($p,[\mathbf{s}]$), where common AR parameters are used to describe the autocorrelation at every node. In particular, for each time-lag $j$, the regression coefficients fulfil $\alpha_{i,j} = \alpha_{j}$ for all $i=1,\ldots,N$, and the network time series vary according to the underlying graph structure. 

%-------------------------------------------------------------------
%  NEW BIT
%-------------------------------------------------------------------
\section{Long Memory Network Time Series Models}\label{sec:proposal}
While GNAR processes effectively capture connections in network-structured time series, they are not suitable to represent data with long memory. Following the approach of \cite{lobato1997consistency}, we address this limitation by extending Model \eqref{eq:gnar} to long memory network time series, as follows.

\subsection{Proposed Models}\label{subsec:models}
\begin{definition}
Let $\G = (\K, \E)$ be a graph with $N$ nodes, let $\{\vveps_t\}$ be a $N$-dimensional white noise process with covariance matrix $\Sigma_{\vveps} = \operatorname{diag}\left(\sigma_{1}^{2}, \ldots, \sigma_{N}^{2} \right)$, and let $\XX_{t} = \left(X_{1,t}, \ldots, X_{N,t} \right)$ be a $N$-dimensional vector of nodal time series at time $t =1,\ldots, T$. 

\noindent Additionally, let $\gA(L)= I + \sum_{j=1}^{p} \tilde{A}_{j} L^{j}$ be the GNAR autoregressive filter with order $p\in \Nat$ and matrices $\{\tilde{A}_j\}_{j=1}^p$ as in \eqref{eq:gnarfilter}, maximal neighbour dependence stage $\ss \in \Nat^{p} \setminus \set{\mathbf{0}}$, and let $\dd=(d_1,\ldots,d_N)$ be the differencing coefficients such that $0 < d_i < \frac{1}{2}$ for each  node $i=1,\ldots,N$. 

\noindent With this notation, we define the following two long memory process models. \\

\noindent The Fractionally Integrated GNAR, $\mathbf{FIGNAR(p,[\ss],\dd)}$, model is defined for each time $t$ as
\begin{equation}\label{eq:fignar}
	\gA(L) \left( 1-L\right)^{\dd} \XX_{t} = \vveps_t.
\end{equation}
The GNAR with Fractionally Integrated noise, $\mathbf{GNARFI(p,[\ss],\dd)}$, model is defined as
%for each time $t \in \Z$ as
\begin{equation}\label{eq:gnarfi}
	\left( 1-L \right)^{\dd} \gA(L) \XX_{t} = \vveps_t. 
\end{equation}
\end{definition}

\medskip 

Note that, similar to their non-network vector counterparts, models \eqref{eq:fignar} and \eqref{eq:gnarfi} contain two sets of parameters. The short-memory component, inherited from the GNAR process, is characterised by the usual AR parameters $\aalpha$ and the network AR parameters $\boldsymbol{\beta}$. The long memory component is specified by the differencing coefficients $\dd$.

In addition to the global- or individual-$\alpha$ specifications for the AR part, the models can impose either global or individual differencing. In the global-$d$ FIGNAR and GNARFI specifications $d_i = d$ for all $i=1,\ldots,N$, otherwise, they are referred to as individual-$d$ models. Likewise, if the white noise covariance matrix is a scalar multiple of the identity, $\Sigma_{\vveps} = \sigma^2 I$ with $\sigma^2>0$, we say we are in the global-$\sigma^2$ setting. If instead variances are node-specific, i.e., $\sigma_i^2$ for $i=1,\ldots,N$, we have individual-$\sigma^2$ models.

One of the main advantages of our long memory network time series models is parameter parsimony. Let $\ttheta = \left( \aalpha, \bbeta, \dd, \ssigma^{2} \right)$ denote the parameter vector, with $\ssigma^2 = \left(\sigma_{1}^{2}, \ldots, \sigma_{N}^{2} \right)$. The dimension of $\ttheta$ depends on the specification. In the individual-$\alpha$ case, $\left( \aalpha, \bbeta \right)$ accounts for $Np + C \sum_{j=1}^{p} s_{j}$ parameters, and it reduces to $p + C \sum_{j=1}^{p} s_{j}$ parameters under global-$\alpha$ specifications. Similarly, individual-$d$ models use $N$ differencing parameters, whereas global-$d$ models have a single value. For the innovation variances, the global-$\sigma^2$ setting estimates one variance in place of the full diagonal, in contrast to the $N$ variances of the individual-$\sigma^2$ case.
Hence, in the most parsimonious specification, global-$\set{\alpha, d, \sigma^2}$, the dimension of $\ttheta$ is independent of the dimension of the process, $N$, namely $p + C \sum_{j=1}^{p} s_{j} + 2$. In all other cases, the parameter count grows linearly with $N$, with the maximal count occurring for the individual-$\set{\alpha, d, \sigma^2}$ model, with $Np + C \sum_{j=1}^{p} s_{j} + 2N$ parameters.
Such parameter efficiency makes network models more computationally tractable for naturally high-dimensional problems when compared to the FIVAR and VARFI models of \cite{lobato1997consistency} (see also~\eqref{fivar-varfi}), where the number of parameters grows quadratically with the dimension of the process, $N$. %\newline
A sufficient condition for the (second-order) stationarity of long memory network models is presented in the following proposition.
\begin{proposition}
If
\begin{equation}\label{eq:stationarity}
    \sum_{j=1}^p \left( |\alpha_{i, j}| + \sum_{c=1}^C \sum_{r=1}^{s_j} |\beta_{j, r, c}| \right) < 1 \quad \forall i=1,\ldots, N,
\end{equation}
then the FIGNAR($p,[\mathbf{s}],\dd$) model \eqref{eq:fignar} and the GNARFI($p,[\mathbf{s}],\dd$) model \eqref{eq:gnarfi} are stationary.
\end{proposition}
\medskip
\begin{proof}
Since $d_1,\ldots,d_N \in (0,\frac{1}{2})$ by definition, stationarity follows when the roots of $\gA(L)$ lie outside the unit circle \citep{sowell1989maximum}. Condition \eqref{eq:stationarity} guarantees this, as shown by \cite{knight2020generalized} for the static network GNAR process.
\end{proof}

\paragraph{Remark (Differences between the proposed FIGNAR and GNARFI models).} The two proposed models, defined in equations \eqref{eq:fignar} and \eqref{eq:gnarfi}, differ not only in their outcomes but also in their interpretation. 

A FIGNAR($p,[\ss],\dd$) time series is the order-$\dd$ fractional integration of a GNAR($p,[\ss]$) process $\set{\YY_{t}}_{t}$, and can be explicitly rewritten as follows
\begin{equation}\label{eq:fignar_equiv}
\begin{dcases}
    \left( 1-L \right)^{\dd} \XX_t = \YY_t, \\
    \gA(L) \YY_{t} = \vveps_t.
\end{dcases}
\end{equation}
Node-wise, the above expression, equivalent to equation \eqref{eq:fignar}, specifies at each node $i$
\begin{equation*}
    X_{i, t} = (1-L)^{-d_i} \left[ \sum_{j=1}^p \left(\alpha_{i, j} Y_{i, t-j} + \sum_{c=1}^C \sum_{r=1}^{s_j} \beta_{j, r, c} \sum_{l \in \N_t^{(r)}(i)} \omega_{i, l, c}^{(t)} Y_{l, t-j}\right) + \varepsilon_{i, t} \right].
\end{equation*}

In contrast, a GNARFI($p,[\ss],\dd$) model is a GNAR($p,[\ss]$) model driven by a fractionally integrated white noise FIWN($\dd$) process, rather than by the usual white noise, as demonstrated by the following expression
\begin{equation}\label{eq:gnarfi_equiv}
\begin{dcases}
    \tilde{A}(L) \XX_{t} = \ZZ_t, \\
    \left( 1-L \right)^{\dd} \ZZ_t = \vveps_t.
\end{dcases}
\end{equation}
Hence, at each node $i = 1,\ldots, N$, the equivalent to equation \eqref{eq:gnarfi} is
\begin{equation*}
    X_{i, t} = \sum_{j=1}^p \left(\alpha_{i, j} X_{i, t-j} + \sum_{c=1}^C \sum_{r=1}^{s_j} \beta_{j, r, c} \sum_{l \in \N_t^{(r)}(i)} \omega_{i, l, c}^{(t)} X_{l, t-j}\right) + (1-L)^{-d_i}\varepsilon_{i, t}.
\end{equation*}

\paragraph{Remark (Process equivalence).} The models in \eqref{eq:fignar_equiv} and \eqref{eq:gnarfi_equiv} are equivalent if and only if the $\left( 1-L \right)^{\dd}$ and the $\tilde{A}(L)$ operators commute. This would occur when $\left( 1-L \right)^{\dd}$ is a scalar multiple of the identity matrix, namely if $d_i=d$ for all $i=1,\ldots,N$, or if both $\tilde{A}(L)$ and $\Sigma_{\vveps}$ correspond to diagonal matrices. However, this latter case is not pertinent to network time series, as the graph structure characterising these processes is preserved within the off-diagonal terms of the AR matrices. %\newline
\medskip

\begin{figure}[!ht]
	\centering
	\includegraphics[width=0.45\textwidth]{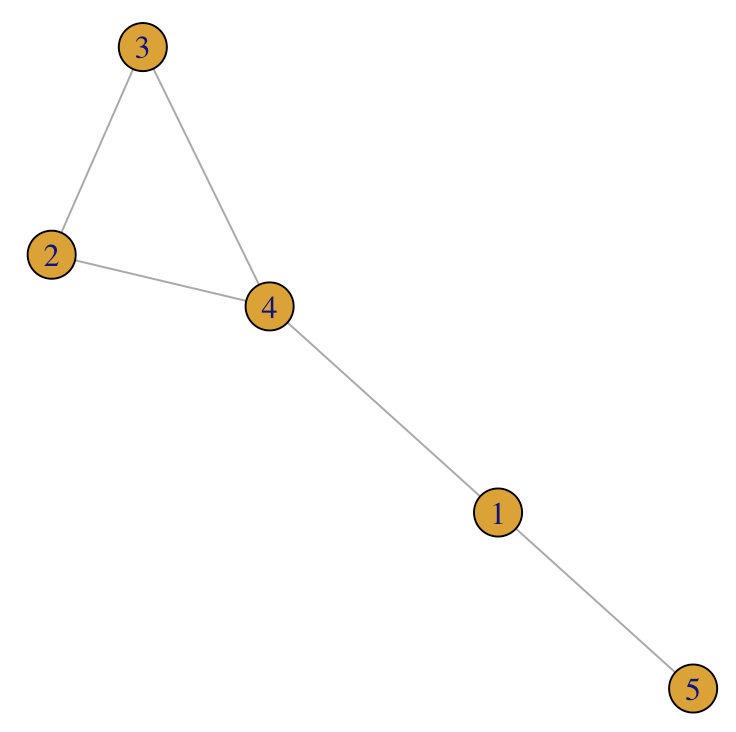}
	\captionof{figure}{Undirected, unweighted graph introduced as {\tt fiveNet} by \cite{knight2020generalized}.}\label{img:fivenet}
\end{figure}

\paragraph{Example (DGP1).} As an illustrative example to demonstrate the proposed long memory network models \eqref{eq:fignar} and \eqref{eq:gnarfi}, as well as the associated characteristics of the data they represent, we consider the unweighted, undirected five-node network shown in Figure \ref{img:fivenet} and simulate a global-$\alpha$ FIGNAR(1,[1],$\dd$) and a global-$\alpha$ GNARFI(1,[1],$\dd$) time series, both with parameters $\alpha = 0.35$, $\beta = 0.2$, and $\dd = (0.05, 0.15, 0.25, 0.35, 0.45)$ that satisfy the stationarity condition \eqref{eq:stationarity}. We will henceforth refer to these types of \underline{d}ata \underline{g}enerating \underline{p}rocesses as DGP1. We assume both processes have independent and identically distributed standard normal innovations, $\Sigma_{\vveps} = I$. This setup will be revisited in experiments in Section \ref{sec:simstudy}.
% Realisations of proposed long memory network models \eqref{eq:fignar} and \eqref{eq:gnarfi}.)

Realisations of length $T = 500$ for DGP1 under both FIGNAR and GNARFI models are shown in Appendix \ref{app:realisations}, Figures \ref{img:fignar-ts-acf}-\ref{img:gnarfi-ts-acf}. Even though both realisations are drawn from the same white noise process, note they correspond to different dependence behaviours: as expected with both models, the slow decay in the autocorrelation becomes more evident as the differencing parameter $d_i$ increases, but the FIGNAR model exhibits higher autocorrelation at larger lags. This behaviour is also reflected when plotting the network autocorrelation function of \cite{nason2023new}, compared with the GNAR case (Figures \ref{img:nacfgnar}--\ref{img:nacfgnarfi}). 

%-------------------------------------------------------------------
\subsection{Parameter Estimation}\label{sec:parest}
In this section we describe our proposed parameter estimation procedures for the new FIGNAR and GNARFI models introduced in the Section~\ref{subsec:models}.  In what follows, we take a maximum likelihood approach. 

\paragraph{FIGNAR Likelihood.}
By definition, given a sample of $T$ observations from a stationary time series $\set{\XX_{t}}$ distributed according the FIGNAR model \eqref{eq:fignar} with Gaussian white noise process $\set{\vveps_t}$, the corresponding likelihood function is
\begin{equation}\label{eq:fignar-like}
    \mathcal{L}_{F}(\ttheta; \xx) = (2\pi)^{-\frac{TN}{2}} \left| \Sigma_{F} \right|^{-\frac{1}{2}}  \expfun{-\frac{1}{2} \xx^{\top} \Sigma^{-1}_{F} \xx},
\end{equation}
where $\ttheta = \left( \aalpha, \bbeta, \dd, \ssigma^{2} \right)$ is the vector of all unknown parameters, namely the autoregressive, network dependence, long memory and the covariance noise parameters, to be estimated as described next. Here, $\xx = (\xx_{1}, \ldots, \xx_{N})^{\top}$ is an $NT$-dimensional vector, with $\xx_{i} = \left( x_{i,1}, \ldots, x_{i,T} \right)^{\top}$ for each $i=1,\ldots,N$, and $|\mathbf{A}|$ denotes the determinant of matrix $\mathbf{A}$. 

The overall covariance matrix of the FIGNAR process is denoted in~\eqref{eq:fignar-like} by $\Sigma_{F}$, where $\Sigma_{F} = \Sigma_{F}(\ttheta) = \left( \Omega_{F}(t-s)\right)_{t,s=1}^{T}$ is an $(NT \times NT)$-dimensional block matrix. Notice that the $(i,k)$-th block of $\Sigma_{F}$ is the $(T \times T)$-dimensional Toeplitz covariance matrix corresponding to the $i$-th and $k$-th series components, and $\Omega_{F}(h)$ is the $(N \times N)$-dimensional autocovariance matrix at lag $h$, $h=-(T-1),\ldots,T-1$.

Due to the presence of the AR component, the autocovariance matrix function does not have a closed-form solution, making the evaluation of $\Sigma_{F}$ a challenging task. 
In our optimisation scheme, we evaluate the covariance matrix $\Sigma_{F}$ using the explicit formula derived by \cite{sela2009computationally}. Specifically, the autocovariance matrix at lag $h$ is
\begin{equation}\label{eq:acv-fignar}
    \Omega_{F}(h) = \sum_{m=0}^{+\infty} \xi(h+m) \odot \phi(h),
\end{equation}
where $\odot$ denotes the Hadamard product, $\xi(h)$ is the autocovariance matrix of the GNAR component at lag $h$, and $\phi(h)$ is the $(N\times N)$-dimensional matrix with elements
\begin{equation}\label{eq:fiwn-acv}
    \phi_{i,k}(h) = \frac{\Gamma(1-d_{i}-d_{k}) \Gamma(h+d_{k})}{\Gamma(d_{k}) \Gamma(1-d_{k}) \Gamma(h+1-d_{i})},
\end{equation}
for each $i,k=1,\ldots,N$, and $\Gamma(\cdotp)$ is the gamma function.
Since the autocovariance sequence of the GNAR process decays exponentially fast, we set the elements of $\xi(m)$ to 0 for all $|m|>M$, where $M$ is selected according to the criterion in \cite{sela2009computationally}.

\paragraph{GNARFI Likelihood.}
Similarly, the likelihood given a sample from a stationary GNARFI process \eqref{eq:gnarfi} is 
\begin{equation}\label{eq:gnarfi-like}
    \mathcal{L}_{G}(\ttheta; \xx) = (2\pi)^{-\frac{TN}{2}} \left| \Sigma_{G} \right|^{-\frac{1}{2}}  \expfun{-\frac{1}{2} \xx^{\top} \Sigma^{-1}_{G} \xx},
\end{equation}
with $\ttheta$ denoting the vector of parameters to be estimated,  the sample $\xx$ is the $NT$-dimensional vector of GNARFI data, and $\Sigma_{G} = \Sigma_{G}(\ttheta) = \left( \Omega_{G}(t-s)\right)_{t,s=1}^{T}$ the $(NT \times NT)$-dimensional covariance matrix of the GNARFI time series.

The exact formula of the $\Sigma_{G}$ stems from the autocovariance matrix of the GNARFI($1,[s],\dd$) model. As derived by \cite{sela2009computationally}, for each lag $h=-T+1, \ldots, T-1$, such an autocovariance matrix can be expressed as
\begin{equation}\label{eq:acv-gnarfi}
    \Omega_{G}(h) = V \left[ \sum_{r,s=0}^{+\infty} \Lambda^{r}  V^{-1} \eta(r-s-h) (V^{H})^{-1} (\Lambda^{H})^{s} \right] V^{H}.
\end{equation}
This formula is based on the spectral decomposition of the AR coefficient matrix of the GNARFI($1,[s],\dd$) process. Specifically, it assumes that there exist an orthogonal matrix of eigenvectors $V$ and a diagonal matrix of eigenvalues $\Lambda$ such that $A = V \Lambda V^{H}$. The term $\eta(\cdot)$ is the autocovariance of the corresponding FIWN($\dd$) process, with $\eta(h) = \Sigma_{\vveps} \odot \phi(h)$, where $\phi(h)$ is defined in \eqref{eq:fiwn-acv}.
The infinite sum in \eqref{eq:acv-gnarfi} is then truncated based on a the tolerance error and decay rate of its elements \citep{sela2009computationally}. 
Finally, $\Sigma_{G}$ for a general GNARFI($p,[\ss],\dd$) process can be obtained by rewriting it as a GNARFI($1,[s],\dd^{\#}$) process, with $\dd^{\#} = (\dd, 0, \ldots, 0)$ the corresponding $Np$-dimensional vector. \newline

Computing $\Sigma_{G}$ from the exact GNARFI covariances involves matrix inversions which can cause numerical instability, making the parameter estimation process more difficult. For this reason, we also propose maximising the conditional likelihood \citep{tsay2010maximum, kechagias2020modeling}. This approach leverages the relation between a GNARFI model and its FIWN component noted in \eqref{eq:gnarfi_equiv} and the conditional likelihood principle \citep{box2015time}.

Let $\set{\xx_{t}}_{t=1}^{T+p}$ be $(T+p)$ observations from a GNARFI($p,[\ss],\dd$) time series. By definition, $\ZZ_{t} = \gA(L)\XX_{t}$, so we can construct $\set{\zz_{t}}_{t=1}^{T}$ from the original data and the GNAR coefficient matrices ${\gA}_1,\ldots, {\gA}_p$. Since $\set{\zz_{t}}_{t=1}^{T}$ arises from a FIWN($\dd$) process, the corresponding likelihood is
\begin{equation}\label{eq:condgnarfi-like}
    \mathcal{L}_{CG}(\ttheta; \zz) = (2\pi)^{-\frac{TN}{2}} \left| \Sigma_{FIWN} \right|^{-\frac{1}{2}}  \expfun{-\frac{1}{2} (\zz - \hat{\zz})^{\top} \Sigma^{-1}_{FIWN} (\zz - \hat{\zz})}.
\end{equation}
Here, $\hat{\zz} = (\hat{\zz}_1, \ldots, \hat{\zz}_{N})^{\top}$, and each $\hat{\zz}_{i} = (\hat{z}_{i,1}, \ldots, \hat{z}_{i,T})^{\top}$ is the one-step ahead predictor vector of $\zz_{i} = (z_{i,1}, \ldots, z_{i,T})^{\top}$.
The choice of conditioning values depends on the series \citep{box2015time}, and extra care is needed when parameters are near the stationarity boundary. In line with using unconditional means for initialisation, we set the first $p$ observations $\set{\xx_{t}}_{t=1}^{p}$ to zero.

The advantage of this method is that the likelihood \eqref{eq:condgnarfi-like} depends only on the FIWN autocovariance matrix, which is straightforward to evaluate:
\begin{equation*}
    \Sigma_{FIWN} = (\eta(h))_{h=-T+1}^{T-1} = (\Sigma_{\vveps} \odot \phi(h))_{h=-T+1}^{T-1}.
\end{equation*}
As noted by \cite{tsay2010maximum}, this yields a one-step procedure in which all parameters in $\ttheta$ are estimated simultaneously. Moreover, it remains an exact method: the likelihood uses no estimated covariance matrix, and the transformation to $\set{\zz_{t}}_{t=1}^{T}$ relies only on initial values of the GNAR coefficients. In what follows, we refer to this estimation approach as the {\em conditional-GNARFI} method.

\paragraph{Quadratic Form and Determinant Evaluation.}
Once $\Sigma_{F}$, $\Sigma_{G}$, or $\Sigma_{FIWN}$ is estimated, calculating the quadratic term and the determinant in \eqref{eq:fignar-like}, \eqref{eq:gnarfi-like} or \eqref{eq:condgnarfi-like} become computationally demanding as the network dimension ($N$) and process size ($T$) increase. As a result, deriving exact maximum likelihood estimates is difficult, and various optimisation strategies have been proposed in the literature over the years to circumvent this. 

We propose to use the multivariate preconditioned conjugate gradient method to compute the quadratic form exactly \citep{sela2009computationally, pai2009multivariate}. This approach allows us to avoid the Cholesky factorisation of $\Sigma$ and the computation of its inverse, which is particularly challenging in high-dimensional settings. 

For computing the determinant, we employ \cite{sowell1989decomposition}'s exact formula. This method is based on the Durbin-Levinson algorithm, which provides a decomposition of the autocovariance matrix. By leveraging this approach, the computational cost is reduced to $\mathcal{O}(T^2)$, in contrast to the standard $\mathcal{O}(T^3)$ required for computing the determinant. In a similar fashion to \cite{sela2009computationally}, we also propose a spline interpolation to approximate the determinant. However, in contrast to their regression-based curve-fitting method, our spline-based approximation is a non-parametric approach that provides precise control over the approximation error. Despite its theoretical advantages, we observe empirically that this approximation does not always reduce computational time as the number of nodes increases, primarily due to the complexity of the preconditioned conjugate gradient method. For this reason, and because it ensures exact evaluation, we prefer \cite{sowell1989decomposition}'s formula for computing the determinant. See Appendix \ref{app:det-approx} for further details. \newline

To summarise, the only approximation we employ is the truncation of the infinite sum in the autocovariance formulas \eqref{eq:acv-fignar} and \eqref{eq:acv-gnarfi}. Naturally, this approximation can be made negligible by allowing the truncation index to approach infinity, albeit at the expense of increased computational time in practice. Therefore, our method constitutes an exact maximum likelihood estimation. This implies that the theoretical properties, such as asymptotic normality and consistency of the estimator, would follow from general maximum likelihood theory when $\set{\XX_{t}}_{t}$ is a stationary and stable Gaussian process. 

%-------------------------------------------------------------------
\subsection{Order and Model Selection}\label{sec:orderselection}
Before fitting network models \eqref{eq:gnar}, the network order must be identified. In the GNAR setting, \cite{knight2020generalized} propose using information criteria, BIC and AIC, to jointly select the maximal AR lag $p$ and the maximal neighbourhood set at each lag $\ss$.
We adopt the same strategy for the long memory network models \eqref{eq:fignar} and \eqref{eq:gnarfi}. The optimal order minimise the BIC \citep{schwarz1978estimating}, defined as
\begin{equation*}
    \text{BIC}(M) = -2\log(\hat{\mathcal{L}}) + M \log T,
\end{equation*}
where $\hat{\mathcal{L}}$ is the maximised likelihood, and $M$ is the number of parameters. We remind the reader that $M$ ranges from $M = Np + C \sum_{j=1}^{p} s_j + 2N$ for the individual-$\{\alpha,d,\sigma^2\}$ specification down to $M = p + C \sum_{j=1}^{p} s_j + 2$ for the global-$\{\alpha,d,\sigma^2\}$ case.
An alternative order selection criterion is the AIC (see e.g. \cite{akaike1974new}), namely
\begin{equation*}
    \text{AIC}(M) = -2\log(\hat{\mathcal{L}}) + 2M,
\end{equation*}
with $\hat{\mathcal{L}}$ and $M$ assuming the same values as for BIC.

We also use these criteria to distinguish between FIGNAR and GNARFI. Since the AIC and BIC depend on the log-likelihood, which differs by framework-  \eqref{eq:fignar-like} for FIGNAR and \eqref{eq:gnarfi-like} (or \eqref{eq:condgnarfi-like}) for GNARFI-fitting alternative specifications, selecting the smallest criterion value identifies the preferred model for the considered data.

\paragraph{Example (DGP1, continued).} To empirically assess our selection methods, we apply the criteria to $K=100$ replicates, each of length $T=200$, simulated under the DGP1 parameter setup on the {\tt fiveNet} network. Tables \ref{tab:data-model-mismatch} and \ref{tab:order} in Appendix \ref{app:extrasims} report how often BIC and AIC are minimised for both FIGNAR and GNARFI models, with GNARFI being also estimated via the conditional method.
Table \ref{tab:data-model-mismatch} concerns model selection: we evaluate the criteria for FIGNAR models of different orders fitted to data simulated from FIGNAR(1,[1],$\dd$) and GNARFI(1,[1],$\dd$) with DGP1 parameters. Overall, BIC selects the correct model 57 times and AIC 58 times, and the correct order is identified 48 and 30 times, respectively.
We repeat the same exercise for GNARFI models fitted to DGP1-FIGNAR(1,[1],$\dd$) and DGP1-GNARFI(1,[1],$\dd$) data. In this case, both BIC and AIC always favour GNARFI specifications over FIGNAR ones, with BIC detecting the correct order more often than AIC. The criteria show strong performance in identifying the correct order, (1,[1],$\dd$), particularly when parameters are estimated using the exact likelihood method. This may reflect the fact that, as previously noted, the exact likelihood approach is more sensitive to numerical instability and fails to converge more frequently for alternative orders or the alternative model.

If the correct model is fixed, we see in Table \ref{tab:order} that the BIC selects the correct order most of the time, and with better performance than the AIC, which sometimes picks the corresponding individual-$\alpha$ case (recall that DGP1 is a global-$\alpha$ model). We notice both selection methods perform worse with the conditional estimation method for the GNARFI model, as the global-$\alpha$ with (2,[1,0]) order is frequently chosen as the preferred model, and according to the AIC, it should be the correct order. In fact, if we examine the estimated parameters, we observe that $\hat{\alpha}_{2}$ is equal to $0.004$ on average, indicating that the chosen model includes a very weak autoregressive component at lag $p=2$. 

Empirically, the BIC appears to be more reliable than the AIC, reflecting the consistency of the BIC in univariate long memory settings, unlike that of the AIC \citep{beran1998unified, huang2022consistent}.

%-------------------------------------------------------------------
\section{Simulation Study}\label{sec:simstudy}
In this section, we assess the efficiency and quality of our estimation framework through several simulation studies. For the results here, as well as for the data analysis in Section \ref{sec:application}, simulation and estimation for our long memory network time series models were implemented in the {\em R} statistical computing environment \citep{Rcore}, using modifications to the code from the {\tt GNAR} package (\cite{gnar}, see also \cite{knight2020generalized}).  
For the estimation scheme, we use the \texttt{Rsolnp} \citep{Rsolnp} package to numerically solve the maximisation problem. Specifically, we implement the popular iterative optimisation algorithm Broyden-Fletcher-Goldfarb-Shanno (BFGS) quasi-Newton method while ensuring the feasibility region of the parameters is controlled. \newline

Our simulation study considers three data generating processes, each defined on the {\tt fiveNet} network in Figure \ref{img:fivenet}, with i.i.d. standard normal innovations at each node and diagonal covariance $\Sigma_{\vveps} = I$.
For all simulations, we fix the correct order at $(p,[s]) = (1,[1])$ and use a proper model specification.

The first case, DGP1, involves a global-$\alpha$ GNAR(1,[1]) with $\alpha_{1}=0.35$ and $\beta_{1,1}=0.2$, and long memory coefficients $\dd = (0.05, 0.15, 0.25, 0.35, 0.45)$. This corresponds to global-$\alpha$, individual-$d$, and individual-$\sigma^2$ specification, and is designed to evaluate empirical performance under heterogeneous differencing parameters.
Secondly, DGP2 shares the same GNAR(1,[1]) parameters as DGP1, but with a flat long memory profile, i.e. $d_i=0.25$ for all $i=1,\ldots,5$. Here we compare estimation under global-$\alpha$ and individual-$\sigma^2$ for FIGNAR and GNARFI models, using either global-$d$ or individual-$d$ option, to assess robustness when allowing more flexibility in the long memory component.
Finally, DGP3 specifies $\aalpha_{1}=(-0.4, 0.3, 0.3, 0.2, -0.3)$, $\beta_{1,1}=0.4$, and $\dd = (0.05, 0.15, 0.25, 0.35, 0.45)$. This is the most flexible scenario due to the individual-$\alpha$ and individual-$d$ specifications, and it is used to compare performance under global-$\sigma^2$ versus individual-$\sigma^2$.

For each proposed model and parameter specification, we simulate $K = 100$ realisations with different lengths, $T=200, 500, 1000$. We evaluate the performance of the estimations by the average mean squared error metric 
\begin{equation*}
    \operatorname{AMSE}(\ttheta)= \frac{1}{K} \sum_{k=1}^{K} \operatorname{MSE}_{k}(\ttheta) = \frac{1}{KM} \sum_{k=1}^{K} \sum_{m=1}^{M} (\theta_{m} - \hat{\theta}_{m,k})^{2}
\end{equation*}
where $\hat{\ttheta}_k$ is the $M$-dimensional estimated vector of parameters for the $k$-th iteration. \newline

Table \ref{tab:mse-fignar} reports the average errors in the FIGNAR estimations of their corresponding parameters across the various DGPs. 
Similarly, Table \ref{tab:mse-gnarfi} shows the estimation errors for the GNARFI model, using either the standard likelihood \eqref{eq:gnarfi-like} or the conditional likelihood \eqref{eq:condgnarfi-like} approach.
As expected, the performance of the estimates worsens as the number of parameters increases, evidenced by comparing the errors of the global (`gl.') and individual models (`ind.'). 

\begin{table}[!ht]
\centering
\begin{NiceTabular}{c||>{\centering\arraybackslash}p{2.0cm}||>{\centering\arraybackslash}p{2.0cm}>{\centering\arraybackslash}p{2.0cm}||>{\centering\arraybackslash}p{2.0cm}>{\centering\arraybackslash}p{2.0cm}}
    & DGP1 & \Block{1-2}{DGP2} && \Block{1-2}{DGP3} \\
    $T$ & & gl. $d$ & ind. $d$ & gl. $\sigma^2$ & ind. $\sigma^2$ \\
    \hline\hline
    200 & 8.415 & 7.898 & 7.911 & 12.386 & 16.940 \\
    500 & 3.842 & 3.355 & 3.248 & 6.629 & 9.821 \\
    1000 & 2.524 & 2.027 & 1.890 & 4.977 & 7.588 \\
\end{NiceTabular}
\caption{AMSE ($\times 10^3$) for maximum likelihood estimates of FIGNAR model under the various parameter settings.}\label{tab:mse-fignar}
\end{table}

\begin{table}[!ht]
\centering
\begin{subtable}[t]{\linewidth}
\centering
\begin{NiceTabular}{cc||>{\centering\arraybackslash}p{2.0cm}||>{\centering\arraybackslash}p{2.0cm}>{\centering\arraybackslash}p{2.0cm}||>{\centering\arraybackslash}p{2.0cm}>{\centering\arraybackslash}p{2.0cm}}
    & & DGP1 & \Block{1-2}{DGP2} && \Block{1-2}{DGP3} \\
    Method & $T$ & & gl. $d$ & ind. $d$ & gl. $\sigma^2$ & ind. $\sigma^2$ \\
    \hline\hline
    Standard & 200 & 8.743 & 7.791 & 7.913 & 12.162 & 18.936 \\
    & 500 & 3.590 & 3.125 & 3.106 & 6.137 & 11.869 \\
    & 1000 & 2.213 & 1.770 & 1.720 & 3.271 & 8.939 \\
& & & & & & \\
% \hline
% & & & & & \\
    Conditional & 200 & 9.416 & 7.660 & 8.350 & 8.034 & 10.425 \\
    & 500 & 3.570 & 2.867 & 3.064 & 3.338 & 4.289 \\
    & 1000 & 1.931 & 1.484 & 1.557 & 1.435 & 1.928 \\
\end{NiceTabular}
\end{subtable}
\caption{AMSE ($\times 10^3$) for maximum likelihood estimates of GNARFI model under the various parameter settings, using either the standard (top) or the conditional likelihood (bottom) approach.}\label{tab:mse-gnarfi}
\end{table}

\begin{figure}[!ht]
\centering
% Row 1
\begin{subfigure}[t]{0.48\textwidth}
    \includegraphics[width=\linewidth]{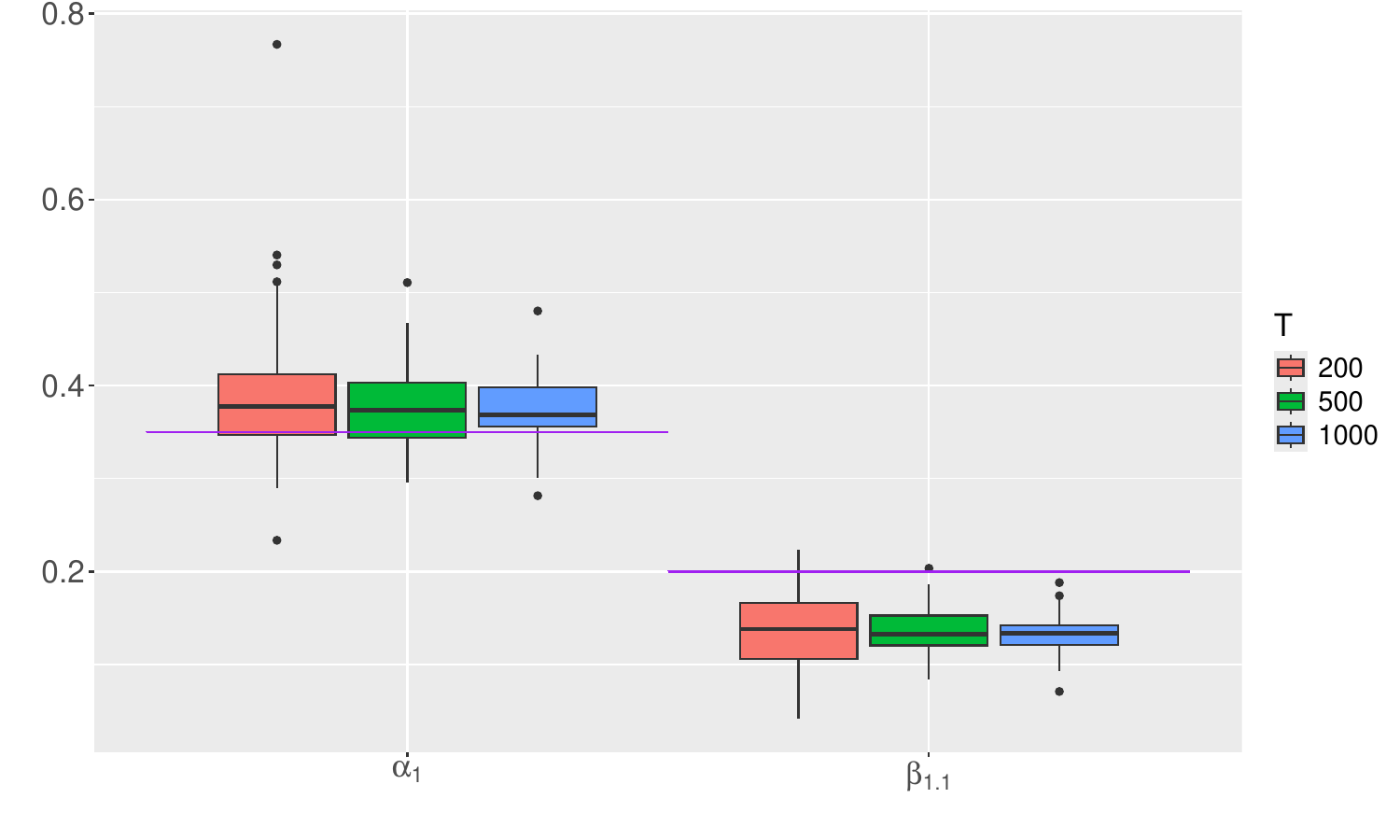}
\end{subfigure}\hfill
\begin{subfigure}[t]{0.48\textwidth}
    \includegraphics[width=\linewidth]{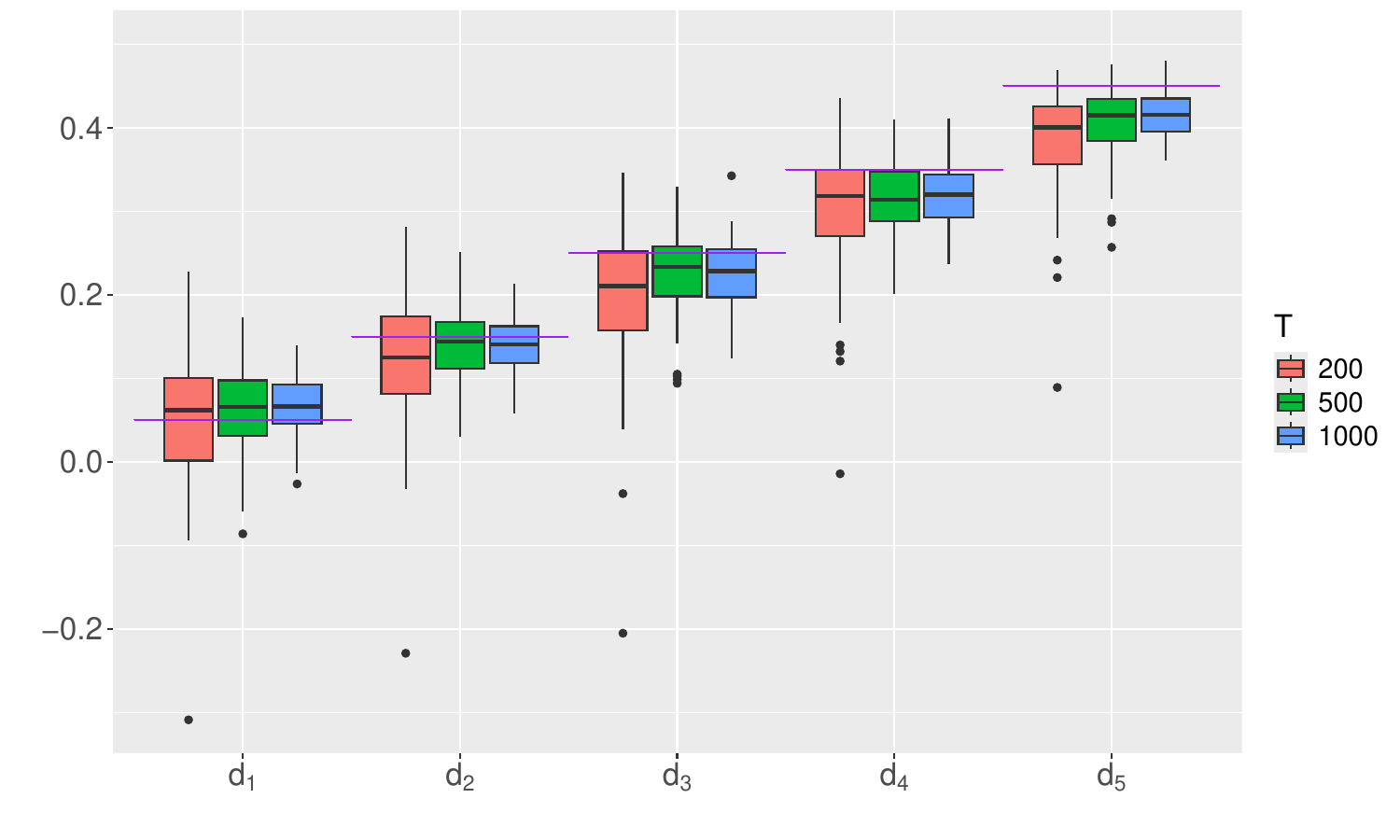}
\end{subfigure}

\medskip
% Row 2
\begin{subfigure}[t]{0.48\textwidth}
    \includegraphics[width=\linewidth]{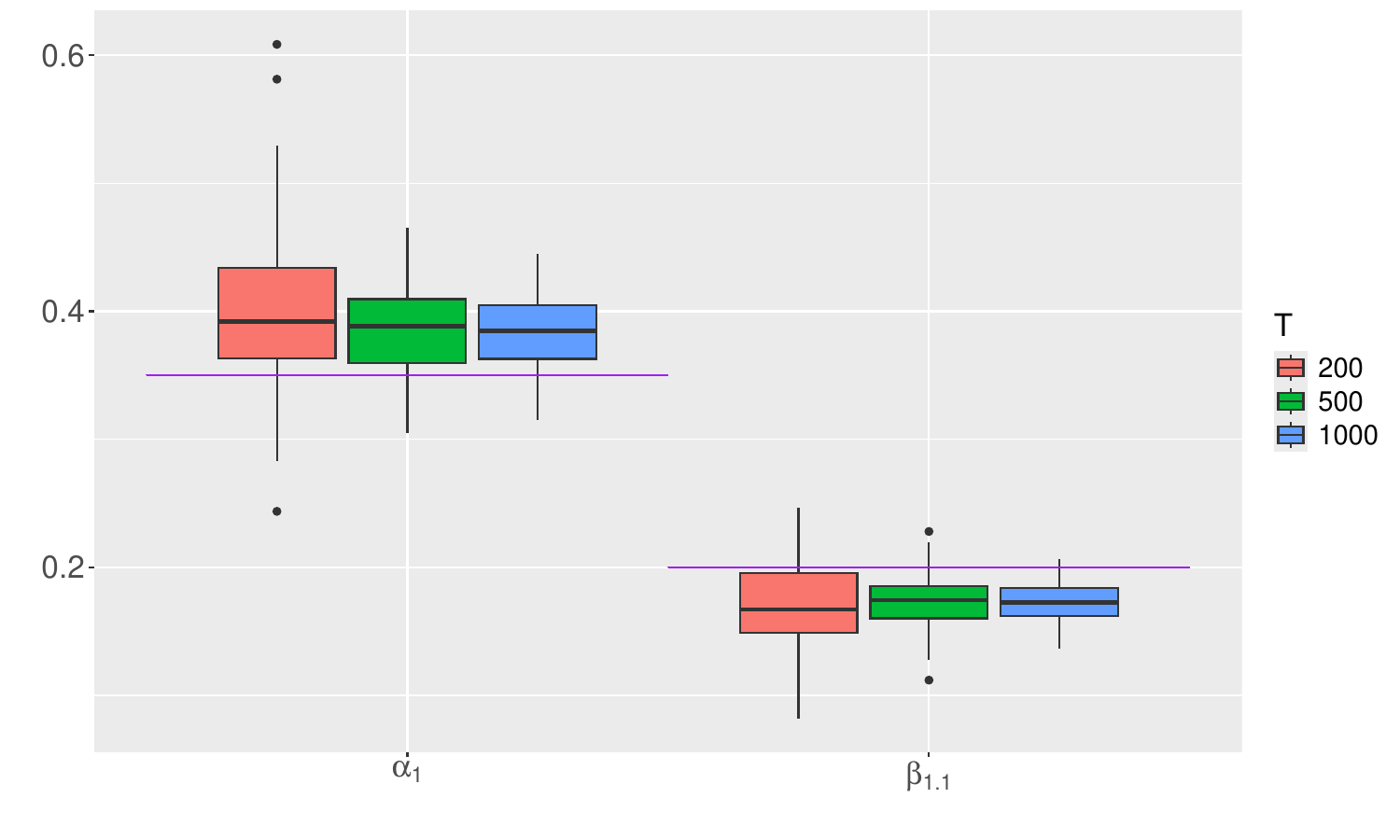}
\end{subfigure}\hfill
\begin{subfigure}[t]{0.48\textwidth}
    \includegraphics[width=\linewidth]{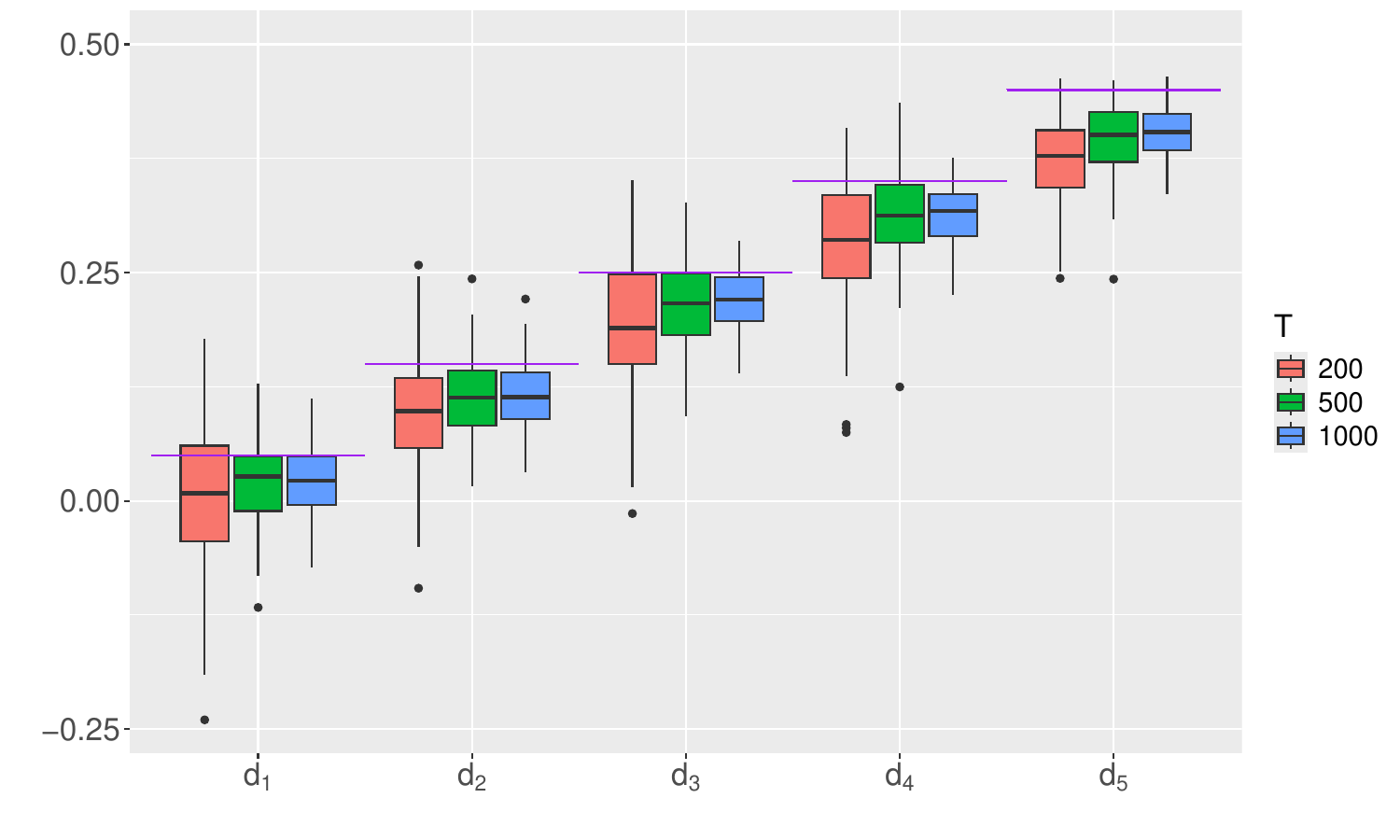}
\end{subfigure}

\medskip
% Row 3
\begin{subfigure}[t]{0.48\textwidth}
    \includegraphics[width=\linewidth]{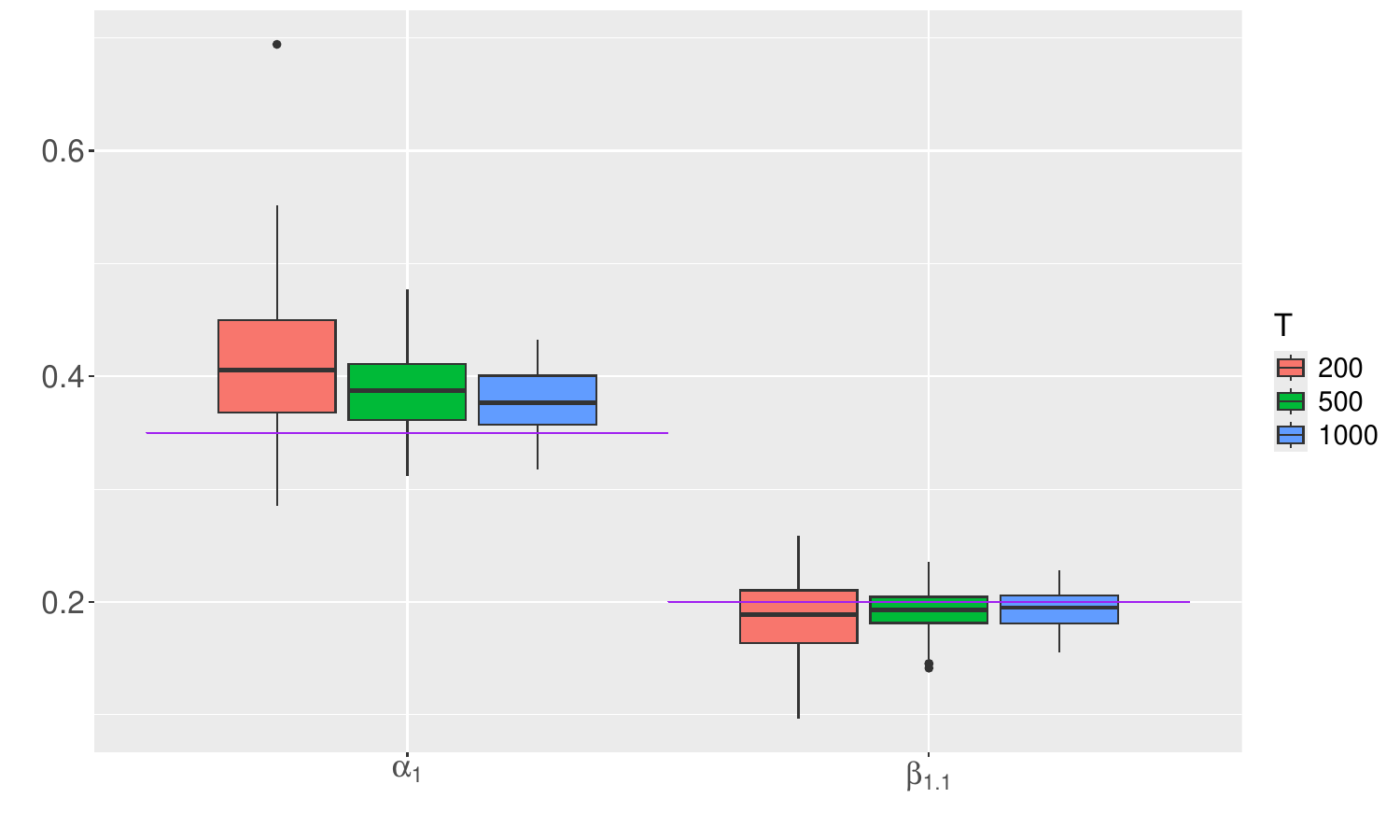}
\end{subfigure}\hfill
\begin{subfigure}[t]{0.48\textwidth}
    \includegraphics[width=\linewidth]{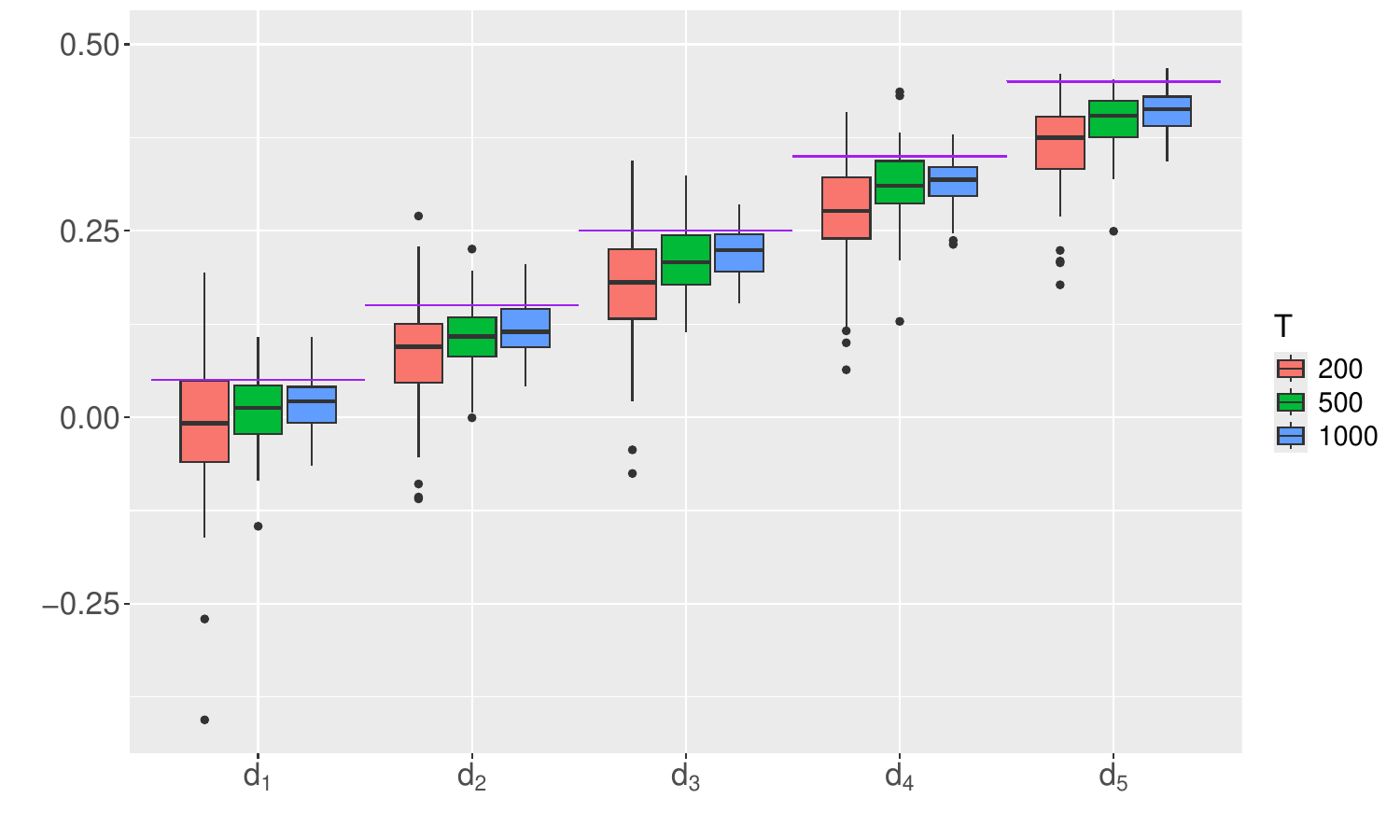}
\end{subfigure}
\caption{Boxplots of the estimated $\{\aalpha,\bbeta\}$ (right) and $\dd$ (left) for DGP1. Top: FIGNAR(1,[1],$\dd$) parameter estimation; middle: GNARFI(1,[1],$\dd$) parameter estimation with standard method; bottom: GNARFI(1,[1],$\dd$) parameter estimation with conditional method. The purple horizontal lines indicate the true values.}\label{img:DGP1}
\end{figure}

Similarly, the error magnitude decreases as the number of observations $T$ increases. Figure \ref{img:DGP1} shows boxplots of the short and long memory parameter estimation over the $K=100$ realisations for DGP1. We indeed observe that the boxplots become progressively narrower as the sample size increases, suggesting that the usual asymptotic properties of MLEs may hold for our estimators.

Boxplots of the estimation for the noise covariance parameters shown in Figure \ref{img:DGP1sig} in Appendix \ref{app:extrasims} indicate that the conditional GNARFI estimation approach is more stable than the other estimation scheme. Similar patterns can be seen from estimation of the parameters of DGP2 and DGP3; the respective boxplots are shown in Appendix \ref{app:extrasims}, Figures \ref{img:DGP2-gld}-\ref{img:DGP3}.

We also assess the model in a larger setting, namely on a 10-dimensional network (Figure \ref{img:tenNet}), by simulating $K = 50$ realisations under the three parameter specifications, with lengths $T = 200, 500, 1000$. The long memory parameters $\dd$ of DGP1 and DGP3 are again increasing with respect to the node index, $d_{i} = 0.05 + 2(i-1)/45$, and the GNAR parameters $\aalpha_1$ and $\beta_{1,1}$ of DGP3 are randomly generated so that the stationarity condition \eqref{eq:stationarity} is not violated. As expected, computation time increases, but the numerical optimisation still converges. Table \ref{tab:mse-ga-10d} reports the AMSEs, which remain small and decrease as $T$ increases.
Note that estimations for DGP3 when $T=200$ are worse than in the other cases, due to the relatively large number of parameters ($M=22$) and the short time series length. However, they improve substantially once more observations are available, for example at $T=1000$.

\begin{table}[!ht]
\centering    
\begin{NiceTabular}{ccc||>{\centering\arraybackslash}p{2cm}>{\centering\arraybackslash}p{2cm}>{\centering\arraybackslash}p{2cm}}
    Model & Method & T & DGP1 & DGP2 & DGP3 \\
    \hline\hline
    FIGNAR(1,[1],$\dd$) & Standard & 200 & 9.115 & 7.962 & 21.565 \\
    & & 500 & 3.399 & 3.079 & 9.047 \\
    & & 1000 & 2.189 & 1.886 & 3.652\\
    \hline
    GNARFI(1,[1],$\dd$) & Standard & 200 & 9.047 & 7.537 & 17.983 \\
    & & 500 & 3.696 & 3.171 & 6.534 \\
    & & 1000 & 2.222 & 1.911 & 3.110 \\
    & & & & \\
    & Conditional & 200 & 9.225 & 7.476 & 17.393 \\
    & & 500 & 3.398 & 2.863 & 4.575 \\
    & & 1000 & 1.793 & 1.543 & 2.219 \\
\end{NiceTabular}
\caption{AMSE ($\times 10^3$) for maximum likelihood estimates of for the 10-dimensional FIGNAR and GNARFI models of the three data generating processes.}\label{tab:mse-ga-10d}
\end{table} 
\medskip
     
We finally compare the empirical computational times of the estimation approaches and how they scale with the number of parameters and the time-series length. Specifically, we simulate 5-dimensional network data under DGP2 and estimate the models for all combinations of global and individual ${\alpha, d, \sigma^2}$ specifications. This allows us to compare runtimes not only as the number of parameters increases, but also according to which parameters are being estimated. Results are reported in Table \ref{tab:comp times} in Appendix \ref{app:extrasims}.
In general, GNARFI models are estimated faster than FIGNAR models, with the conditional likelihood approach being the fastest. As expected, the computational cost depends more strongly on the time length $T$ than on the number of parameters $M$. Moreover, even when $M$ is the same, the individual-$\alpha$ and individual-$d$ specifications require more time than the individual-$\sigma^{2}$ specification, indicating that computational burden depends not only on network size but also on which parameters are estimated.

%-------------------------------------------------------------------
%  FORECASTING
%-------------------------------------------------------------------
\section{Forecasting Long Memory Processes}\label{sec:forecasting}

\subsection{Forecasting When the Network is Known}
In many application fields, a task of particular interest is forecasting.  For our proposed FIGNAR and GNARFI models, various strategies can be employed to compute $h$-step ahead predictions, based on the estimated parameters. In what follows, we describe our approach to forecasting long memory network time series data via three prediction avenues, as follows.
Recall that for any time lag $h > 0$, and given all available data $\XX$ arranged as an $NT$-dimensional vector, the best linear predictor is given by $\hat{\XX}_{T+h} := \mathbb{E}(\XX_{T+h} | \XX)$. 

\noindent{\bf (1- DLF).} A first approach is based on the Durbin-Levinson (DL) algorithm. Given the estimated autocovariance matrix, we can use the DL algorithm to compute the coefficients $\Phi^{(s)}_{1}, \ldots,\Phi^{(s)}_{s}$ for the projection of $\XX_{t}$ onto the previous $s$ lagged values $\XX_{t-1}, \ldots, \XX_{t-s}$. Consequently, the one-step-ahead predictor is given by
\begin{equation}\label{DL formula}
    \hat{\XX}_{T+1} = \sum_{s=1}^{T} \Phi^{(T)}_{s} \XX_{T+1-s}.
\end{equation}
As in \cite{kechagias2020modeling}, we can recursively apply the one-step ahead predictors to obtain a similar expression for $\hat{\XX}_{T+1}$.
Although it is more efficient than the innovation algorithm, which is commonly used in forecasting tasks, the Durbin-Levinson algorithm still remains computationally demanding since the coefficient matrices $\{\Phi_{s}^{(T)}\}_s$ are computed in $\mathcal{O}(T^2)$ steps. We denote this prediction type by DLF for the \underline{D}urbin-\underline{L}evinson \underline{F}ormula \eqref{DL formula}.  

\noindent{\bf (2- EF).} As an alternative, recall that for any Gaussian time series, we have 
\begin{equation}\label{expectation formula}
    \mathbb{E}\left(\XX_{T+h} | \XX \right) = \mathrm{Cov}\left(\XX, \XX_{T+h}\right) \Sigma^{-1}\XX, 
\end{equation}
where $\Sigma$ is the $(NT\times NT)$-dimensional covariance matrix of $\XX$.
Hence, we obtain $\hat{\XX}_{T+h}$ using the above relation, where we employ the preconditioned conjugate gradient algorithm to evaluate $\Sigma^{-1} \XX$ \citep{sela2009computationally}. In what follows, we shall refer to this forecasting approach as EF, for the \underline{E}xpectation \underline{F}ormula \eqref{expectation formula}.  

\noindent{\bf (3- RF).} Finally, explicit approximate recursive formulas can be derived by exploiting the power series representation of the long memory filter. Given that $(1-L)^{\dd} = \sum_{j=0}^{\infty} D_j L^{j}$ with $D_0 = I$ and $D_j=(-1)^{j} \operatorname{diag}\left( \frac{\Gamma(d_{1}+1)}{\Gamma(j+1) \Gamma(j-d_{1}+1)}, \ldots, \frac{\Gamma(d_{N}+1)}{\Gamma(j+1) \Gamma(j-d_{N}+1)} \right)$ for $j\ge 1$, the FIGNAR($1,[\ss], \dd$) model can be rewritten as 
\begin{align*}
    (1 - L)^{\dd}(I - A_{1} L)\XX_{t} = \vveps_{t} & \Leftrightarrow \left(\sum_{j=0}^{\infty} D_j L^{j}\right) (I - A_{1} L)\XX_{t} = \vveps_{t} \\
    & \Leftrightarrow \left(\sum_{j=0}^{\infty} D_j L^{j} - \sum_{j=0}^{\infty} D_j A_{1} L^{j+1}\right) \XX_{t} = \vveps_{t} \\
    & \Leftrightarrow \left( I + \sum_{j=1}^{\infty} D_j L^{j} - \sum_{j=1}^{\infty} D_{j-1} A_{1} L^{j}\right) \XX_{t} = \vveps_{t} \\
    & \Leftrightarrow \left( I + \sum_{j=1}^{\infty} (D_j - D_{j-1} A_{1}) L^{j}\right) \XX_{t} = \vveps_{t} \\
    & \Leftrightarrow \XX_{t} = \sum_{j=1}^{\infty} (D_{j-1} A_{1} - D_j) \XX_{t-j} + \vveps_{t},
\end{align*}
where we set $D_{-1}$ to be the zero matrix. Then, based on the past observations and the estimated parameters, we propose as the $h$-step ahead predictor of the FIGNAR($1,[s],\dd$) process
\begin{equation}\label{recursive formula}
    \hat{\XX}_{T+h} = \sum_{j=1}^{T+h-1} (\hat{D}_{j-1} \hat{A}_{1} - \hat{D}_j) \XX_{T+h-j}.
\end{equation}
Note this expression holds similarly for the GNARFI($1,[s],\dd)$ model, just with swapped ordering $\hat{A}_{1}\hat{D}_{j-1}$. We refer to this approach as RF, the \underline{R}ecursive \underline{F}ormula \eqref{recursive formula}.

\paragraph{Forecasting example (DGP1, continued).} To evaluate the relative performance of the three proposed forecasting approaches, we compare the prediction errors in a one-step ahead forecasting task on simulated realisations of our proposed long memory network time series models. Data are simulated under the three data generating processes in Section \ref{sec:simstudy}, but we report results for DGP1 only, as the other two yield similar findings.
We proceed as follows. For each series simulated under either the FIGNAR or GNARFI framework, we estimate parameters by maximising the FIGNAR likelihood \eqref{eq:fignar-like} or, for GNARFI, the exact likelihood \eqref{eq:gnarfi-like} and the conditional likelihood \eqref{eq:condgnarfi-like}. 
Using the estimated parameters, we produce one-step-ahead forecasts with each approach and compute mean squared prediction errors (MSPEs). This is repeated for $K=100$ replications. Table \ref{tab:forecast} reports the average MSPEs across replications. 
For comparison, we also include the forecast errors based on the vector models (FIVAR and VARFI). In this case, parameters are estimated via the parametric methods of \cite{sela2009computationally} and \cite{tsay2010maximum}, and we obtain the one-step-ahead forecasts and respective errors. Note Table \ref{tab:forecast} reports the `-Stand' and `-Cond' notation to distinguish the standard and the conditional-likelihood GNARFI estimation methods, respectively.
Table \ref{tab:forecast} shows that the three approaches deliver similar prediction accuracy, and the vector models’ errors are comparable to those of the network models. However, the vector-based estimators recover the long-memory parameters poorly. The AMSEs of $\boldsymbol{\hat{d}}$ for FIVAR are 0.0725, 0.0371, and 0.0352 for $T = 200, 500, 1000$, respectively, compared to 0.0072, 0.0030, and 0.0017 for the network models. Additionally, the computational time for vector methods is approximately 10 times greater than that of network methods. A similar pattern is observed in forecasts within the GNARFI framework. Furthermore, we note that vector methods encounter difficulties in accurately estimating the parameters, particularly the conditional likelihood method, which either fails to converge or does not properly estimate the AR component. Specifically, it struggles to explore the parameter space thoroughly and frequently returns the same values. 

Because of the similar performances, we focus the remaining forecast analysis on the EF method. 
   
\begin{table}[!ht]
\setlength{\tabcolsep}{13pt} % Adjusts space between columns
 %%% FIGNAR %%%
\centering
\begin{NiceTabular}{cl||lll}%{@{}c!{\quad}cccccc!{\quad}cccccc@{}}
    && \Block{1-3}{Forecasting Method}  \\
    T & Method & DLF & EF & RF \\
    \hline\hline
%%% FIGNAR - FIVAR %%%
    % 200 obs
    200 & FIGNAR & 0.911 & 0.911 & 0.912 \\  
    & FIVAR & 0.951 & 0.951 & 0.950  \\
    \hline
    % 500 obs  
    500 & FIGNAR & 1.044 & 1.046 & 1.046 \\  
    & FIVAR & 1.054 & 1.056 & 1.055 \\  
    \hline
    % 1000 obs  
    1000 & FIGNAR & 0.935 & 0.937 & 0.937 \\
    & FIVAR & 0.967 & 0.967 & 0.967 \\
    \hline\hline
%%% GNARFI %%%
    % 200 obs  
    200 & GNARFI-Stand & 0.870 & 0.870 & 0.872 \\ 
    & GNARFI-Cond & 0.875 & 0.875 & 0.877 \\  
    & VARFI-Stand & 1.066$^{\star\star}$ & 1.066$^{\star\star}$ & 1.057$^{\star\star}$ \\ 
    & VARFI-Cond & 0.985$^{\star}$ & 0.985$^{\star}$ & 0.981$^{\star}$ \\ 
    \hline
    % 500 obs  
    500 & GNARFI-Stand & 1.003 & 1.005 & 1.005 \\ 
    & GNARFI-Cond & 1.004 & 1.006 & 1.006 \\ 
    & VARFI-Stand & 1.057$^{\star\star}$ & 1.058$^{\star\star}$ & 1.058$^{\star\star}$ \\ 
    & VARFI-Cond & 1.079$^{\star\star}$ & 1.080$^{\star\star}$ & 1.080$^{\star\star}$ \\ 
    \hline
    % 1000 obs  
    1000 & GNARFI-Stand & 0.949 & 0.949 & 0.949 \\ 
    & GNARFI-Cond & 0.945 & 0.944 & 0.944 \\ 
    & VARFI-Stand & 0.988$^{\star}$ & 0.988$^{\star}$ & 0.988$^{\star}$ \\ 
    & VARFI-Cond & 0.968$^{\star\star}$ & 0.968$^{\star\star}$ & 0.968$^{\star\star}$ \\  
\end{NiceTabular}
\caption{MSPE of the 1-step ahead predictions of data generated according to DGP1; $\star$ and $\star\star$ indicate if some simulations did not converge (less and more than $5\%$, respectively).} \label{tab:forecast}
\end{table}
\bigskip

We also conduct a rolling-window forecast analysis. A comparison of the errors presented in Table \ref{tab:forecast-rw} and the corresponding boxplots (Figure \ref{img:forecast-rw}) in Appendix \ref{app:extraforecast} shows the stability of our proposed prediction methods. Moreover, the predictive accuracy corresponding to GNARFI data is generally better than that of the FIGNAR data.

\begin{table}[!ht]
\centering
{\setlength{\tabcolsep}{6pt}\small
\begin{NiceTabular}{l||*{10}{c}}
    & \Block{1-9}{Predicted observation index} \\%{T of predicted observation} \\
    Method & 201 & 202 & 203 & 204 & 205 & 206 & 207 & 208 & 209 & 210 \\
    \hline\hline
    FIGNAR & 0.911 & 1.167 & 1.002 & 1.016 & 1.090 & 0.908 & 0.931 & 0.918 & 1.037 & 1.162 \\
    FIVAR & 0.951 & 1.225 & 1.041 & 1.074 & 1.159 & 0.983 & 0.976    & 0.947 & 1.098 & 1.230 \\
    \hline\hline
    GNARFI-Stand & 0.870 & 0.910 & 0.968 & 1.029 & 0.983 & 0.994 & 1.040 & 0.966 & 0.930 & 1.115 \\
    GNARFI-Cond & 0.875 & 0.908 & 0.974 & 1.026 & 0.987 & 0.985 & 1.042 & 0.966 & 0.929 & 1.120 \\
    VARFI-Stand$^{\star\star}$ & 1.066 & 0.865 & 1.067 & 1.087 & 1.091 & 1.157 & 0.861 & 1.003 & 1.144 & 1.123 \\
    VARFI-Cond$^{\star}$ & 0.985 & 0.943 & 1.060 & 1.037 & 1.098 & 1.105 & 0.965 & 1.012 & 1.099 & 1.108 \\
\end{NiceTabular}}
\caption{MSPE of rolling window forecasts across $100$ replicates each of length $T=200$, generated according to DGP1; $\star$ and $\star\star$ indicate if some simulations did not converge (less and more than $5\%$, respectively).}\label{tab:forecast-rw} 
\end{table}

Forecasts for $h > 1$ also yield similar outcomes, with network predictions generally proving more accurate than vector-based ones. In Table \ref{tab:forecast-long} and Figure \ref{img:forecast-long}, we report results for prediction horizons $h=1$ to $10$ for DGP1 with $200$ observations only, since similar conclusions can be drawn for $T = 500$ or $T = 1000$. Overall, the network models are not only numerically more stable, but also tend to correspond to better forecasts than their counterparts which do not include network information. When comparing individual node estimates, network models perform better, particularly in predicting the value at node 5, which corresponds to the variable with the largest long memory coefficient. This is expected, as vector methods struggle to accurately estimate the long memory component and, consequently, perform worse when the long memory effect is significant. 

\begin{table}[!ht]
\centering
{\setlength{\tabcolsep}{6pt}\small
\begin{NiceTabular}{l||*{10}{c}}
    & \Block{1-9}{Predicted observation index} \\%{T of predicted observation} \\
    Method & 201 & 202 & 203 & 204 & 205 & 206 & 207 & 208 & 209 & 210 \\
    \hline\hline
    FIGNAR & 0.911 & 1.489 & 1.603 & 1.638 & 1.715 & 1.759 & 1.814 & 1.865 & 2.019 & 1.946 \\
    FIVAR & 0.951 & 1.577 & 1.712 & 1.724 & 1.833 & 1.854 & 1.865 & 1.931 & 2.064 & 1.988 \\
    \hline\hline
    GNARFI-Stand & 0.870 & 1.274 & 1.543 & 1.777 & 1.803 & 2.064 & 2.142 & 2.035 & 2.098 & 2.083 \\
    GNARFI-Cond & 0.875 & 1.278 & 1.547 & 1.779 & 1.814 & 2.071 & 2.156 & 2.040 & 2.094 & 2.081 \\
    VARFI-Stand$^{\star\star}$ & 1.066 & 1.384 & 1.727 & 1.961 & 2.001 & 2.258 & 2.368 & 2.261 & 2.349 & 2.302 \\
    VARFI-Cond$^{\star}$ & 0.985 & 1.276 & 1.560 & 1.775 & 1.816 & 2.068 & 2.097 & 1.993 & 2.046 & 2.071 \\
\end{NiceTabular}}
\caption{MSPE of $h$-step ahead predictions with $h=1,\ldots,10$, across $100$ replicates each of length $T=200$, generated according to DGP1; $\star$ and $\star\star$ indicate if some simulations did not converge (less and more than $5\%$, respectively).} \label{tab:forecast-long} 
\end{table}

%-------------------------------------------------------------------
\subsection{Forecasting When the Network is Unknown}\label{sec:varforecast}
The primary challenge in estimating vector models is the large number of parameters, which can cause numerical instability and long computation times. Consequently, vector models do not always results in good forecasting performances.
By contrast, simpler models with fewer parameters can yield better predictions. Prior works \citep{knight2020generalized, nason2025forecasting}) show that the network framework simplifies forecasting, and GNAR models perform well even when there is no explicit underlying graph $\G$. 
Motivated by this, we also evaluate our proposed long memory network models on data randomly generated from long memory vector autoregressive models. Thanks to the parameter parsimony of \eqref{eq:fignar} and \eqref{eq:gnarfi}, which substantially reduces the complexity of models, we aim to model higher-dimensional data effectively and still obtain strong predictive performance.

By definition, the network setting requires to specify an underlying graph. When one is not available, a straightforward approach is to fit the long memory network models on a fully connected graph; here, sparsity is controlled by the global- or individual-$\alpha$ specification and by the first-stage neighbour parameters.

If one is interested in finding an appropriate graph $\G$, a brute-force strategy would involve fitting network models of varying orders across different graphs, comparing the prediction errors for each combination, and selecting the optimal order; this approach has been used to achieve good forecasting performance in standard GNAR models \citep{knight2020generalized, nason2025forecasting}. However, it may be too computationally intensive due to the extensive computations required for the long memory setting.
On the other hand, since standard AR time series estimations are  less prohibitive, high-order AR models have been used to approximate long memory models \citep{ray1993modeling, lewis1985prediction, beran2013long}. Consequently, we propose to use the brute-force strategy on several high-order GNAR models on randomly generated graphs. Although this method may offer insights regarding the neighbour dependence and choosing between individual or global-$\alpha$ models, it is not suitable for selecting the model order. Based on an empirical analysis, we notice the prediction error tends to decrease as the maximal lag $p$ gets large. In what follows, we refer to the graph found with this strategy as `GNAR($\infty$)-approx'.

Another approach to network selection involves sparse estimation of the inverse spectral density function. Specifically, the graphical lasso has been employed to estimate sparse undirected graphical models by applying an $\ell_1$-penalty to the inverse covariance matrix \citep{friedman2008sparse}. This is motivated by the fact that zero off-diagonal elements in the inverse covariance matrix indicate conditional independence between the corresponding variables, given all other variables. Graphical Lasso estimation has also been extended to the multivariate time series context \citep{jung2015graphical, tugnait2022sparse, dallakyan2022time}. Recently, \cite{baek2023local} considered sparse estimation of the long-run covariance matrix for long memory semiparametric models. Although the theory of sparse covariance estimation for multivariate long memory parametric models remains an area for further research, we propose using the so-called {\tt TSglasso} method of \cite{dallakyan2022time} to estimate the inverse spectral density function and subsequently construct the network $\G$ based on the non-zero pairs. 

\paragraph{Example.} We study the performance of \eqref{eq:fignar} and \eqref{eq:gnarfi} with individual-$\alpha$, individual-$d$, and individual-$\sigma^2$ specification when fitted to vector data. Specifically, we simulate $T=200$ observations from a 5-dimensional stationary long memory vector process of order $p=1$ with a full matrix $A_1$, a full noise covariance matrix $\Sigma_{\vveps}$, and $\dd=(0.05, 0.15, 0.25, 0.35, 0.45)$. As described above, we either impose sparsity on the parameter space with a fully-connected graph or we identify a graph using the GNAR($\infty$) approximation or {\tt TSglasso}. We fit network models with different combinations of neighbourhood dependence, and compare the MSPE based on predictions using the EF method with the MSPE of the true vector models, FIVAR(1,$\dd$) and VARFI(1,$\dd$).
Note that no graph is needed for long memory network models with order (1,[0],$\dd$), as well as for the vector models. Similarly, the fully connected case accounts for first-stage neighbours only.

As shown in Table \ref{tab:forecast-vector}, network models experience some convergence issues, particularly when fitted to the (1,[2],$\dd$) order, though only in a moderate number of cases. GNARFI estimation produces larger errors than FIGNAR cases, and the GNAR($\infty$) graph leads to the smallest errors. Furthermore, the network conditional maximisation is more stable than the vector counterpart. Overall, network predictions yield promising results, outperforming the MSPEs of the vector counterparts (first row) and, when combined with the GNAR($\infty$)-approximation graph, they results in the smallest prediction errors.

\begin{table}[!ht]
\setlength{\tabcolsep}{13pt}
\centering
\begin{NiceTabular}{l|l||l|ll}
& & & \Block{1-2}{GNARFI}\\
Order & Graph & FIGNAR & Stand & Cond\\
\hline\hline
(1,[0],$\dd$) & No graph & 8.972 & 10.441 & 10.450 \\
(1,[1],$\dd$) & Fully-connected & 8.376 & 9.559 & 9.755 \\
\textbf{(1,[1],$\dd$)} & \textbf{GNAR($\infty$) approx} & \textbf{8.146} & \textbf{9.169} & \textbf{8.941} \\
(1,[1],$\dd$) & TSglasso & 8.454 & 9.610 & 9.858 \\
(1,[2],$\dd$) & GNAR($\infty$) approx & 8.294$^{\star}$ & 9.313$^{\star}$ & 9.101$^{\star}$ \\
(1,[2],$\dd$) & TSglasso & 8.509$^{\star}$ & 11.367$^{\star}$ & 10.460$^{\star}$ \\
\hline
Vector-(1,$\dd$) & No graph & 9.146 & 16.226 & 11.338$^{\star\star}$ 
\end{NiceTabular}
\caption{Mean squared 1-step ahead prediction errors; $\star$ and $\star\star$ indicate if some simulations did not converge (less and more than $5\%$, respectively). Last row shows the MSPEs of the FIVAR and VARFI cases, and bold font highlights the smallest MSPEs.} \label{tab:forecast-vector}
\end{table}

%-------------------------------------------------------------------
%  REAL DATA
%-------------------------------------------------------------------
\section{Application to Real Data}\label{sec:application}
In this section, we illustrate the practical use of the long memory network models \eqref{eq:fignar} and \eqref{eq:gnarfi}. We consider datasets arising from two scientific applications; firstly, we analyse wind speed time series from 12 Irish meteorological stations, followed by realised volatility data corresponding to 14 global stock markets. We refer to Appendix \ref{app:extrarealdata} for additional results. 

%------------------------------------------------------------------
\subsection{Irish Wind Data}
\begin{figure}[!ht]
    \centering
    \includegraphics[width=0.9\textwidth]{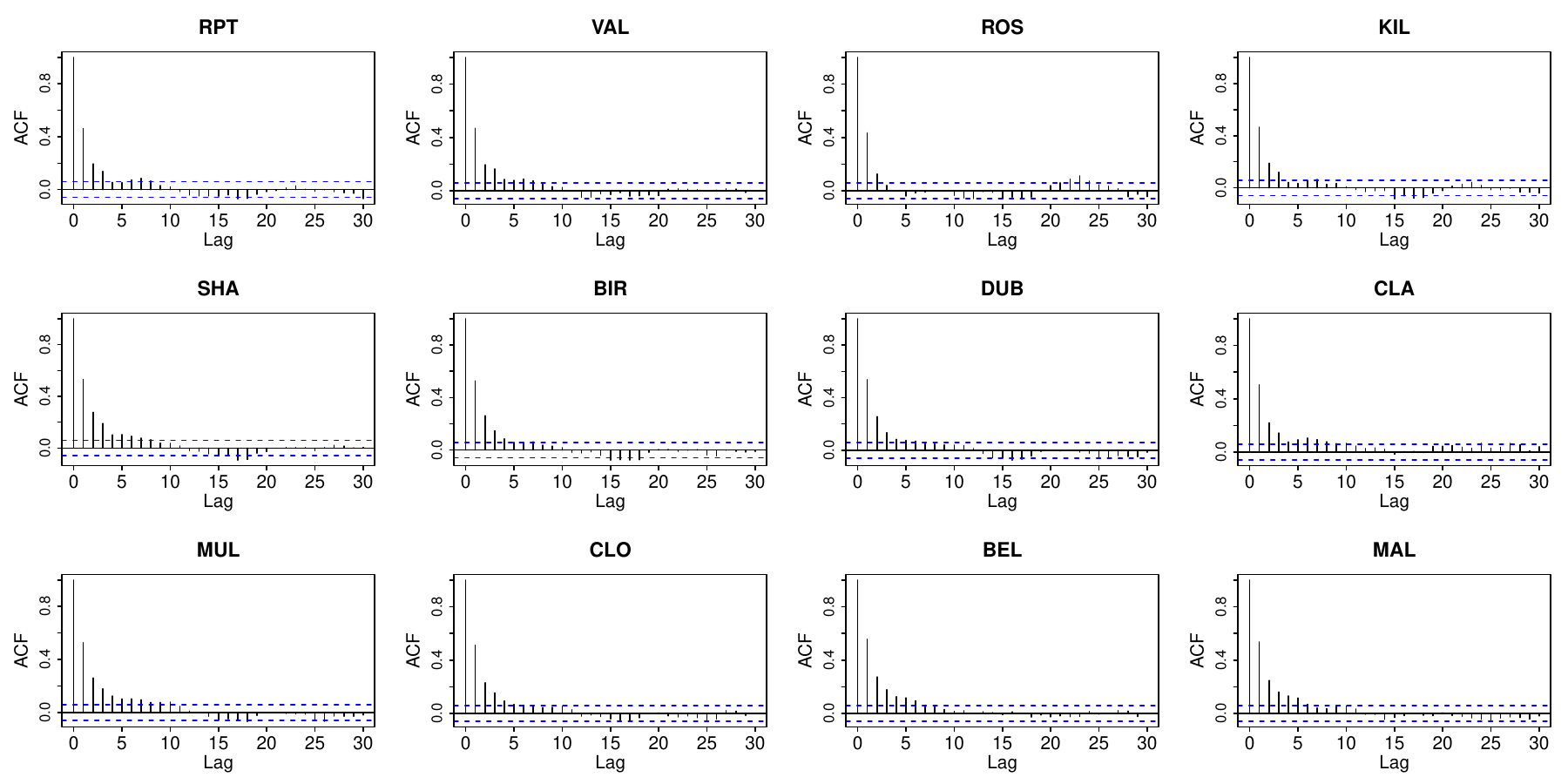}
    \captionof{figure}{Empirical autocorrelation function plots at each meteorological station.}\label{img:irishwind-acf}
\end{figure}

The Irish wind dataset comprises average daily wind speeds recorded from 1961 to 1978 at 12 meteorological stations in the Republic of Ireland and is available in the \texttt{R} package \texttt{gstat} \citep{gstat, gstatrjournal}. We focus on data from the first three available years, yielding a dataset with $T = 1095$ time points and $N = 12$ stations. See Table \ref{tab:irish-wind-data} for summary statistics about the data. Following the procedure outlined by \cite{haslett1989space}, we begin by removing the seasonal component in the data.
A preliminary exploratory analysis reveals that every station is associated to high inter-station correlations. Moreover, empirical autocorrelation function plots suggest the presence of long memory behaviour, of varying strengths. For instance, as shown in Figure \ref{img:irishwind-acf}, the long memory properties at the DUB (Dublin) station appear to be relatively weak, whereas BEL (Belmullet) and MAL (Malin Head) display a slower autocorrelation decay. Nevertheless, the overall long memory characteristics do not appear to be particularly pronounced.

Based on the spatial coordinates of each meteorological station, we define a minimal spanning tree (MST) network whose edges are weighted according to the various distances provided, with which to fit our proposed models. We also identify other suitable graphs based on the smallest prediction error given by an approximate GNAR($\infty$) fit or based on the non-zero elements of the estimated inverse spectral density matrix (as described in Section \ref{sec:varforecast}). Finally, we also fit the data to a fully connected graph. 

We fit global and individual-$\alpha$ FIGNAR models with network AR orders (1,[1]), (1,[2]), (1,[3]), (2,[1,0]), and (2,[1,1]). Following \cite{ferreira2017spatiotemporal}, who use a spatial framework with a global long memory parameter, we also estimate FIGNAR models with a global-$d$ specification.
Each fitted model is then used to predict the next seven days \citep{ferreira2017spatiotemporal}; the MSPEs averaged across the seven forecast days are reported in Table~\ref{tab:irish1}. 
A clear difference emerges between $p=1$ and $p=2$. Some FIGNAR(2,[$\ss$],$\dd$) fits converge to unreliable solutions, especially in the global-$d$ specifications, and when they yields stable estimates, the forecasts are worse than their $(1,[s],\dd)$ counterparts. This suggests that $p=2$ may be too complex for these data and that a first-order AR is sufficient.

In general, the individual-$d$ models yield the smaller errors than the global-$d$ ones, indicating that a global approach may not be suitable for capturing the long memory features of the Irish wind data. The best predictions are given by the individual-$\{\alpha,d\}$ models, with the FIGNAR(1,[3],$\dd$) with MST being the best forecasting model. Note also that FIGNAR(1,[2],$\dd$) on the MST or the {\tt TSglasso} graph are also performing well. Notice models fitted on the MST produce errors of similar magnitude to those from the other data-driven graphs, which is reasonable given the spatial nature of the data.

Overall, network models offer sufficient flexibility to distinguish between the short- and long memory components without becoming overly complex, unlike the network FIVAR(1,$\dd$) model which failed to converge, thus precluding a direct comparison. \newline

\begin{table}[!ht]
\setlength{\tabcolsep}{13pt}
\centering
\begin{NiceTabular}{l|l||llll}
 & & ind-$\alpha$ & gl-$\alpha$ & ind-$\alpha$ & gl-$\alpha$ \\
Order & Graph & ind-$d$ & ind-$d$ & gl-$d$ & gl-$d$ \\
\hline\hline
(1,[0],$\dd$) & No graph & $-$ & $-$ & $-$ & $-$ \\
(1,[1],$\dd$) & Fully-connected & 0.590 & 0.591 & 0.592 & 0.593 \\
& MST & 0.589 & 0.592 & 0.593 & 0.593 \\
& GNAR($\infty$) approx & 0.590 & 0.591 & 0.592 & 0.592 \\
& TSglasso & 0.588 & 0.590 & 0.590 & 0.590 \\
(1,[2],$\dd$) & MST & \textbf{0.586} & 0.590 & 0.590 & 0.590 \\
& GNAR($\infty$) approx & 0.589 & 0.593 & 0.591 & 0.595 \\
& TSglasso & \textbf{0.587} & 0.589$^{\star}$ & 0.590 & 0.591 \\
(1,[3],$\dd$) & MST & \textbf{0.584} & 0.589 & 0.590 & 0.590 \\
& GNAR($\infty$) approx & 0.591$^{\star}$ & 0.594 & 0.595 & 0.593 \\
& TSglasso & 0.589 & 0.589$^{\star}$ & 0.597$^{\star}$ & 0.590 \\
(2,[0,0],$\dd$) & No graph & 0.605 & 0.606 & 0.610 & 0.608 \\
(2,[1,0],$\dd$) & Fully-connected & 0.603 & 0.605 & 0.609 & 0.608 \\
& MST & 0.601 & 0.607 & 0.591 & 0.606 \\
& GNAR($\infty$) approx & 0.602 & 0.604 & 0.608 & 0.608 \\
& TSglasso & 0.599 & 0.602 & 0.605 & 0.608 \\
(2,[1,1],$\dd$) & Fully-connected & 0.602 & 0.609 & 0.585$^{\star}$ & 0.597$^{\star}$ \\
& MST & 0.576$^{\star}$ & 0.604 & 0.607$^{\star}$ & 0.606 \\
& GNAR($\infty$) approx & 0.585$^{\star}$ & 0.608 & 0.585$^{\star}$ & 0.602$^{\star}$ \\
& TSglasso & 0.599 & 0.601 & 0.590$^{\star}$ & 0.588$^{\star}$ \\
\end{NiceTabular}
\caption{FIGNAR estimation: Irish data MSPEs for one week (7-step) ahead predictions; $\star$ indicates the method converged to unreliable solution, and $-$ indicates the method did not converge. Bold font highlights the three smallest errors.}\label{tab:irish1}
\end{table}

We fit the data using the GNARFI models under the same specifications and estimate parameters via the conditional maximisation approach. As before, GNARFI(2,[$\ss$],$\dd$) framework appears to be unsuitable: it yields unstable estimates or worse predictions than the $(1,[s],\dd)$ cases. In fact, the individual-$\{\alpha,d\}$ GNARFI(1,[3],$\dd$) on the MST attains again the smallest error. On the other hand, the individual-$\{\alpha,d\}$ models with network AR order (1,[3]) using the GNAR($\infty$) graph or with (1,[1]) using the \texttt{TSglasso} graph deliver also strong forecasts. \newline

\begin{table}[!ht]
\setlength{\tabcolsep}{13pt}
\centering
\begin{NiceTabular}{l|l||llll}
 & & ind-$\alpha$ & gl-$\alpha$ & ind-$\alpha$ & gl-$\alpha$ \\
Order & Graph & ind-$d$ & ind-$d$ & gl-$d$ & gl-$d$ \\
\hline\hline
(1,[0],$\dd$) & No graph & 0.592 & 0.596 & 0.595 & 0.595 \\
(1,[1],$\dd$) & Fully-connected & 0.589 & 0.592 & 0.592 & 0.592 \\
& MST & 0.591 & 0.595 & 0.594 & 0.594 \\
& GNAR($\infty$) approx & 0.588 & 0.591 & 0.591 & 0.592 \\
& TSglasso & \textbf{0.587} & 0.590 & 0.590 & 0.590 \\
(1,[2],$\dd$) & MST & 0.589 & 0.593 & 0.592 & 0.593 \\
& GNAR($\infty$) approx & 0.589 & 0.592 & 0.592 & 0.592 \\
& TSglasso & 0.589 & 0.593 & 0.592 & 0.591 \\
(1,[3],$\dd$) & MST & \textbf{0.582} & 0.590 & 0.590 & 0.591 \\
& GNAR($\infty$) approx & \textbf{0.585} & 0.592 & 0.592 & 0.591 \\
& TSglasso & 0.591 & 0.591$^{\star}$ & 0.595$^{\star}$ & 0.592 \\
(2,[0,0],$\dd$) & No graph & 0.603 & 0.604 & 0.595 & 0.606 \\
(2,[1,0],$\dd$) & Fully-connected & 0.592 & 0.604 & 0.599 & 0.607 \\
& MST & 0.602 & 0.605 & 0.602 & 0.606 \\
& GNAR($\infty$) approx & 0.596 & 0.607 & 0.597 & 0.606 \\
& TSglasso & 0.594 & 0.602 & 0.600 & 0.605 \\
(2,[1,1],$\dd$) & Fully-connected & 0.592$^{\star}$ & 0.613 & 0.589$^{\star}$ & 0.597$^{\star}$ \\
& MST & 0.596 & 0.605 & 0.587$^{\star}$ & 0.606 \\
& GNAR($\infty$) approx & 0.594 & 0.608 & 0.588$^{\star}$ & 0.615$^{\star}$ \\
& TSglasso & 0.596 & 0.595$^{\star}$ & 0.588$^{\star}$ & 0.607 \\
\end{NiceTabular}
\caption{GNARFI estimation: Irish data  MSPEs for one week (7-step) ahead predictions; $\star$ indicates the method converged to unreliable solution. Bold font highlights the three smallest errors.} \label{tab:irish2}
\end{table}

When comparing the BIC of the estimated models resulting in the three smallest prediction errors, FIGNAR(1,[2],$\dd$) model on the MST is selected out of the 11 models. 
% top 3 errors in FIGNAR and in GNARFI< with a total of 11 models
The corresponding parameter estimates are reported in Table \ref{tab:irish-estimates-bestBIC}. The component $\hat{d}_i$'s values range from 0.15 to 0.30, with the largest estimates at stations along the northern coast, BEL (Belmullet) and MAL (Malin Head). Interestingly, BIR (Birr), KIL (Kilkenny), and MUL (Mullingar) exhibit strong persistence despite wind speeds are generally low, which is reasonable due to the location in central Ireland. By contrast, VAL (Valentia) on the southern coast and DUB (Dublin) on the eastern coast have the smallest long memory estimates, coupled with larger autoregressive effects, as reflected by their higher $\hat{\alpha}_1$ values. The network autoregression is more pronounced at the second stage than at the first.

\begin{table}[!ht]
\setlength{\tabcolsep}{13pt}
\centering
\begin{NiceTabular}{l||ccccc}
Station & $\hat{\alpha}_1$ & $\hat{\beta}_{1.1}$ & $\hat{\beta}_{1.2}$ & $\hat{d}$ & $\hat{\sigma}^2$ \\
\hline\hline
RPT & 0.129 & -0.034 & 0.163 & 0.193 & 0.255 \\
VAL & 0.251 & -0.034 & 0.163 & 0.124 & 0.219 \\
ROS & 0.179 & -0.034 & 0.163 & 0.164 & 0.242 \\
KIL & 0.104 & -0.034 & 0.163 & 0.233 & 0.225 \\
SHA & 0.180 & -0.034 & 0.163 & 0.188 & 0.202 \\
BIR & 0.137 & -0.034 & 0.163 & 0.251 & 0.216 \\
DUB & 0.285 & -0.034 & 0.163 & 0.095 & 0.224 \\
CLA & 0.233 & -0.034 & 0.163 & 0.135 & 0.241 \\
MUL & 0.143 & -0.034 & 0.163 & 0.233 & 0.214 \\
CLO & 0.307 & -0.034 & 0.163 & 0.027 & 0.204 \\
BEL & 0.190 & -0.034 & 0.163 & 0.252 & 0.233 \\
MAL & 0.150 & -0.034 & 0.163 & 0.279 & 0.269 \\
\end{NiceTabular}
\caption{Estimated parameters of the FIGNAR(1,[2],$\dd$) model with the MST graph. Values of $\hat{\beta}_{1.1}$ and $\hat{\beta}_{1.2}$ have been repeated for each station for completeness.} \label{tab:irish-estimates-bestBIC}
\end{table}

%------------------------------------------------------------------
\subsection{Realised Volatility Data}
As a second dataset, we consider the realised volatility aggregated from 5-minute within-day returns from the Oxford Man Institute of Quantitative Finance (\url{https://oxford-man.ox.ac.uk}). Due to the computational challenges posed by long memory features, typical vector time series models are too complex for practical use, preventing the fitting of models with more than three variables.
Recently, \cite{baek2023local} developed a sparse semiparametric estimation method capable of identifying a sparse structure among worldwide stocks. This suggests that network models may be a promising option for fitting parametric models that simultaneously analyse both the short- and long memory features.

We consider data of 14 European stocks from January 4th 2016 until October 31st 2019, corresponding to 1001 trading days. Additional summary statistics are presented in Table \ref{tab:rel-vol-data}.
Following the approach of \cite{baek2014distinguishing}, we impute missing data with linear interpolation, apply a log-transformation and remove any potential mean changes.

\begin{figure}[!ht]
    \centering
    \includegraphics[width=0.9\textwidth]{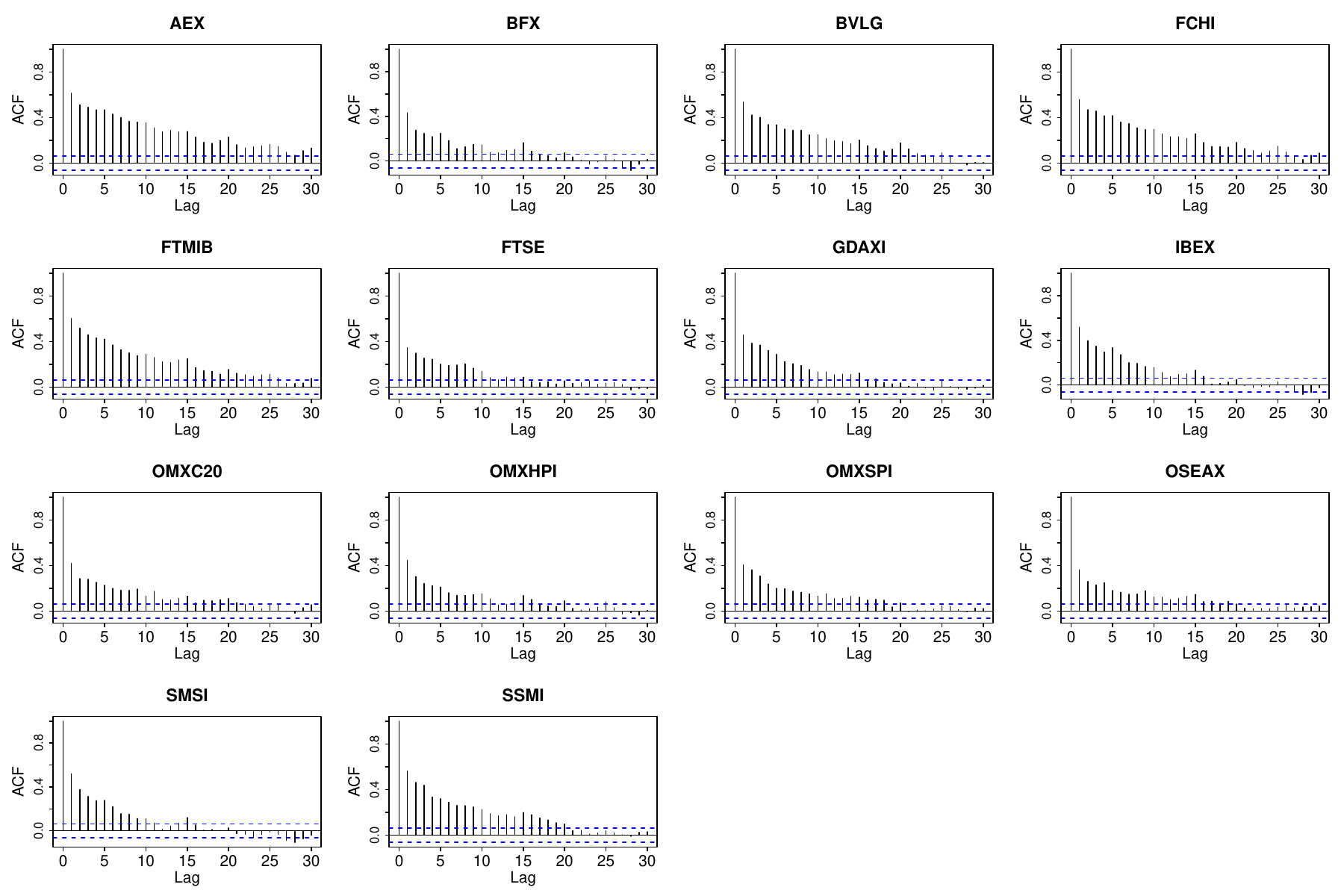}
    \captionof{figure}{Empirical autocorrelation function plots for the 14 European stocks.}\label{fig:acf2}
\end{figure}

The empirical ACF plots in Figure~\ref{fig:acf2} indicate that all stock indices exhibit relatively strong long memory behaviour, albeit with varying rates of decay. For instance, comparing FTMIB and OMXHPI reveals notable differences in their autocorrelation structures.

As before, we fit global-$\alpha$ and individual-$\alpha$ models over different orders and networks, namely the fully connected graph and other graph structures identified using the GNAR($\infty$) approximation or the \texttt{TSglasso} estimation strategies. Additionally, we define our own network based on the geographic characteristics of the dataset. Here, we place an edge between two countries that share a border or are directly separated by the sea (`Border-based'), as in the case of the UK and France or the Netherlands.

GNARFI models are estimated with the conditional likelihood approach, and we note the estimation procedure of FIGNAR(1,[1],$\dd$) over the fully connected graph did not converge. 
% Similar convergence issues occur for the individual-$\alpha$ FIGNAR and GNARFI of order (2,[1,0],$\dd$), and individual-$\alpha$ GNARFI(1,[2],$\dd$) and GNARFI(1,[3],$\dd$) over the GNAR($\infty$)-based graph.
% All estimated models are then used to produce forecasts up to 5 steps ahead. As shown by the average MSPEs in Table~\ref{tab:revol1}, while errors are of similar magnitude, the GNARFI framework generally delivers better forecasting performance, especially for the individual-$\{\alpha,d\}$  specifications. The smallest error is achieved by the individual-$\{\alpha,d\}$ GNARFI(2,[1,0],$\dd$) model on the fully connected graph. The individual-$\{\alpha,d\}$ FIGNAR(1,[3],$\dd$) with the GNAR($\infty$)-based graph also performs well.
All estimated models are then used to produce forecasts up to 5 steps ahead. As shown by the average MSPEs in Table~\ref{tab:revol1}, while the individual-$\{\alpha,d\}$ GNARFI(1,[1],$\dd$) framework generally delivers better forecasting performance, the best performances are reached by the individual-$\{\alpha,d\}$ FIGNAR models with $p=2$. Indeed, the smallest error is achieved by the individual-$\{\alpha,d\}$ FIGNAR(2,[1,1],$\dd$) model on the border-based graph, but the same model on other graphs also performs well.

\begin{table}[!ht]
\setlength{\tabcolsep}{13pt}
\centering
\begin{NiceTabular}{l|l||ll|ll}
& & \Block{1-2}{FIGNAR} & & \Block{1-2}{GNARFI} \\
& & ind-$\alpha$ & gl-$\alpha$ & ind-$\alpha$ & gl-$\alpha$ \\
Order & Graph & ind-$d$ & ind-$d$ & ind-$d$ & ind-$d$ \\
\hline\hline
(1,[0],$\dd$) & No graph & $-$ & 0.331 & 0.327 & 0.331 \\
(1,[1],$\dd$) & Fully-connected & $-$ & $-$ & 0.323 & 0.326 \\
& Border-based & 0.332 & 0.332 & 0.331 & 0.332 \\
& GNAR($\infty$) approx & 0.333 & 0.332 & 0.331 & 0.331 \\
& TSglasso & 0.334 & 0.333 & 0.329 & 0.332 \\
(1,[2],$\dd$) & Border-based & 0.327 & 0.329 & 0.327 & 0.331 \\
& GNAR($\infty$) approx & 0.323 & 0.324 & 0.322 & 0.323 \\
& TSglasso & 0.329 & 0.328 & 0.326 & 0.332 \\
(1,[3],$\dd$) & Border-based & 0.326 & 0.328 & 0.325 & 0.329 \\
& GNAR($\infty$) approx & 0.325 & 0.324 & 0.319 & 0.323 \\
& TSglasso & 0.331 & 0.327 & 0.336 & 0.332 \\
(2,[0,0],$\dd$) & No graph & 0.329 & 0.333 & 0.325 & 0.331 \\
(2,[1,0],$\dd$) & Fully-connected & \textbf{0.315} & 0.328 & 0.317 & 0.339 \\
& Border-based & 0.322 & 0.333 & 0.325 & 0.330 \\
& GNAR($\infty$) approx & 0.328 & 0.333 & 0.325 & 0.329 \\
& TSglasso & 0.332 & 0.333 & 0.327& 0.331 \\
(2,[1,1],$\dd$) & Fully-connected & 0.319 & 0.324 & \textbf{0.316} & 0.331 \\
& Border-based & \textbf{0.302} & 0.329 & 0.319 & 0.326 \\
& GNAR($\infty$) approx & \textbf{0.316} & 0.331 & 0.317 & 0.332 \\
& TSglasso & \textbf{0.316} & 0.331 & 0.334 & 0.329 \\
\end{NiceTabular}
\caption{MSPEs for 5-step ahead predictions of the European stock data; $\star$ indicates the method converged to unreliable solution, and $-$ indicates the method did not converge. Bold font highlights the three smallest errors.} \label{tab:revol1}
\end{table}

We then compare the BIC among the models with the top three forecasting performances (bold entries in Table \ref{tab:revol1}). The individual-$\{\alpha,d\}$ GNARFI(2,[1,1],$\dd$) model on the fully connected graph attains the lowest BIC. Looking at the estimated parameters in Table \ref{tab:relvol-estimates-bestBIC}, the $\hat{\alpha}$'s values at both lags and $\hat{\beta}_{2.1}$ are close to zero, whereas the network parameter $\hat{\beta}_{1.1}$ is positive. On the other hand, a positive first autoregressive component appear to be relevant for SSMI, SMSI, and OMXHPI stocks. This suggests strong cross-stock connectivity: not only BIC selects a fully connected graph, but the network AR appears to be more relevant than the usual AR component. Moreover, the individual-$\{\alpha,d\}$ GNARFI(1,[1],$\dd$) model may be good enough to model the data. Finally, the $\hat{d}_{i}$ estimates vary across stocks, ranging from $0.13$ to $0.37$, indicating pronounced long memory behaviour consistent with Figure \ref{fig:acf2}. In particular, AEX and FTMIB exhibit stronger long memory estimates, whereas the indices FTSE and SMSI show weaker effects.

\begin{table}[!ht]
\setlength{\tabcolsep}{13pt}
\centering
\begin{NiceTabular}{l||cccccc}
Index & $\hat{\alpha}_1$ & $\hat{\beta}_{1.1}$ & $\hat{\alpha}_2$ & $\hat{\beta}_{2.1}$ & $\hat{d}$ & $\hat{\sigma}^2$ \\
\hline\hline
AEX & -0.041 & 0.157 &-0.037 &    0.076 & 0.368 & 0.253 \\
BFX &     0.042 & 0.157 &-0.038 &    0.076 & 0.217 & 0.214 \\
BVLG &    0.006 & 0.157 &-0.029 &    0.076 & 0.293 & 0.146 \\
FCHI &   -0.098 & 0.157 &-0.035 &    0.076 & 0.332 & 0.269 \\
FTMIB &  -0.009 & 0.157 & 0.038 &    0.076 & 0.346 & 0.249 \\
FTSE &    0.064 & 0.157 & 0.106 &    0.076 & 0.129 & 0.339 \\
GDAXI &   0.035 & 0.157 & 0.067 &    0.076 & 0.208 & 0.252 \\
IBEX &    0.087 & 0.157 & 0.044 &    0.076 & 0.227 & 0.222 \\
OMXC20 &  0.062 & 0.157 &-0.031 &    0.076 & 0.237 & 0.368 \\
OMXHPI &  0.113 & 0.157 & 0.018 &    0.076 & 0.171 & 0.215 \\
OMXSPI & -0.042 & 0.157 & 0.031 &    0.076 & 0.248 & 0.231 \\
OSEAX &   0.010 & 0.157 & 0.014 &    0.076 & 0.219 & 0.378 \\
SMSI &    0.206 & 0.157 & 0.064 &    0.076 & 0.124 & 0.197 \\
SSMI &    0.192 & 0.157 & 0.077 &    0.076 & 0.135 & 0.148 \\
\end{NiceTabular}
\caption{Estimated parameters of the GNARFI(2,[1,1],$\dd$) model with the fully-connected graph. Values of $\hat{\beta}_{1.1}$ and $\hat{\beta}_{2.1}$ have been repeated for each index for completeness.} \label{tab:relvol-estimates-bestBIC}
\end{table}

%-------------------------------------------------------------------
%  CONCLUSION
%-------------------------------------------------------------------
\section{Discussion}\label{sec:concs}
%\section{Discussion and Further Work}\label{sec:concs}

This article has proposed two novel models for multivariate time series which exhibit long memory and are potentially network-structured.  The FIGNAR and GNARFI models capture both the short and the long memory features of the data, and they naturally account for interactions among variables. This provides flexibility while promoting sufficient sparsity in the parameter set, making modelling network processes computationally more tractable and enabling the analysis of high-dimensional data that would be unfeasible under traditional vector parametric counterparts.  
Although these models are designed for cases with a known underlying graph, they can also be employed for forecasting when no such structure is available. In particular, we have provided practical guidelines for identifying the suitable networks and demonstrated competitive forecasting performance.
Finally, we applied our framework to two real-world applications and we gained insights into the short and long memory components of the data. 

\paragraph{Disclosure Statement.} The authors report there are no competing interests to declare.

\paragraph{Funding.} CB is supported by a scholarship from the EPSRC Centre for Doctoral Training in Statistical Applied Mathematics at Bath (SAMBa), under the project EP/S022945/1. MAN and MIK gratefully acknowledge support from EPSRC grant EP/X002195/1.

%-------------------------------------------------------------------
%  REFERENCES
%-------------------------------------------------------------------
\newpage
\bibliographystyle{agsm}
\bibliography{LMNTS-references.bib}
\newpage

%-------------------------------------------------------------------
%  APPENDIX
%-------------------------------------------------------------------
\appendix
\setcounter{section}{0}
\setcounter{figure}{0}
\renewcommand{\thefigure}{\thesection.\arabic{figure}}
\setcounter{equation}{0}
\renewcommand{\theequation}{\thesection.\arabic{equation}}
\setcounter{table}{0}
\renewcommand{\thetable}{\thesection.\arabic{table}}

%-------------------------------------------------------------------
\section{Simulated Realisations from FIGNAR and GNARFI DGP1}\label{app:realisations}

\begin{figure}[H]
\centering
\begin{tabular}{cc}
    \includegraphics[width=0.46\textwidth]{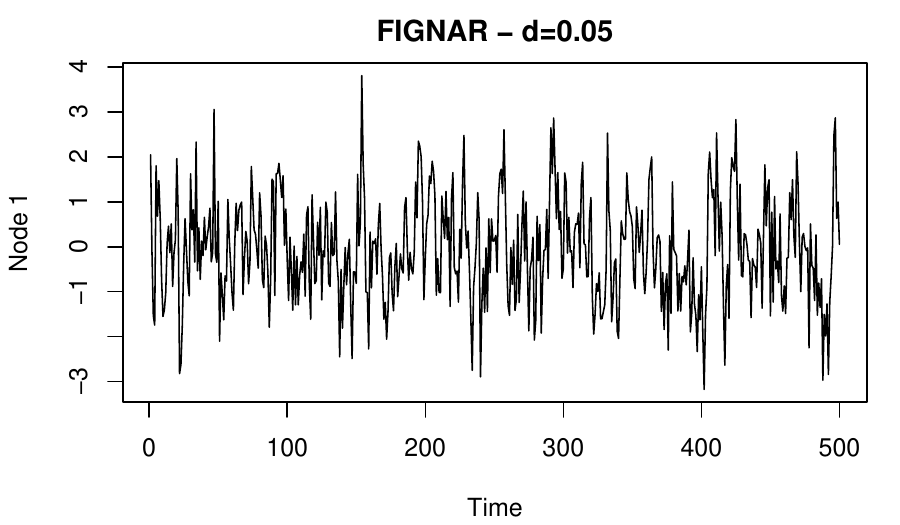} & 
    \includegraphics[width=0.46\textwidth]{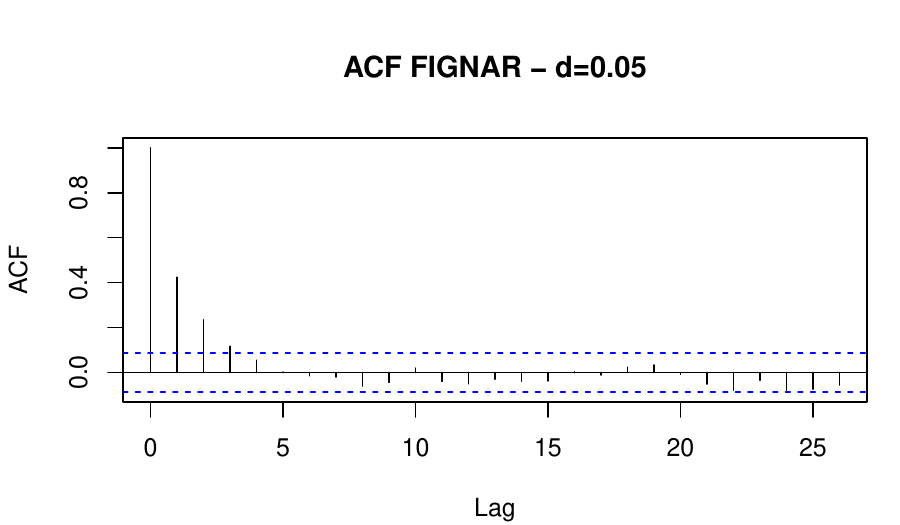} \\
    \includegraphics[width=0.46\textwidth]{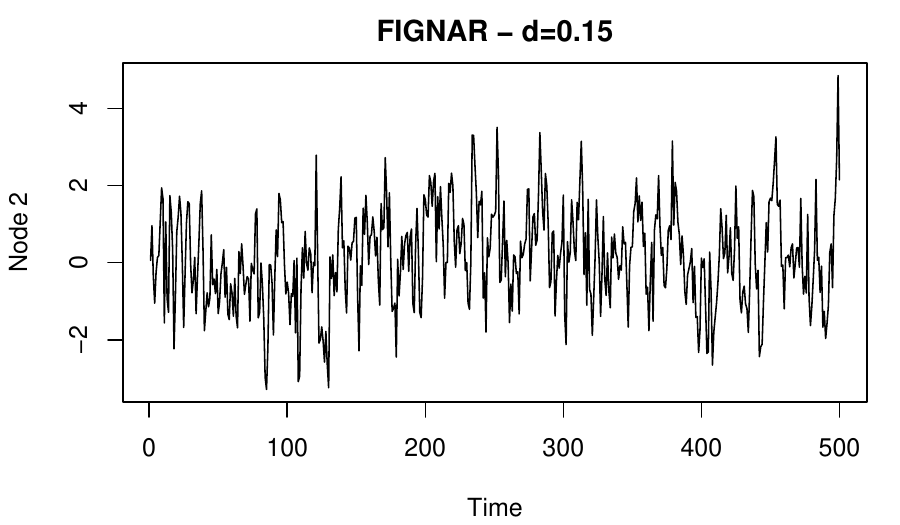} & 
    \includegraphics[width=0.46\textwidth]{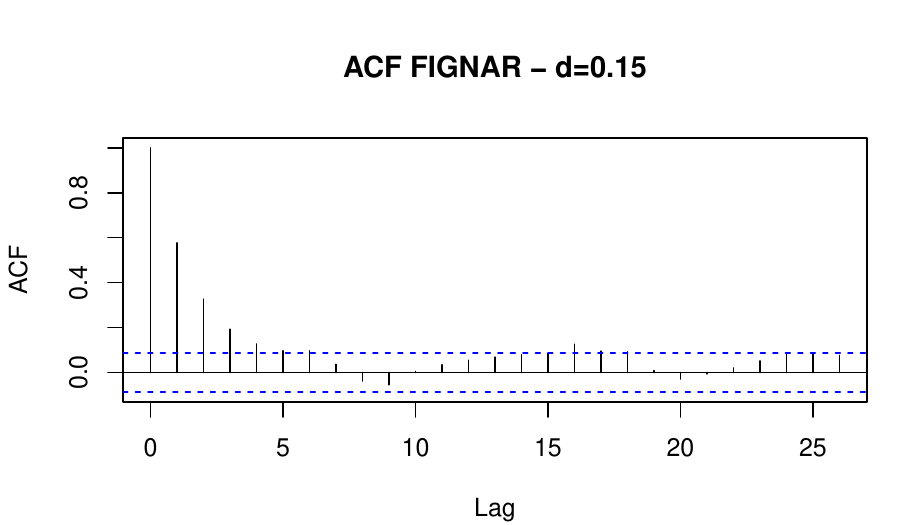} \\
    \includegraphics[width=0.46\textwidth]{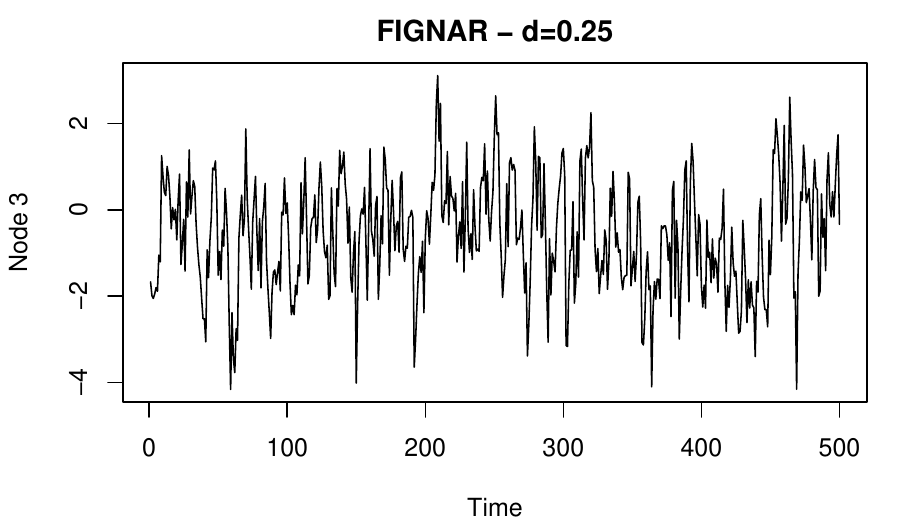} & 
    \includegraphics[width=0.46\textwidth]{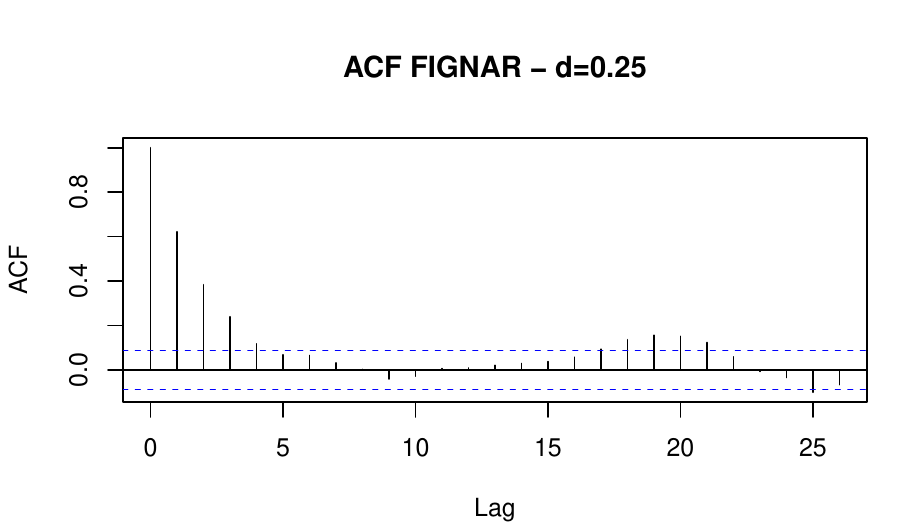} \\
    \includegraphics[width=0.46\textwidth]{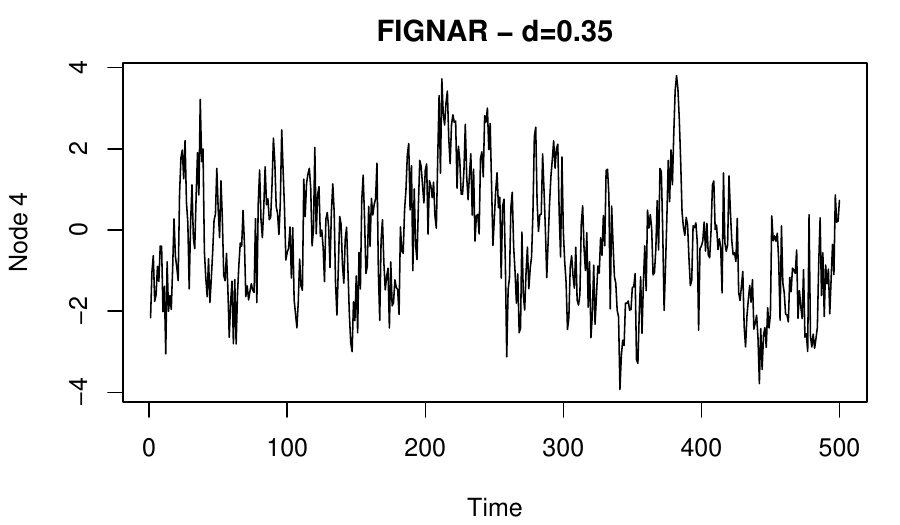} & 
    \includegraphics[width=0.46\textwidth]{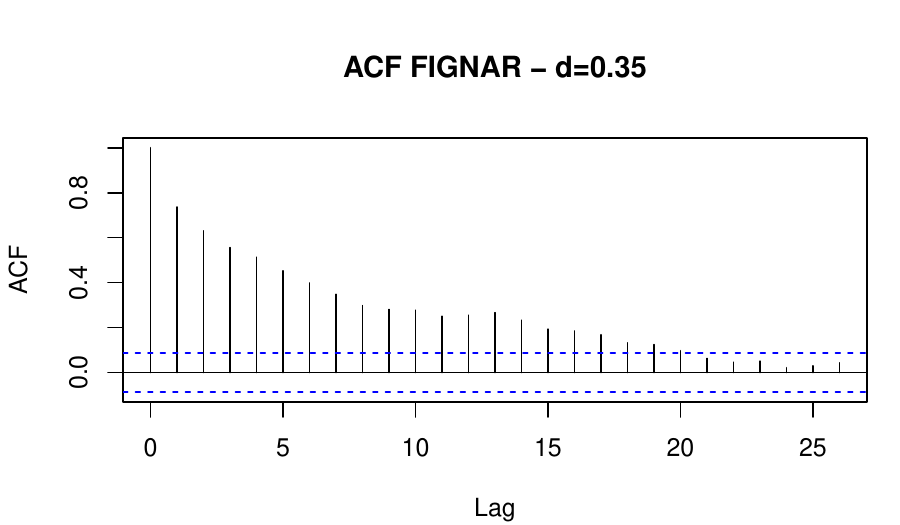} \\
    \includegraphics[width=0.46\textwidth]{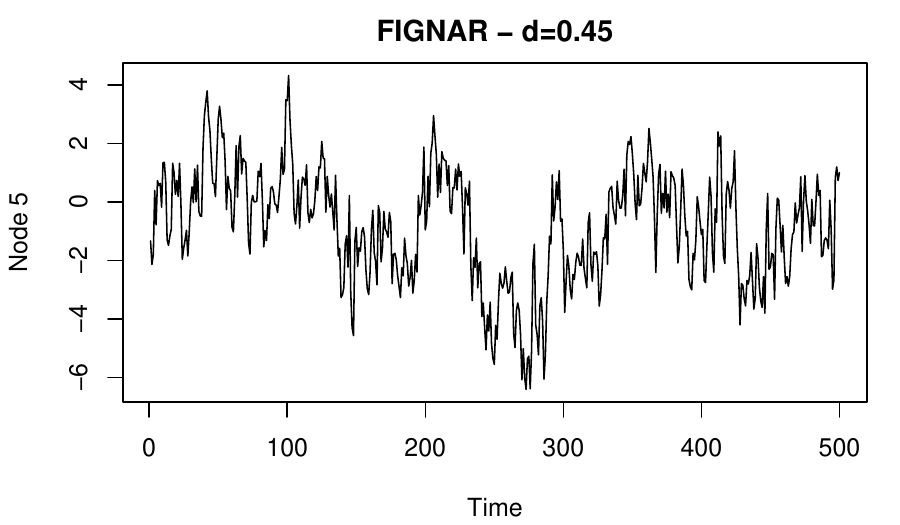} & 
    \includegraphics[width=0.46\textwidth]{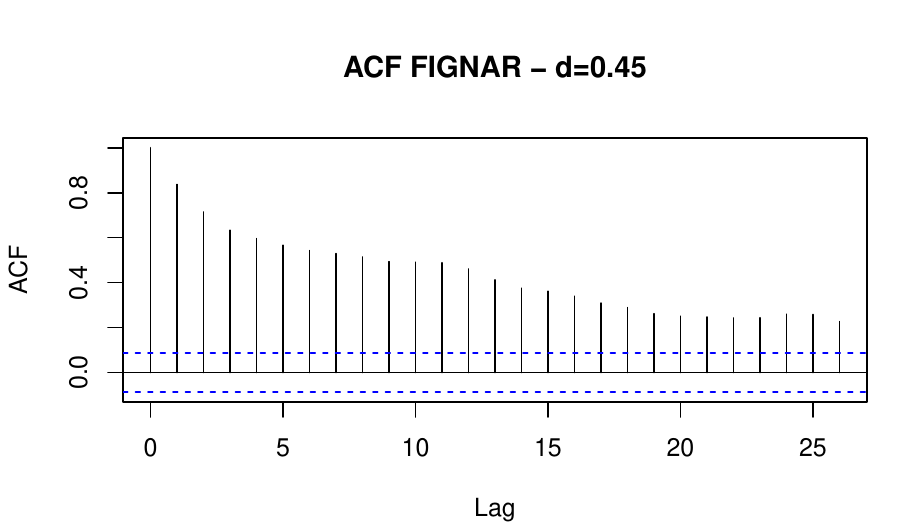} \\
\end{tabular}
\caption{Time series (left) and autocorrelation functions (right) of the FIGNAR model.} \label{img:fignar-ts-acf}
\end{figure}

\begin{figure}[H]
\centering
\begin{tabular}{cc}
    \includegraphics[width=0.46\textwidth]{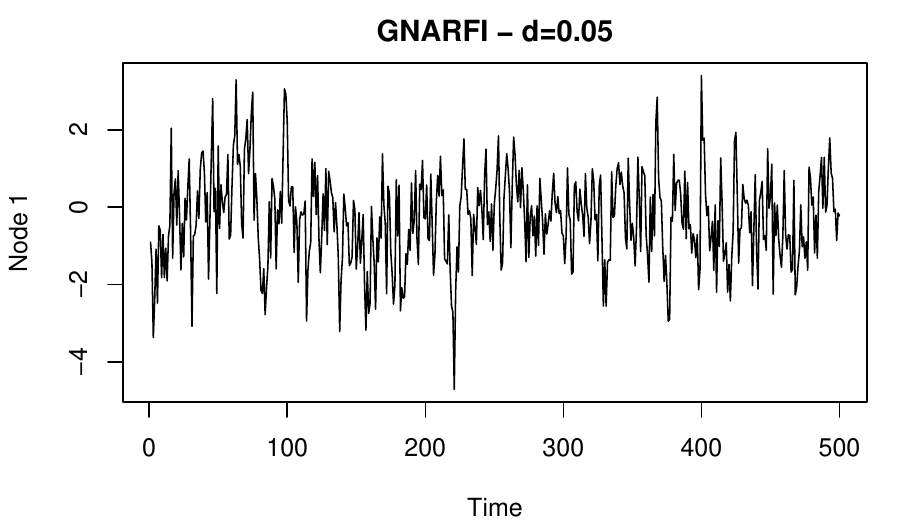} & 
    \includegraphics[width=0.46\textwidth]{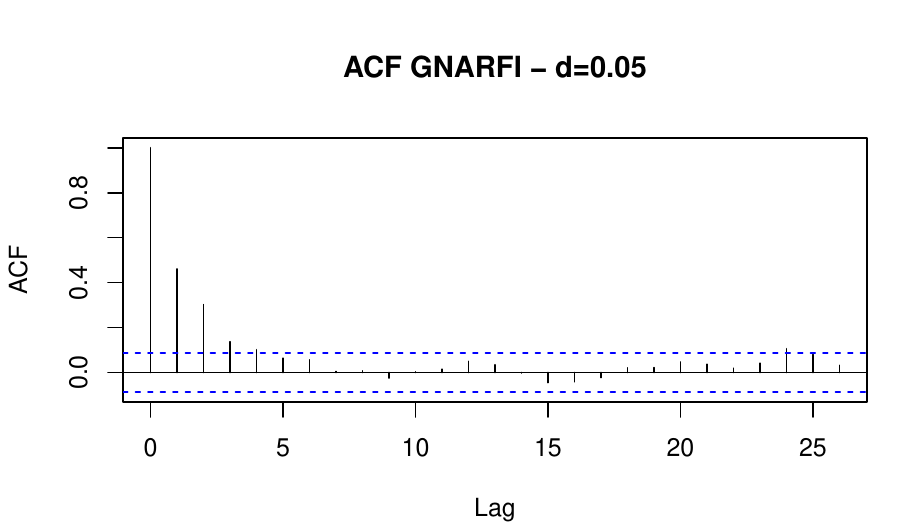} \\
    \includegraphics[width=0.46\textwidth]{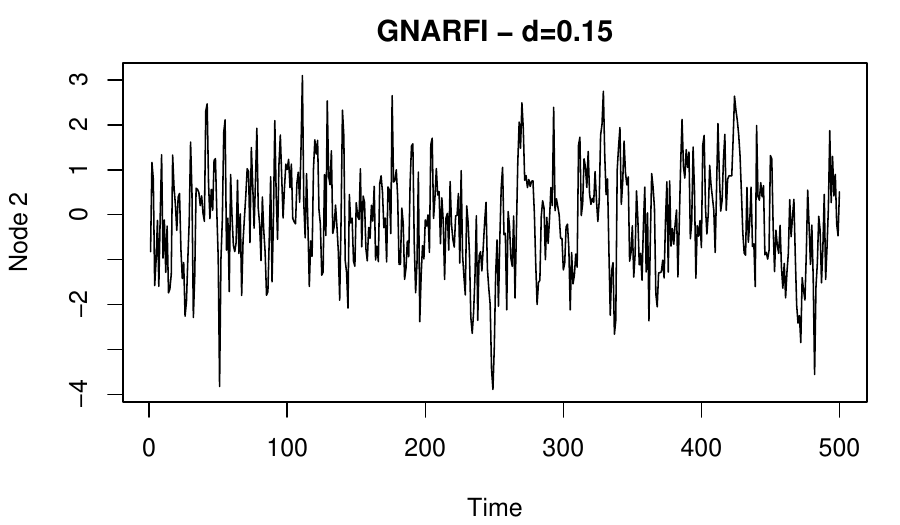} & 
    \includegraphics[width=0.46\textwidth]{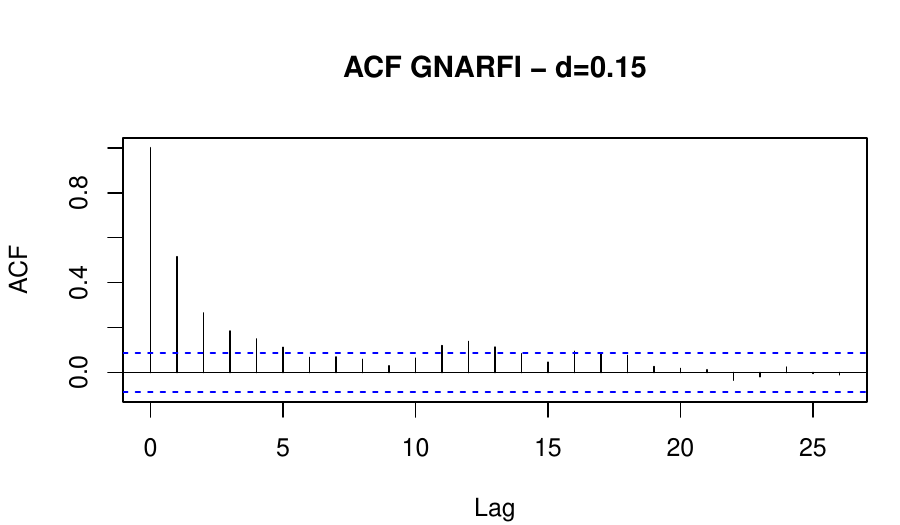} \\
    \includegraphics[width=0.46\textwidth]{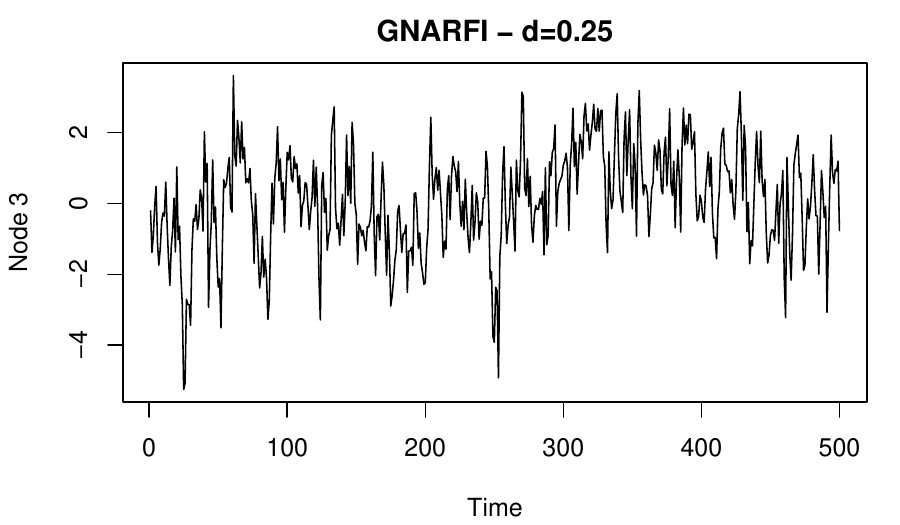} & 
    \includegraphics[width=0.46\textwidth]{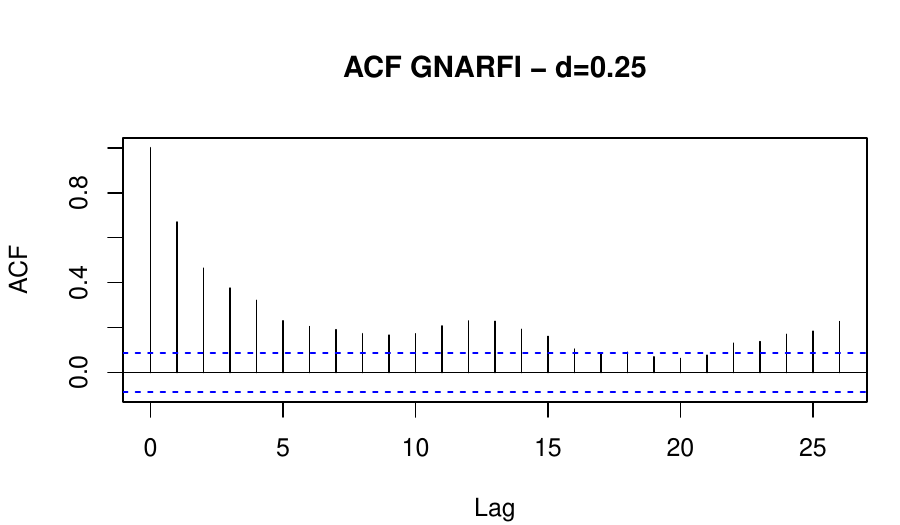} \\
    \includegraphics[width=0.46\textwidth]{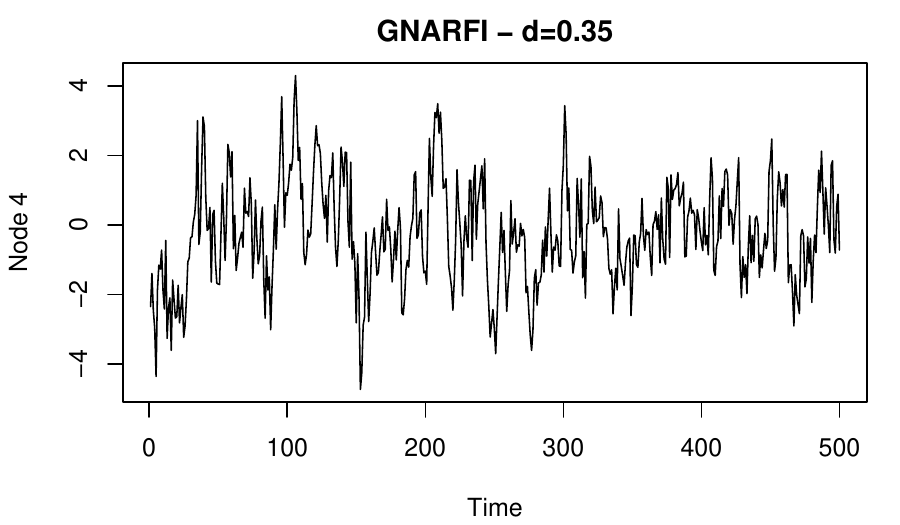} & 
    \includegraphics[width=0.46\textwidth]{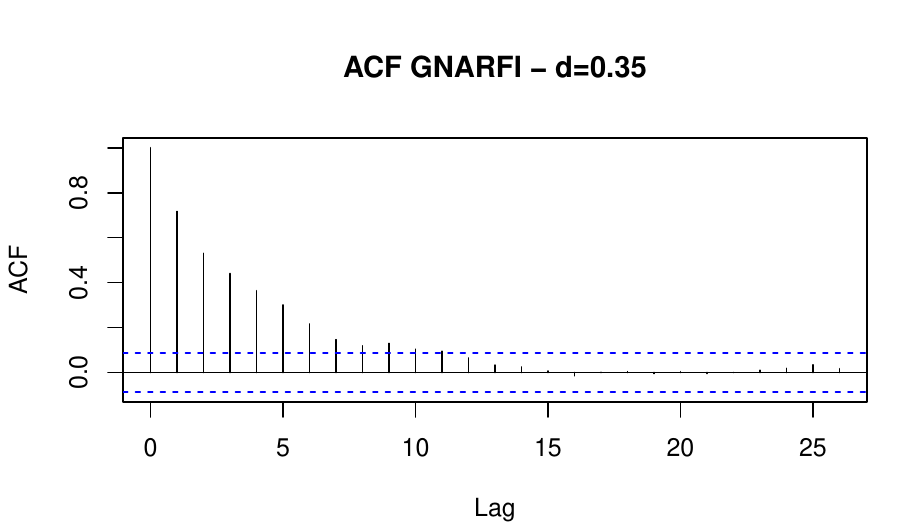} \\
    \includegraphics[width=0.46\textwidth]{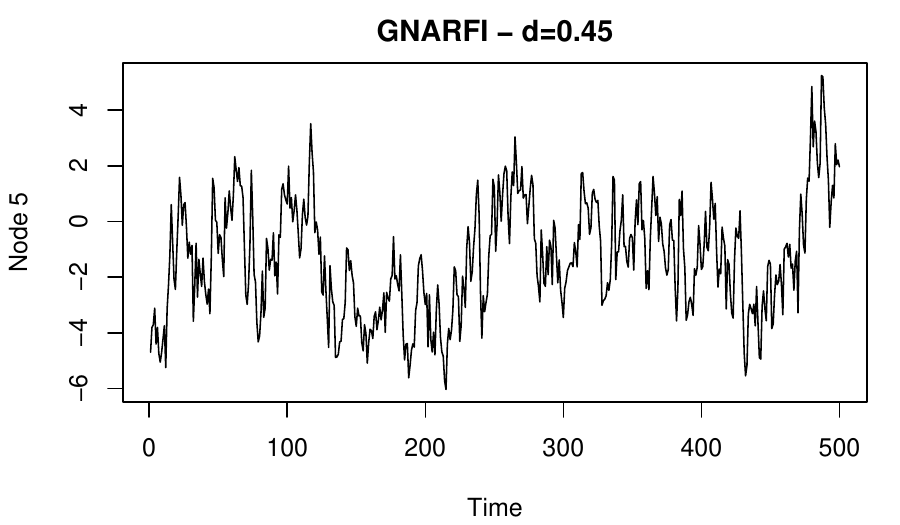} & 
    \includegraphics[width=0.46\textwidth]{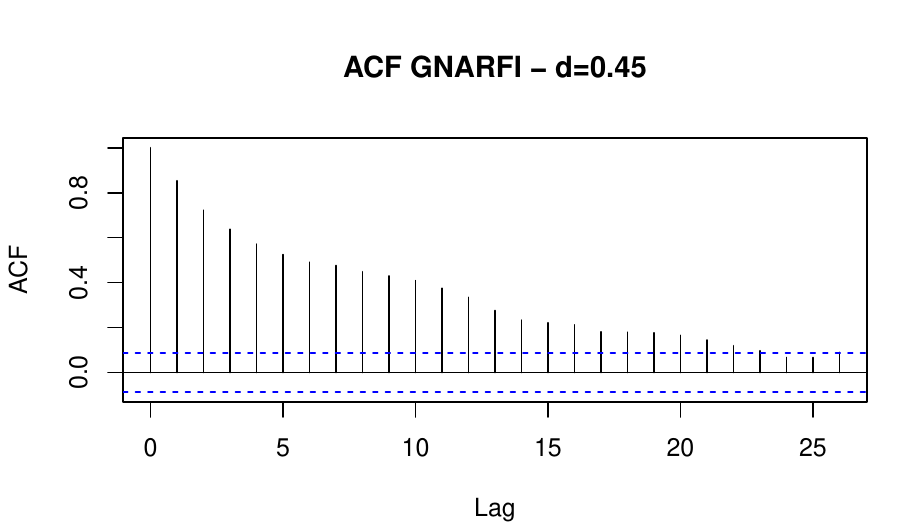} \\
\end{tabular}
\caption{Time series (left) and autocorrelation functions (right) of the GNARFI model.} \label{img:gnarfi-ts-acf}
\end{figure}

\begin{figure}[H]
    \centering
    \includegraphics[width=0.99\textwidth]{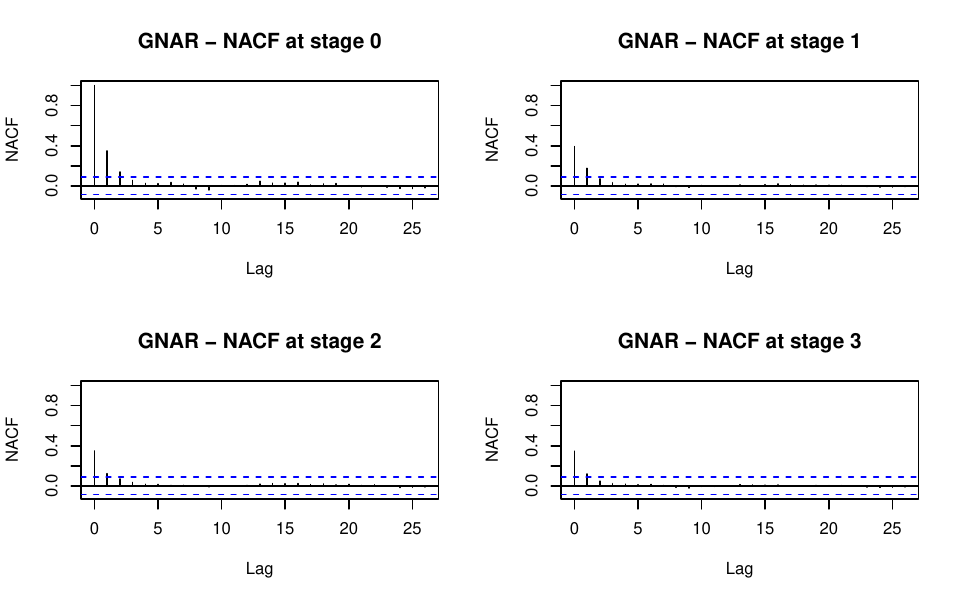}
    \captionof{figure}{Network autocorrelation function of GNAR(1,[1]) process with same AR parameters of DGP1.}\label{img:nacfgnar}
\end{figure}

\begin{figure}[H]
    \centering
    \includegraphics[width=0.99\textwidth]{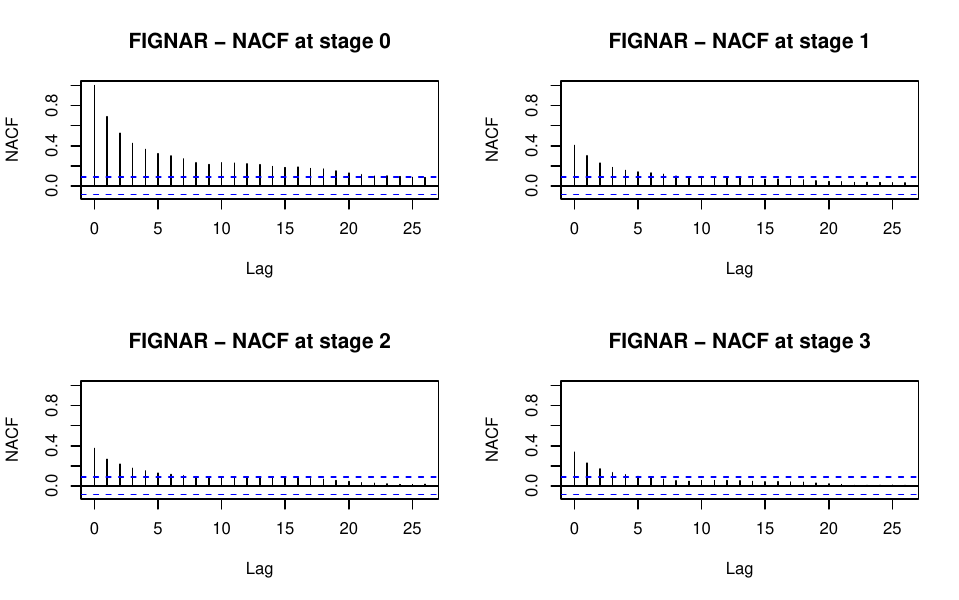}
    \captionof{figure}{Network autocorrelation function of FIGNAR(1,[1],$\dd$) process.}\label{img:nacffignar}
\end{figure}

\begin{figure}[H]
    \centering
    \includegraphics[width=0.99\textwidth]{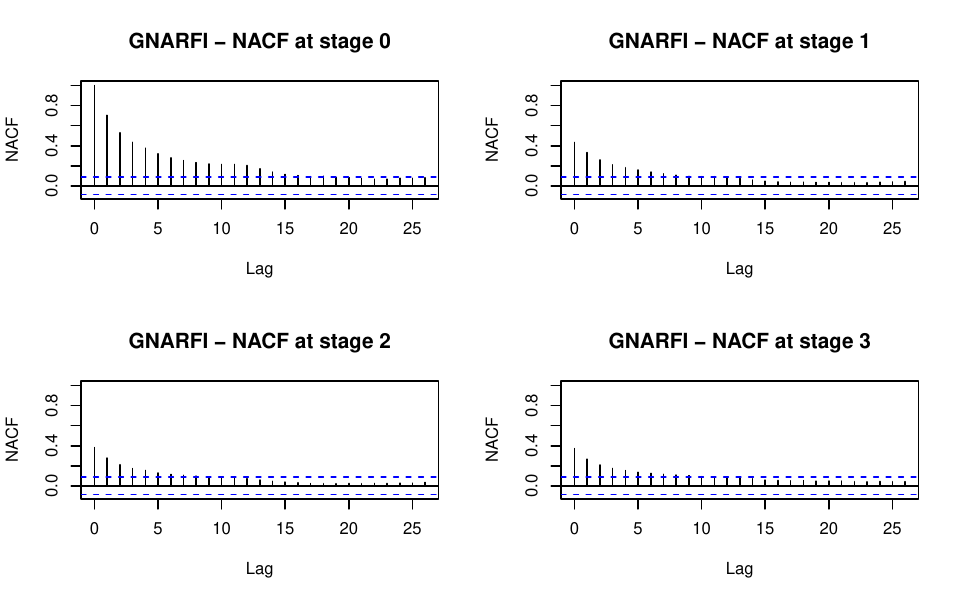}
    \captionof{figure}{Network autocorrelation function of GNARFI(1,[1],$\dd$) process.}\label{img:nacfgnarfi}
\end{figure}

\begin{figure}[H]
    \centering
    \includegraphics[width=0.99\textwidth]{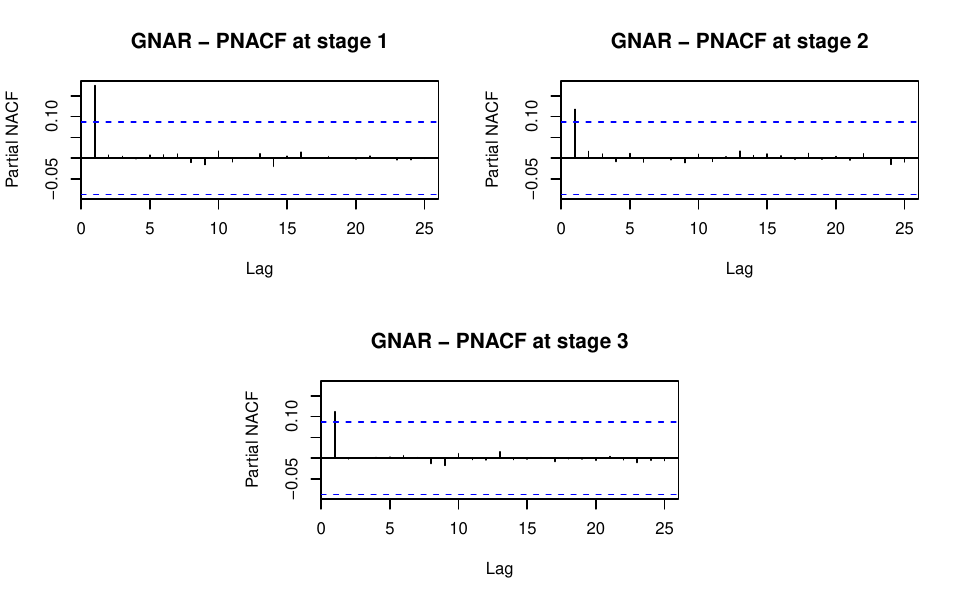}
    \captionof{figure}{Partial network autocorrelation function of GNAR(1,[1]) process with same AR parameters of DGP1.}\label{img:pnacfgnar}
\end{figure}

\begin{figure}[H]
    \centering
    \includegraphics[width=0.99\textwidth]{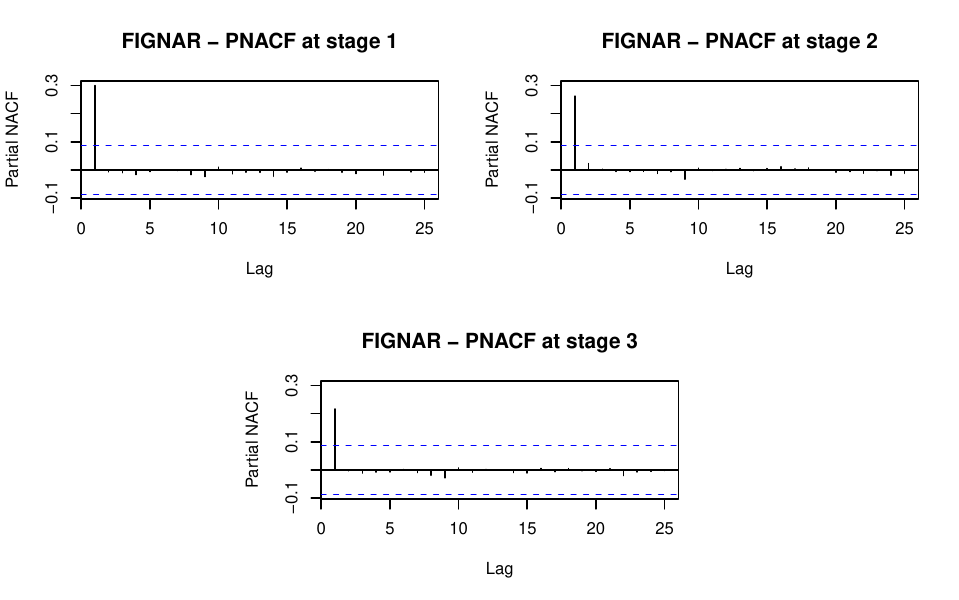}
    \captionof{figure}{Partial network autocorrelation function of FIGNAR(1,[1],$\dd$) process.}\label{img:pnacffignar}
\end{figure}

\begin{figure}[H]
    \centering
    \includegraphics[width=0.99\textwidth]{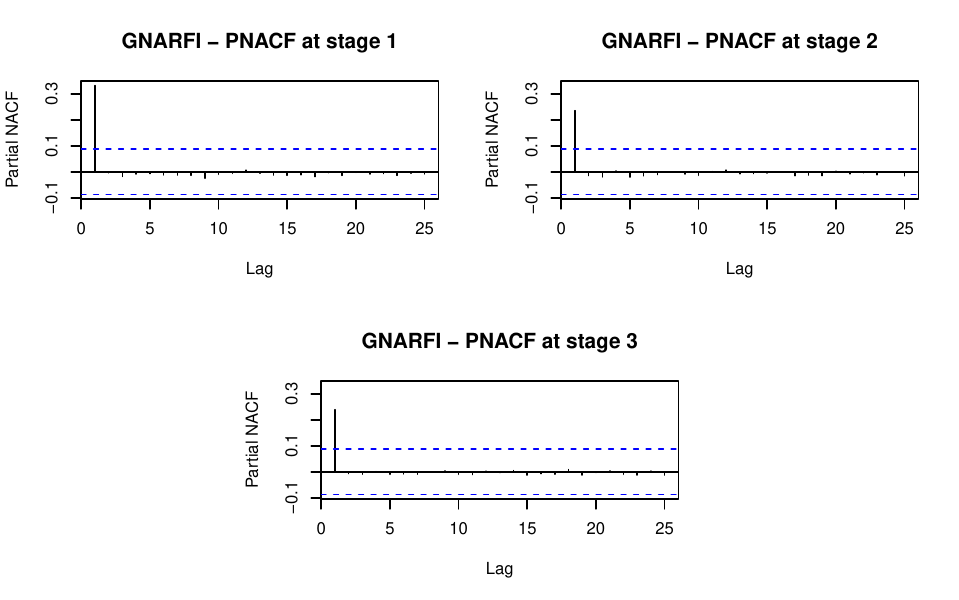}
    \captionof{figure}{Partial network autocorrelation function of GNARFI(1,[1],$\dd$) process.}\label{img:pnacfgnarfi}
\end{figure}

In this appendix, we present simulated examples of the FIGNAR and GNARFI models to illustrate their different behaviours.  As our illustrative example, we simulate a 5-dimensional network time series in the DGP1 setting (Figure \ref{img:fivenet}), whose realisations of length $T = 500$ are showed in Figures \ref{img:fignar-ts-acf}-\ref{img:gnarfi-ts-acf}. On the left-hand side, there are the simulations of both models, and the respective autocorrelation functions are on the right-hand side. %\newline
Figures \ref{img:nacffignar} and \ref{img:nacfgnarfi} present the {\em network} autocorrelation functions of the FIGNAR and GNARFI realisations, respectively, for comparison with the corresponding GNAR case in Figure \ref{img:nacfgnar}. As expected, both long memory models exhibit a decreasing function with a slower decay than the standard network model. Conversely, partial {\em network} autocorrelation plots of both FIGNAR (Figure \ref{img:pnacffignar}) and GNARFI (Figure \ref{img:pnacfgnarfi}) time series indicate the presence of an AR(1) component across the network, aligning with the respective GNAR plots (Figure \ref{img:pnacfgnar}).

\newpage
%-------------------------------------------------------------------
\section{Spline-based Interpolation for Determinant Evaluation}\label{app:det-approx}

In this section, we describe our spline-based approach to approximating the determinant of the autocovariance for the long memory network models proposed in this article.

As mentioned in Section \ref{sec:parest}, \cite{sowell1989decomposition} proposed a formula to compute the determinant of the covariance matrix $\Sigma = \left( \Omega_{\XX}(t-s) \right)_{t,s=1}^{T}$ exactly, based on the Durbin-Levinson decomposition.

We recall that, given the autocovariance function, the Durbin-Levinson algorithm determines the coefficient matrices $\Phi^{(s)}_{1}, \dots, \Phi^{(s)}_{s}$ of the best linear predictor $\hat{\XX}^{(s)}_{t}$ for $\XX_{t}$, based on the previous $s$ observations $\XX_{t-1}, \dots, \XX_{t-s}$. Given the error variances $V(s)$ of the prediction based on $s$ lagged values, namely $V(s) = \mathbb{E}\left[ (\hat{\XX}^{(s)}_{t}-\XX_{t}) (\hat{\XX}^{(s)}_{t}-\XX_{t})^{\top}\right]$, we have
\begin{equation*}
    |\Sigma| = \prod_{s=0}^{T-1} |V(s)|.
\end{equation*}
The overall computational complexity of evaluating the determinant is reduced to $\mathcal{O}(N^{3}T^{2})$, compared to the $\mathcal{O}(N^{3}T^{3})$ required by classical methods. However, as the number of observations $T$ increases, this decomposition becomes impractical, motivating the search for faster alternatives.

\cite{sela2009computationally} proposed a curve-fitting approximation involving three steps. Firstly, for some $S$, find the initial $(S+1)$ prediction variances ($V(0), \ldots, V(S)$) with the Durbin-Levinson recursion, and determine the prediction variance at the final available observation, $V(T-1)$, using the preconditioned conjugate gradient method. Next, the authors fit a linear model to an ad-hoc transformation of their determinants, which is finally used to approximate the determinants at the intermediate $(T-S)$ points.
While significantly faster than \cite{sowell1989decomposition}'s formula, a major drawback of such a curve-fitting method is the lack of control over the approximation error. Specifically, $S$ is manually set to 32, and the fitting is based on a function that empirically resembles the decay of $|V(s)|$ with increasing $s$. Furthermore, when fitting the linear model, the assumption of independence and identically distributed errors is violated, as $V(s)$ recursively depends on previous observations. 

To achieve better control over the numerical approximation error, we propose using spline interpolation instead of a linear model. This approach is motivated by noting the recursion
\begin{equation*}
    V(s) = [I - \Phi^{(s-1)}_{s-1}\bar{\Phi}^{(s-1)}_{s-1}]V(s-1),
\end{equation*}
where $\Phi^{(s-1)}_{s-1}$ and $\bar{\Phi}^{(s-1)}_{s-1}$ are the coefficient matrices of the backward and forward predictions based on $(s-1)$ observations. Taking the logarithm of the determinant on both sides yields
\begin{equation*}
    \epsilon_{s} := \log |V(s)| - \log|V(s-1)| = \log|I - \Phi^{(s-1)}_{s-1}\bar{\Phi}^{(s-1)}_{s-1}|.
\end{equation*}
By definition, $\log |V(s)|$ is a non-increasing function that converges to zero as $s \to \infty$. Consequently, one can select the number $S$ of initial log-determinants to compute exactly such that $\epsilon_{S}$ is sufficiently small, for instance $\epsilon_{S}=10^{-6}$.

Therefore, we retain the exact determinants $|V(s)|$ at $s=0,\ldots,S,T-1$, and we consider the function $f(x):=\log|V(x)|$ over the $(0,1]$ domain, with points $x_{s}=\frac{s}{T-1}$ at $s = 1,\ldots,S, T-1$. The interpolating function $\tilde{f}(x)$ is then defined in terms of monotonic natural cubic splines, such that $\log|V(x)| \approx \tilde{f}(x)$ for $x_{S+1},\ldots,x_{T-2}$. We choose the natural cubic splines because they are the smoothest interpolators, with the approximation error bounded by $\mathcal{O}(h^4)$, where $h=\frac{T-S}{T-1}$.
As in \cite{sela2009computationally}'s strategy, the determinant of the overall covariance matrix is given by
\begin{equation*}
    \widetilde{\log|\Sigma|} = \sum_{s=0}^{S} \log|V(s)| + \sum_{r=S+1}^{T-2} \widetilde{\log|V(s)|} + \log|V(T-1)|.
\end{equation*}
The spline-based determinant implementation is summarised in the pseudocode detailed in Algorithm~\ref{alg:det-approx}.

\begin{algorithm}[ht!]
\vspace{2mm}
\hspace{-5mm}\textbf{Input:}
    $\Sigma = \left( \Omega_{\XX}(t-s) \right){t,s=1,\ldots,T}$, $V(0)=\Omega_{\XX}(0)$, and $\epsilon >0$
\medskip

\hspace{-5mm}\textbf{Exact evaluation:} \\ 
\SetKwBlock{Begin}{While $\log |V(s)| - \log|V(s-1)| < \epsilon$}{end}\Begin{
    Apply Durbin-Levinson algorithm to obtain $V(s)$\\
    Evaluate $\log |V(s)|$ \\
    $s \leftarrow s+1$
}
Compute $V(T-1) = V(0) - \Gamma(T-1)^{\top} \Omega_{\XX}(T-1) \Gamma(T-1)$ using the preconditioned conjugate gradient. $\Gamma(T-1)$ is the array of covariances between $\XX_{T-1-t}$ and $\XX_{T-1}$ for $t=1,\ldots,T-1$ \\
Evaluate $\log |V(T-1)|$ \\
\smallskip
\hspace{-5mm}\textbf{Spline-based evaluation:} \\ 
Find the interpolating spline $\tilde{f}$ for $x_{s}=\frac{s}{T-1}$ at $s = 1,\ldots,S, T-1$ \\
Determine $\widetilde{\log|V(s)|} = \tilde{f}(x)$ for $x_{S+1},\ldots,x_{T-2}$. \\
\medskip
\hspace{-5mm} \textbf{Return:} $\sum_{s=0}^{S} \log|V(s)| + \sum_{r=S+1}^{T-2} \widetilde{\log|V(s)|} + \log|V(T-1)|$
\caption{Evaluation of $\log|\Sigma|$} \label{alg:det-approx}
\end{algorithm}

\newpage

%-------------------------------------------------------------------
\section{Model Estimation: Detailed Results}\label{app:extrasims}
To further explore model estimation, we conducted a simulation study on the model and order selection criteria outlined in Section \ref{sec:orderselection}, using $K=100$ time series replicates, each of length $T=200$, simulated according to DGP1 on the {\tt fiveNet} network, as introduced in Section~\ref{subsec:models}. 
Table \ref{tab:order} reports how often each estimated specification attains the lowest BIC and AIC. Specifically, we estimate global-$\alpha$ FIGNAR models of various orders on data generated from either FIGNAR(1,[1],$\dd$) or GNARFI(1,[1],$\dd$) DGP1, compute the criteria, and count the number of times each is minimal. We repeat the empirical study for GNARFI models, using both the exact and conditional likelihood approaches. Boldface text denotes the theoretically correct model and order.
On the other hand, Table \ref{tab:data-model-mismatch} reports the number of times the order is selected according to BIC and AIC. In this case, the correct models are fitted on the corresponding data, namely the FIGNAR models over the FIGNAR-simulated data, and similarly for the GNARFI case. The bold font represents again the true order and highlights the superior performance of BIC, as discussed in Section~\ref{sec:orderselection}.

\begin{table}[!ht]
\centering
% {\setlength{\tabcolsep}{6pt}\small
\begin{NiceTabular}{ccc||*{6}{c}||*{6}{c}}
    \Block{1-15}{\hspace{5.0cm} Fitted model with global-$\alpha$ and $(p,[\ss])$ order}\\
    &&& \Block{1-6}{FIGNAR} &&&&&& \Block{1-6}{GNARFI} \\
    Data& Method & IC & \rotate{(1,[0])} & \rotate{(1,[1])} & \rotate{(1,[2])} & \rotate{(2,[0,0])} & \rotate{(2,[1,0])} & \rotate{(2,[1,1])} & \rotate{(1,[0])} & \rotate{(1,[1])} & \rotate{(1,[2])} & \rotate{(2,[0,0])} & \rotate{(2,[1,0])} & \rotate{(2,[1,1])} \\
    \hline\hline
    FIGNAR & Standard & BIC & 6 & \textbf{48} & 1 & 0 & 1 & 1 & 8 & 33 & 0 & 0 & 1 & 1 \\
    & Standard & AIC & 1 & \textbf{30} & 9 & 0 & 6 & 12 & 0 & 26 & 7 & 0 & 3 & 6 \\
    \hline
    GNARFI & Standard & BIC & 0 & 0 & 0 & 0 & 0 & 0 & 2 & \textbf{93} & 0 & 0 & 4 & 1 \\
    & Standard & AIC & 0 & 0 & 0 & 0 & 0 & 0 & 0 & \textbf{63} & 15 & 0 & 9 & 13 \\
    & Conditional & BIC & 0 & 0 & 0 & 0 & 0 & 0 & 0 & \textbf{54} & 1 & 0 & 40 & 5 \\
    & Conditional & AIC & 0 & 0 & 0 & 0 &  0 & 0 & 0 & \textbf{11} & 2 & 0 &  57 & 30 \\
    \hline
\end{NiceTabular}
\caption{Number of times each simulation minimised the BIC and the AIC, over 100 repetitions. The bold columns highlight results corresponding to the correct model and order.}\label{tab:data-model-mismatch}
\end{table}

\begin{table}[!ht]
\centering
\begin{NiceTabular}{ccc||cccccc||cccccc}%{@{}c!{\quad}cccccc!{\quad}cccccc@{}}
    \Block{1-15}{\hspace{5.3cm} $(p,[\ss])$ Order}\\
    &&& \Block{1-6}{Global-$\alpha$} &&&&&& \Block{1-6}{Individual-$\alpha$} \\
    Model & Method & IC & \rotate{(1,[0])} & \rotate{\textbf{(1,[1])}} & \rotate{(1,[2])} & \rotate{(2,[0,0])} & \rotate{(2,[1,0])} & \rotate{(2,[1,1])} & \rotate{(1,[0])} & \rotate{(1,[1])} & \rotate{(1,[2])} & \rotate{(2,[0,0])} & \rotate{(2,[1,0])} & \rotate{(2,[1,1])} \\
    \hline\hline
    FIGNAR & Standard & BIC & 14 & \textbf{80} & 1 & 0 & 1 & 2 & 0 & 2 & 0 & 0 & 0 & 0 \\
    & Standard & AIC & 0 & \textbf{34} & 7 & 0 & 7 & 13 & 0 & 26 & 11 & 0 & 2 & 0 \\
    \hline
    GNARFI & Standard & BIC & 2 & \textbf{93} & 0 & 0 & 4 & 1 & 0 & 0 & 0 & 0 & 0 & 0 \\
    & Standard & AIC & 0 & \textbf{54} & 14 & 0 & 7 & 13 & 0 & 9 & 0 & 0 & 3 & 0 \\
    & Conditional & BIC & 0 & \textbf{54} & 1 & 0 & 40 & 5 & 0 & 0 & 0 & 0 & 0 & 0 \\
    & Conditional & AIC & 0 & \textbf{8} & 1 & 0 & 40 & 23 & 0 & 10 & 5 & 0 & 12 & 1 \\
    \hline
\end{NiceTabular}
\caption{Number of times each simulation minimised the BIC and the AIC, over 100 repetitions. The bold column highlights results corresponding to the correct model and order.}\label{tab:order}
\end{table}

Table \ref{tab:comp times} reports the empirical runtimes of the proposed estimation method under various parameter specifications. We simulate 10 realisations according to DGP2 on the 5-dimensional graph, and we record the estimation times under the various global and individual parameter specifications and different time series lengths.

\begin{table}[!ht]
\centering
\begin{subtable}[t]{\linewidth}
\centering
\begin{NiceTabular}{lc||ccc}
Par. specification & N. par. & 200 & 500 & 1000 \\
\hline\hline
& & 8.54 & 34.83 & 127.51 \\
gl-$\{\alpha, d, \sigma^{2}\}$ & 4 & 7.29 & 31.62 & 122.65 \\
& & 5.68 & 30.37 & 111.38 \\
\hline
& & 16.02 & 66.94 & 221.11 \\
gl-$\{d, \sigma^{2}\}$, ind-$\alpha$ & 8 & 14.39 & 59.45 & 215.28 \\
&  & 11.92 & 53.16 & 226.81 \\
\hline
& & 17.03 & 65.87 & 205.41 \\
gl-$\{\alpha, \sigma^{2}\}$, ind-$d$ & 8 & 13.96 & 60.10 & 205.29 \\
&  & 11.66 & 52.61 & 192.48 \\
\hline
& & 13.96 & 57.10 & 198.63 \\
gl-$\{\alpha, d\}$, ind-$\sigma^{2}$ & 8 & 10.82 & 52.06 & 187.71 \\
&  & 10.08 & 47.40 & 180.26 \\
\hline
& & 30.36 & 104.61 & 334.38 \\
gl-$\sigma^{2}$, ind-$\{\alpha, d\}$ & 12 & 24.70 & 84.73 & 312.17 \\
&  & 19.87 & 90.65 & 308.35 \\
\hline
& & 22.38 & 88.16 & 308.08 \\
gl-$d$, ind-$\{\alpha, \sigma^{2}\}$ & 12 & 22.90 & 94.22 & 356.55 \\
&  & 18.09 & 88.51 & 325.96 \\
\hline
& & 22.86 & 85.56 & 272.52 \\
gl-$\alpha$, ind-$\{d, \sigma^{2}\}$ & 12 & 18.52 & 75.26 & 270.69 \\
& & 14.78 & 69.67 & 249.53 \\
\hline
& & 44.54 & 138.57 & 438.65 \\
ind-$\{\alpha, d, \sigma^{2}\}$ & 16 & 29.38 & 116.34 & 400.70 \\
&  & 26.81 & 109.64 & 390.51 \\
\end{NiceTabular}
\end{subtable}
\caption{Computational times (in seconds) for maximum likelihood estimates under the various parameter settings of FIGNAR model (first rows) and GNARFI model using either the standard (second rows) or the conditional likelihood (third rows) approach.}\label{tab:comp times}
\end{table}

\begin{figure}[H]
    \centering
    \includegraphics[width=0.4\textwidth]{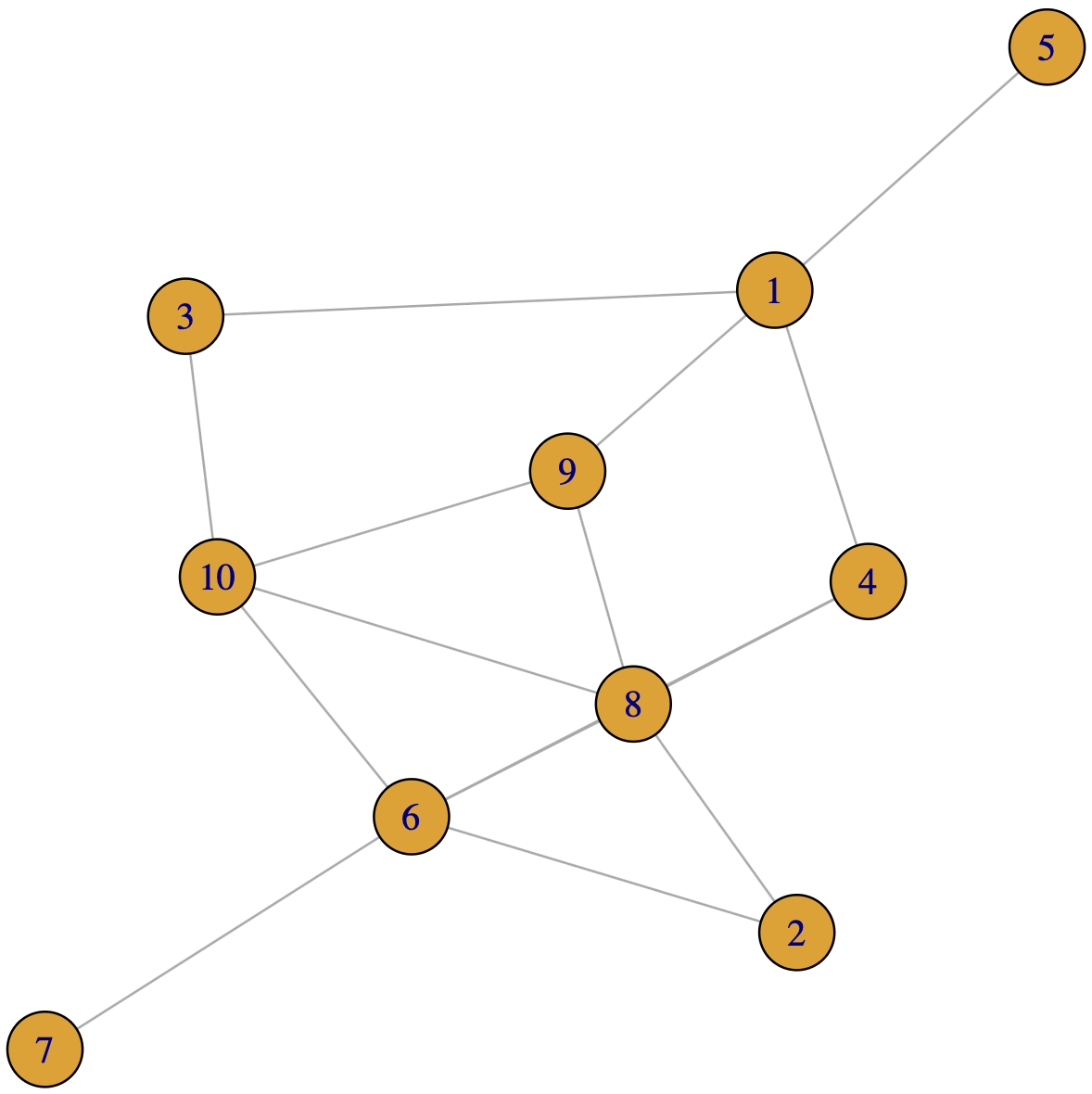}
    \captionof{figure}{Undirected, unweighted graph with ten nodes.}\label{img:tenNet}
\end{figure}

Figure~\ref{img:tenNet} visualises the 10-node graph used to test the performance of our estimation method for the simulation setup in Section \ref{sec:simstudy}.

\begin{figure}[H]
    \centering
    % Row 1
    \begin{subfigure}[t]{0.48\textwidth}
        \includegraphics[width=\linewidth]{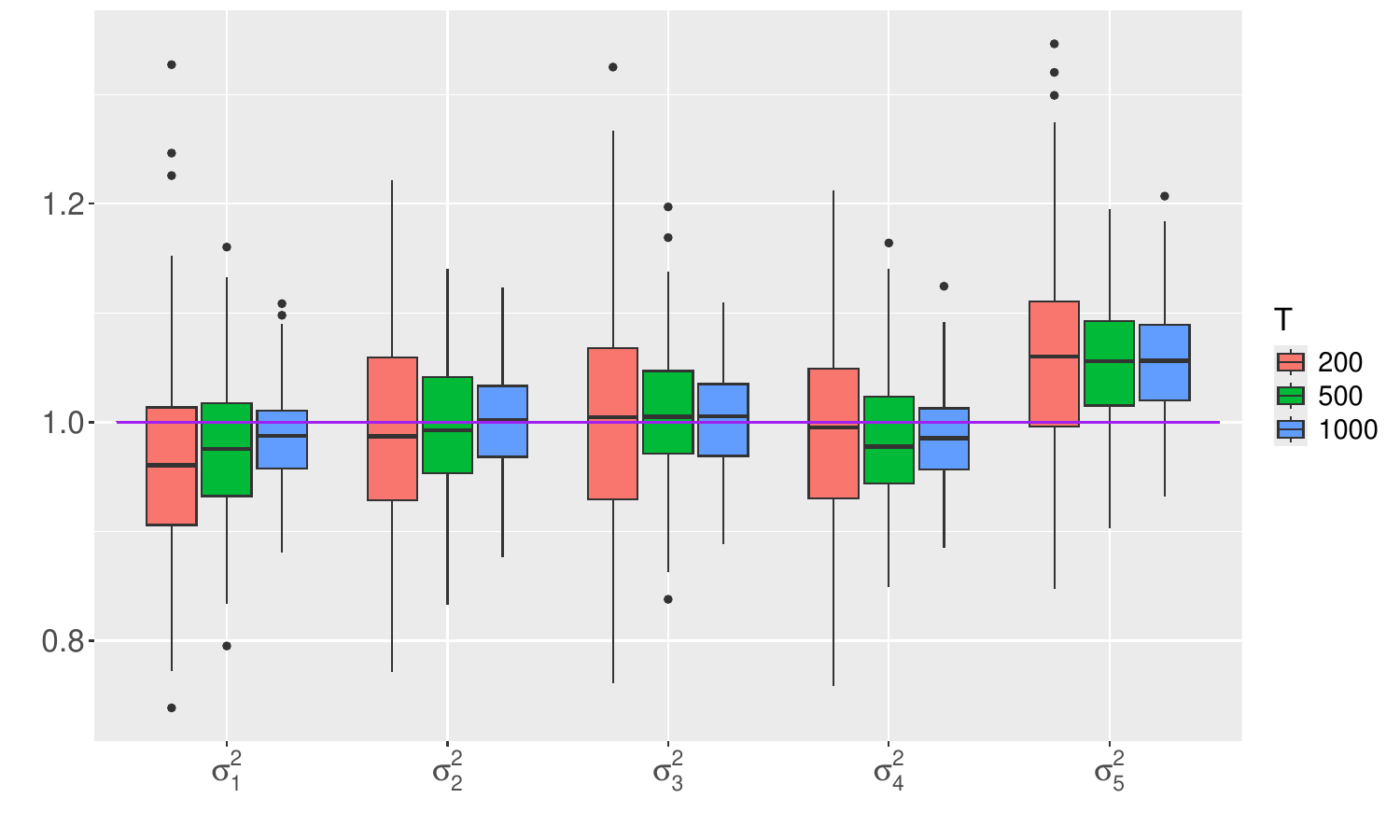}
    \end{subfigure}

    \medskip
    % Row 2
    \begin{subfigure}[t]{0.48\textwidth}
        \includegraphics[width=\linewidth]{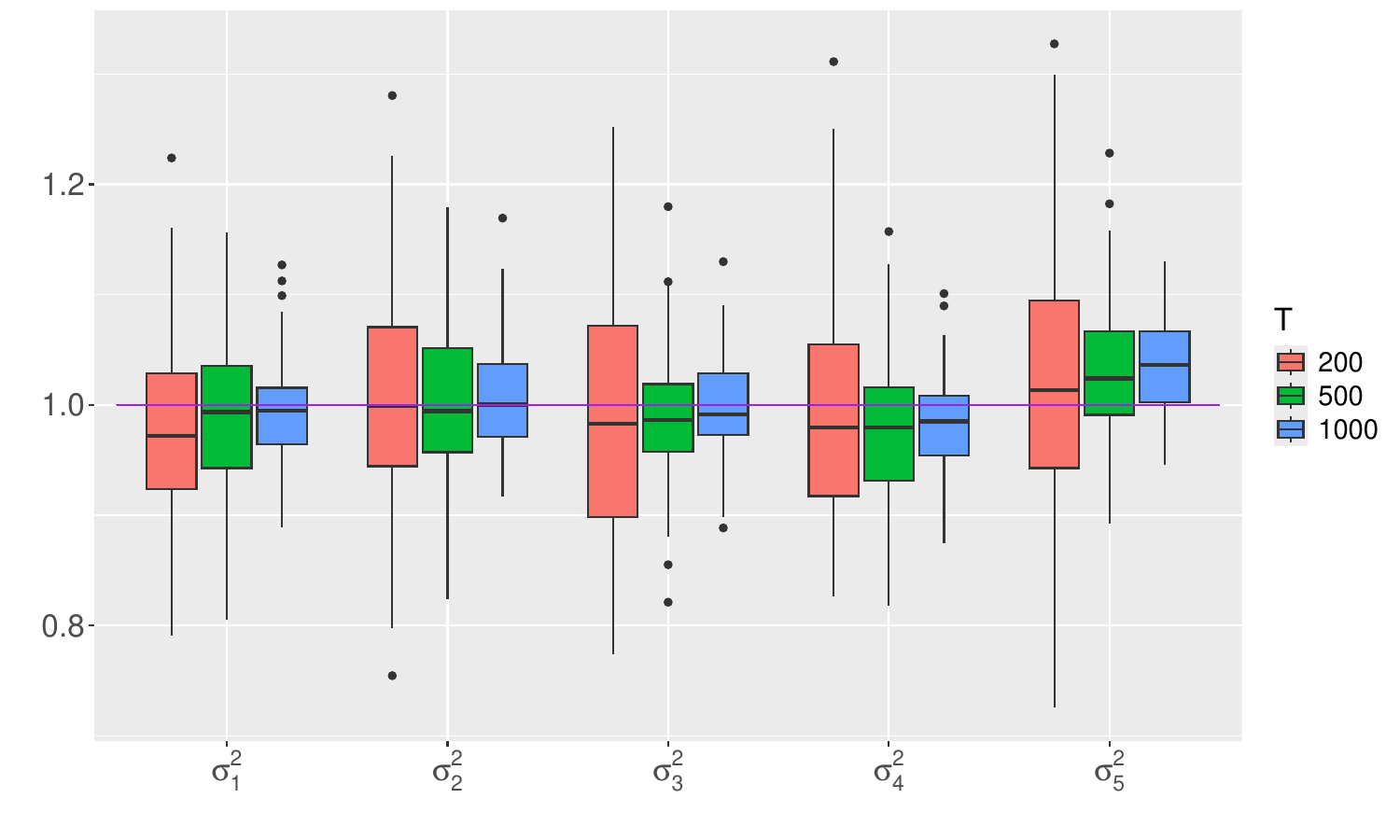}
    \end{subfigure}\hfill
    \begin{subfigure}[t]{0.48\textwidth}
        \includegraphics[width=\linewidth]{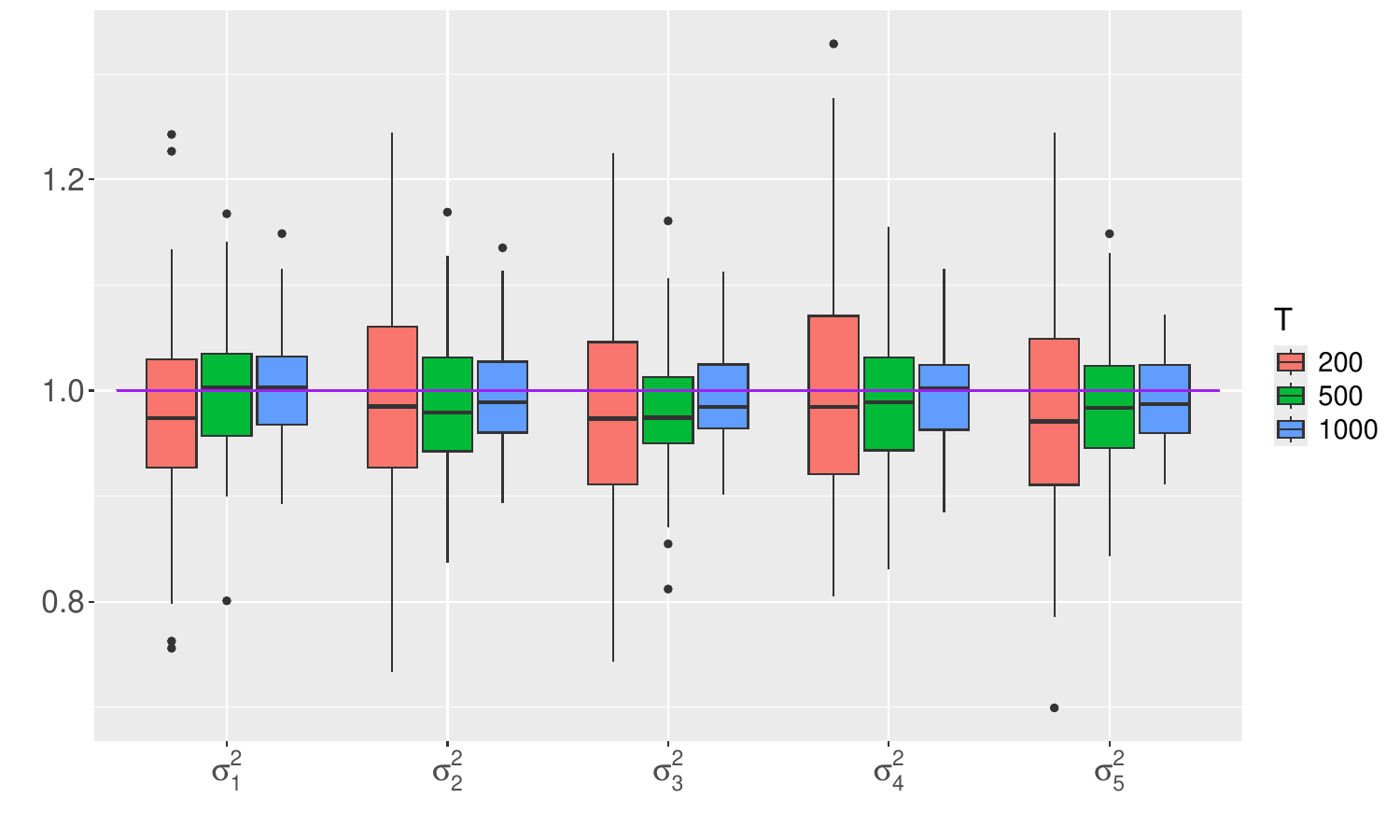}
    \end{subfigure}

    \caption{DGP1 - Boxplots of the estimated $\ssigma_{\vveps}^{2}$. Parameter estimations of FIGNAR(1,[1],$\dd$) (top), GNARFI(1,[1],$\dd$) with standard method (bottom-left), and GNARFI(1,[1],$\dd$) with conditional method (bottom-right).}\label{img:DGP1sig}
\end{figure}

\begin{figure}[H]
\centering
% Row 1
\begin{subfigure}[t]{0.48\textwidth}
    \includegraphics[width=\linewidth]{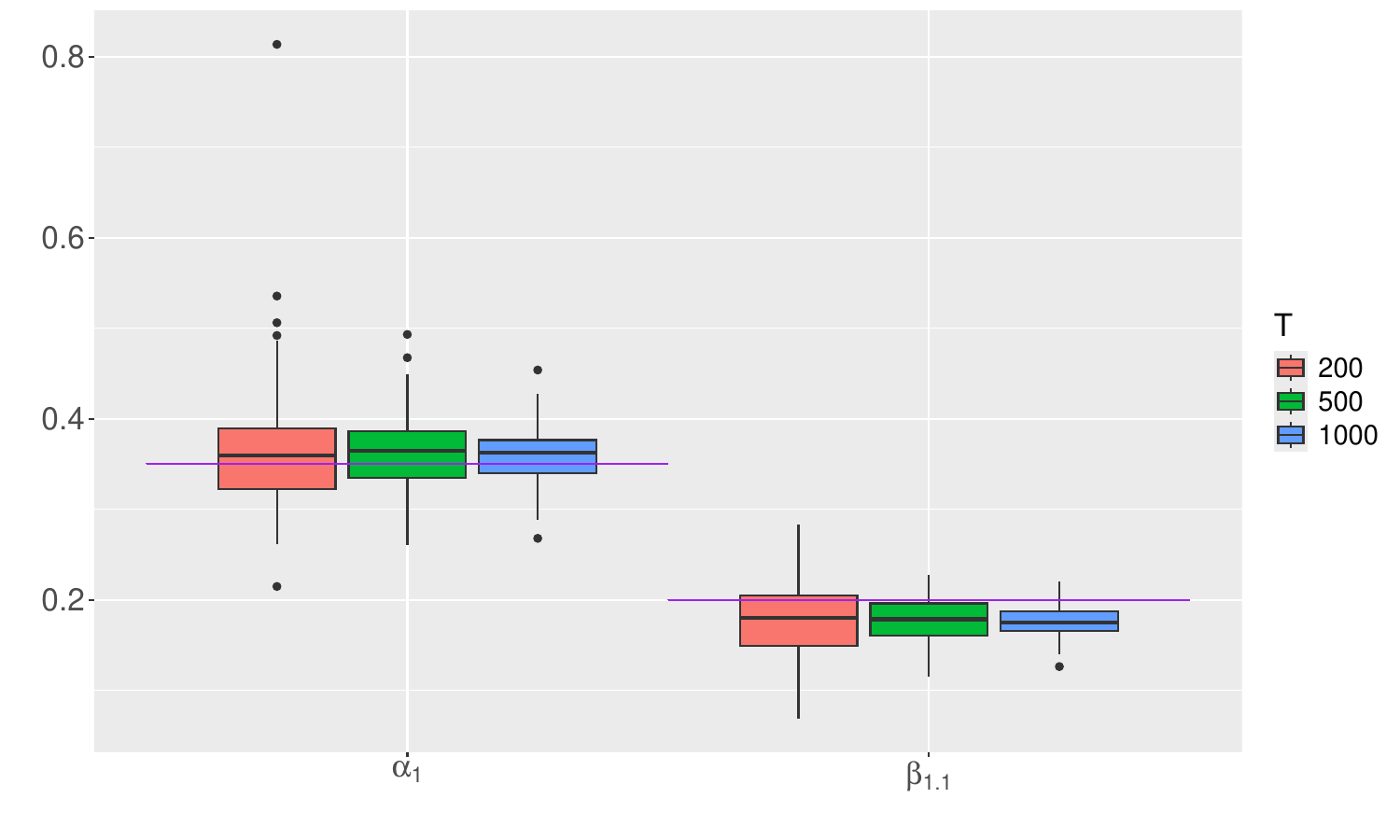}
\end{subfigure}\hfill
\begin{subfigure}[t]{0.48\textwidth}
    \includegraphics[width=\linewidth]{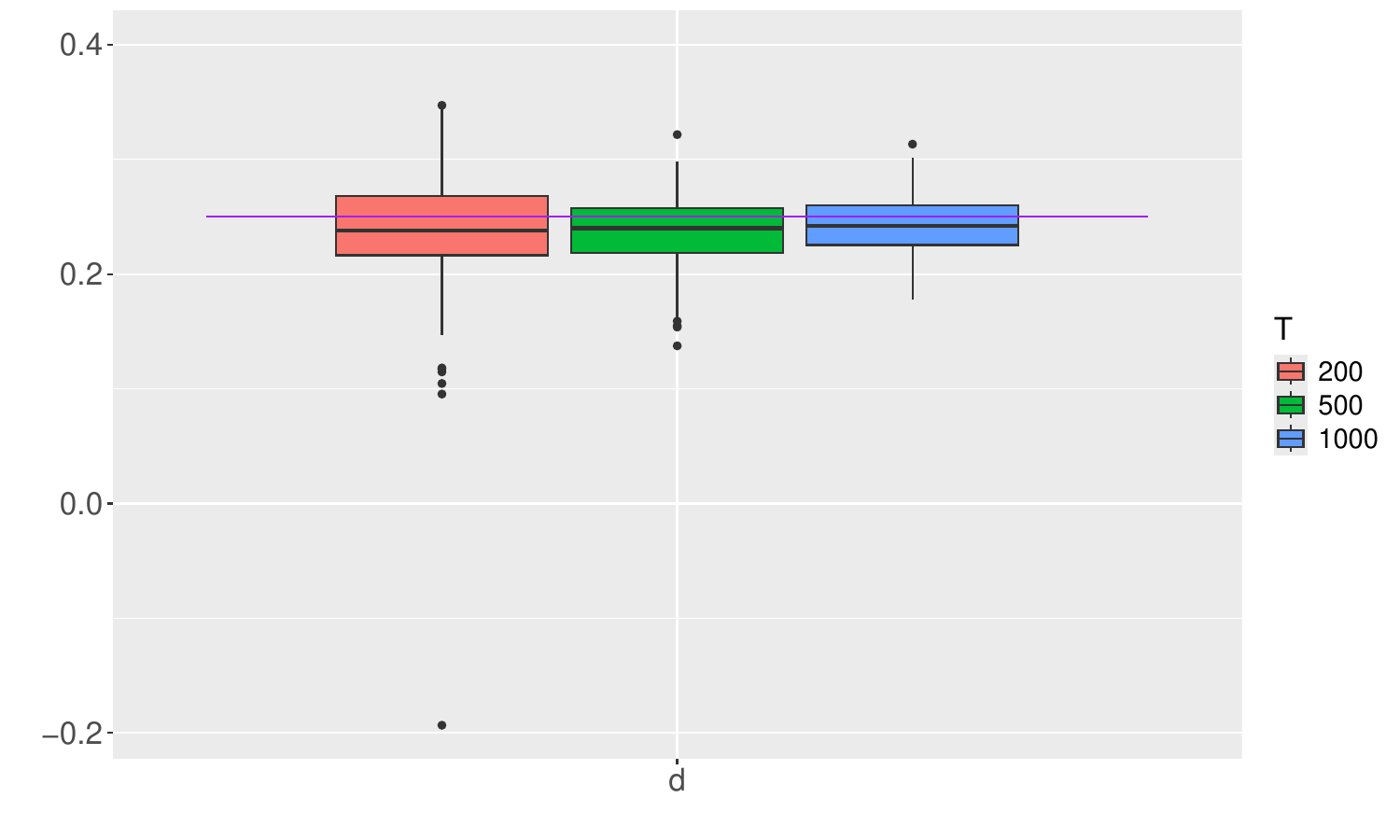}
\end{subfigure}
\begin{subfigure}[t]{0.48\textwidth}
    \includegraphics[width=\linewidth]{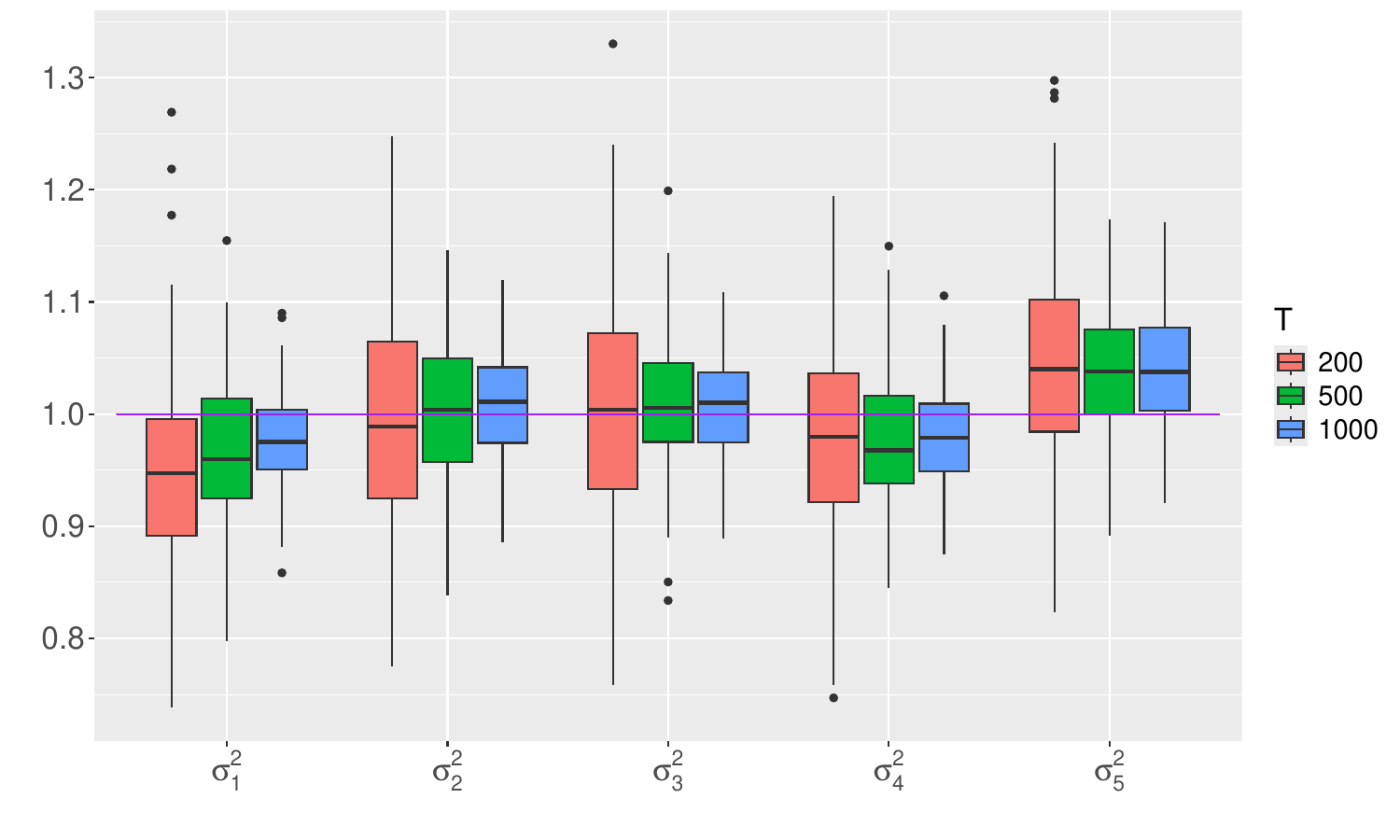}
\end{subfigure}
\caption{DGP2 - FIGNAR(1,[1],$\dd$), global-$d$ specification: boxplots of the estimated ${\aalpha,\bbeta}$ (top left), $\dd$ (top right), and $\ssigma_{\vveps}^{2}$ (bottom).}\label{img:DGP2-gld}
\end{figure}

\begin{figure}[H]
\centering
% Row 1
\begin{subfigure}[t]{0.48\textwidth}
    \includegraphics[width=\linewidth]{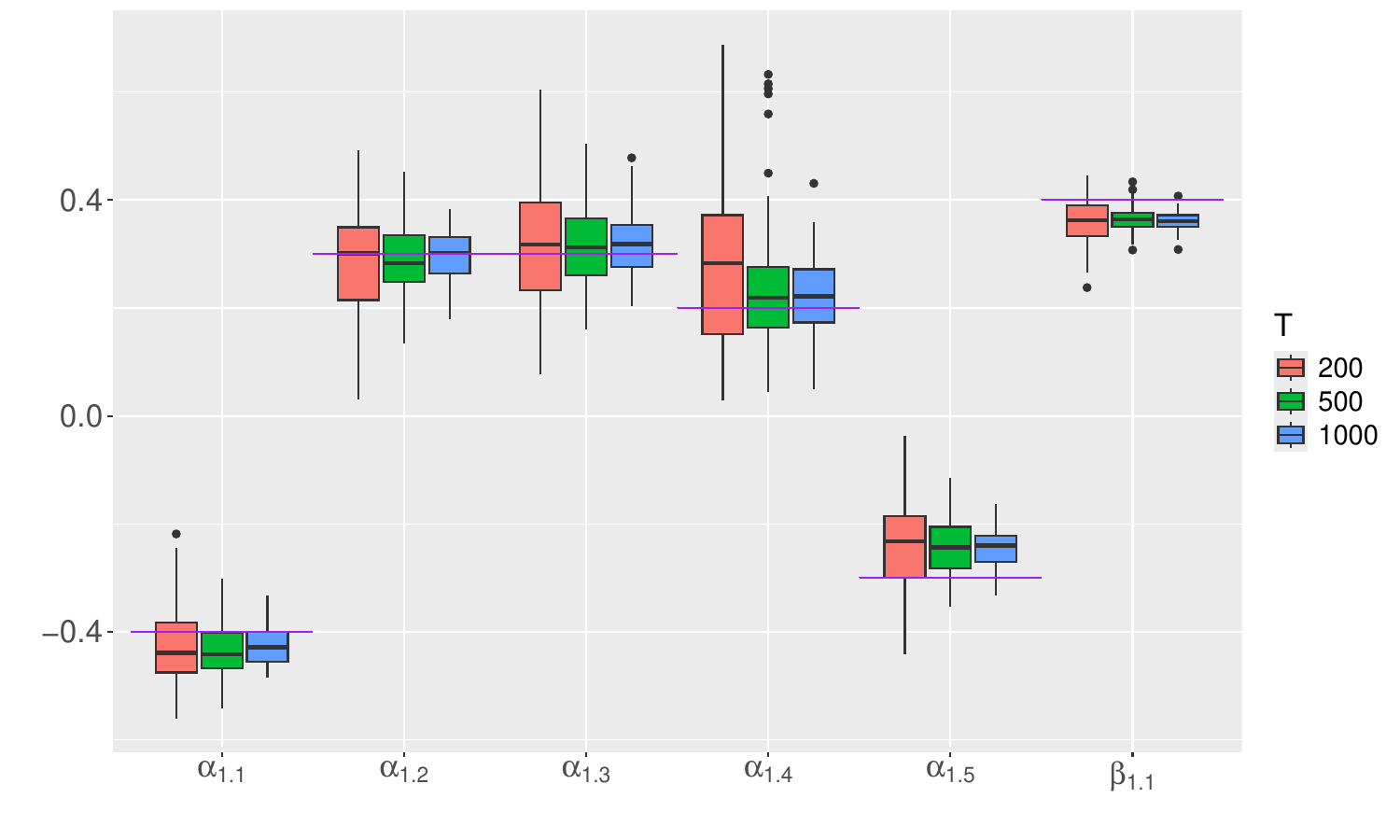}
\end{subfigure}
\hfill
\begin{subfigure}[t]{0.48\textwidth}
    \includegraphics[width=\linewidth]{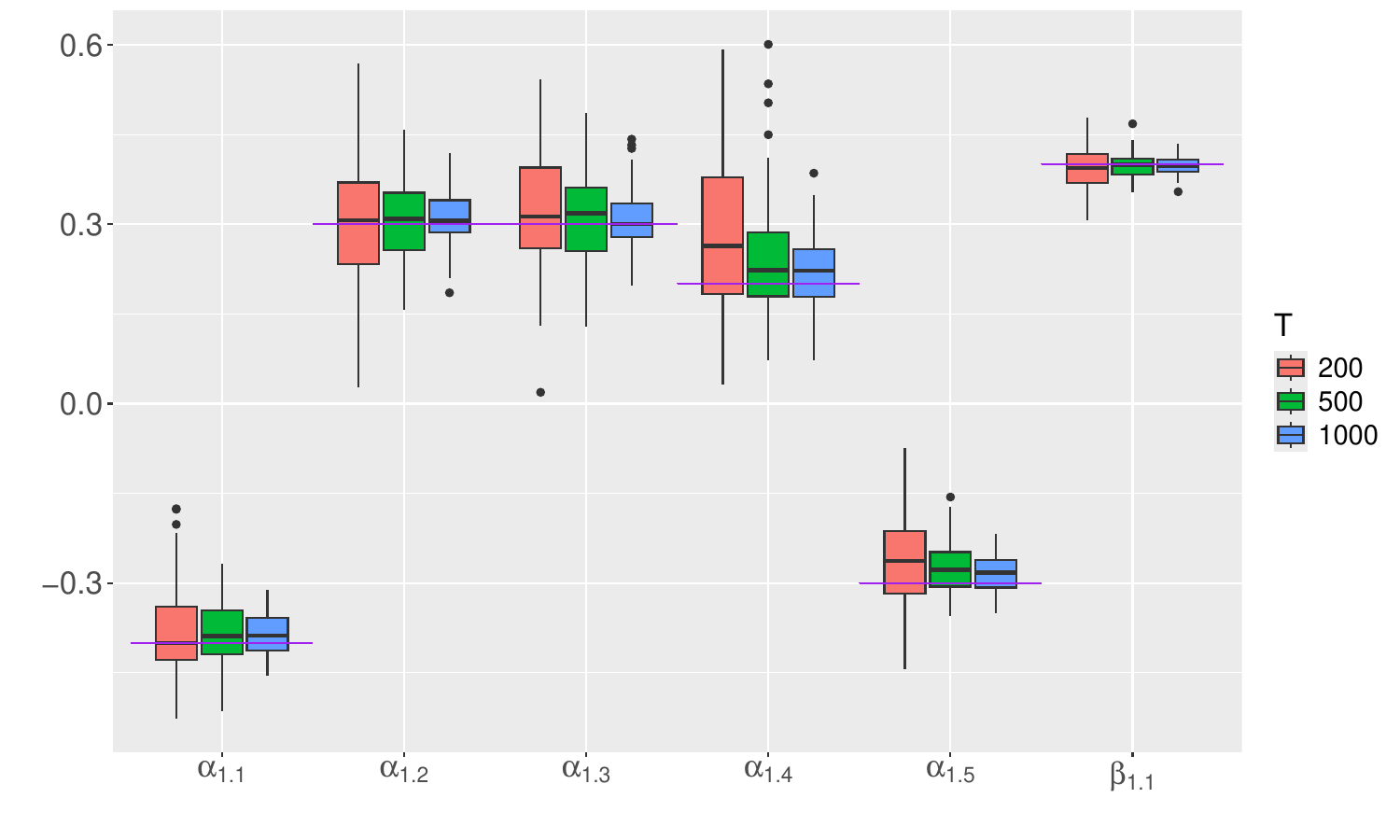}
\end{subfigure}

% Row 2
\begin{subfigure}[t]{0.48\textwidth}
    \includegraphics[width=\linewidth]{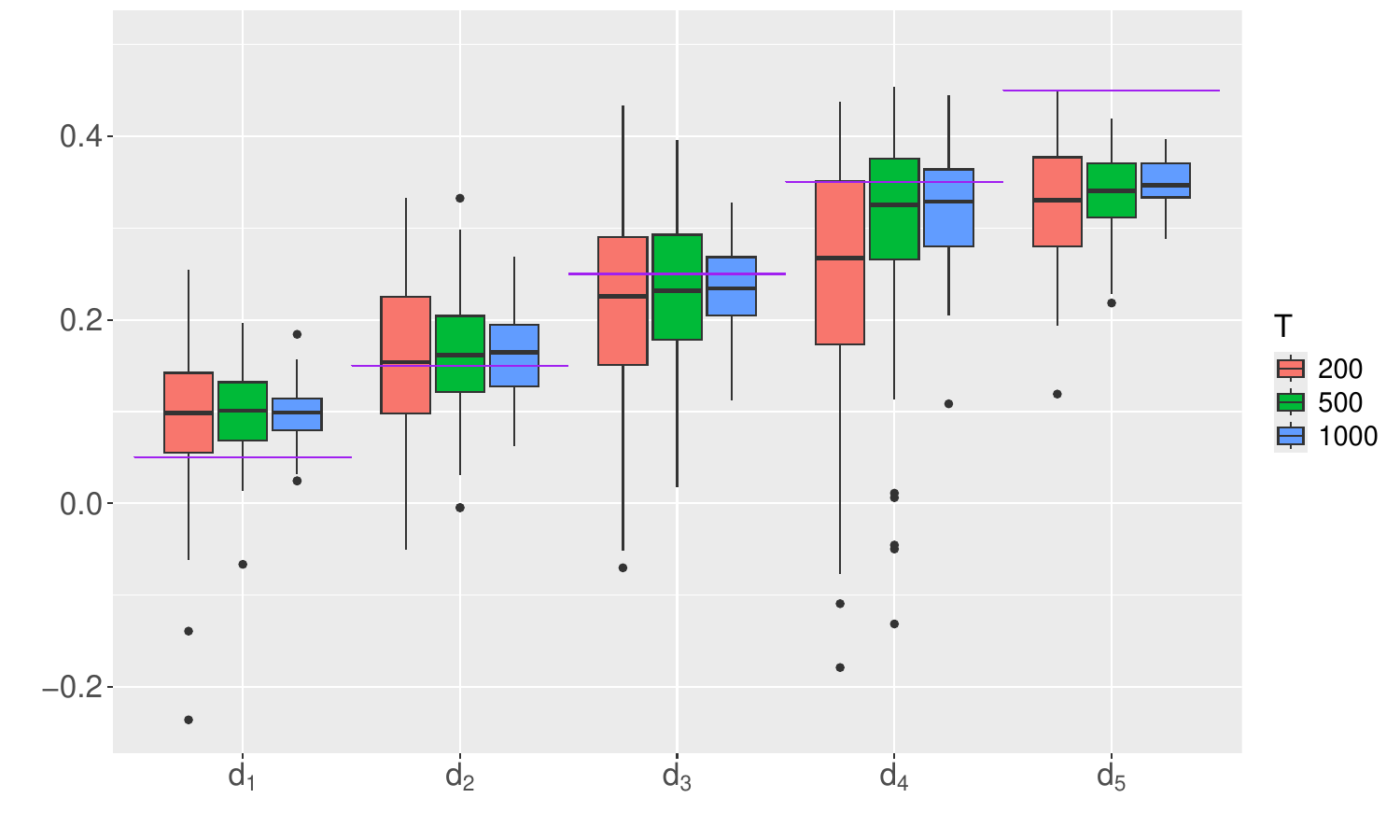}
\end{subfigure}
\hfill
\begin{subfigure}[t]{0.48\textwidth}
    \includegraphics[width=\linewidth]{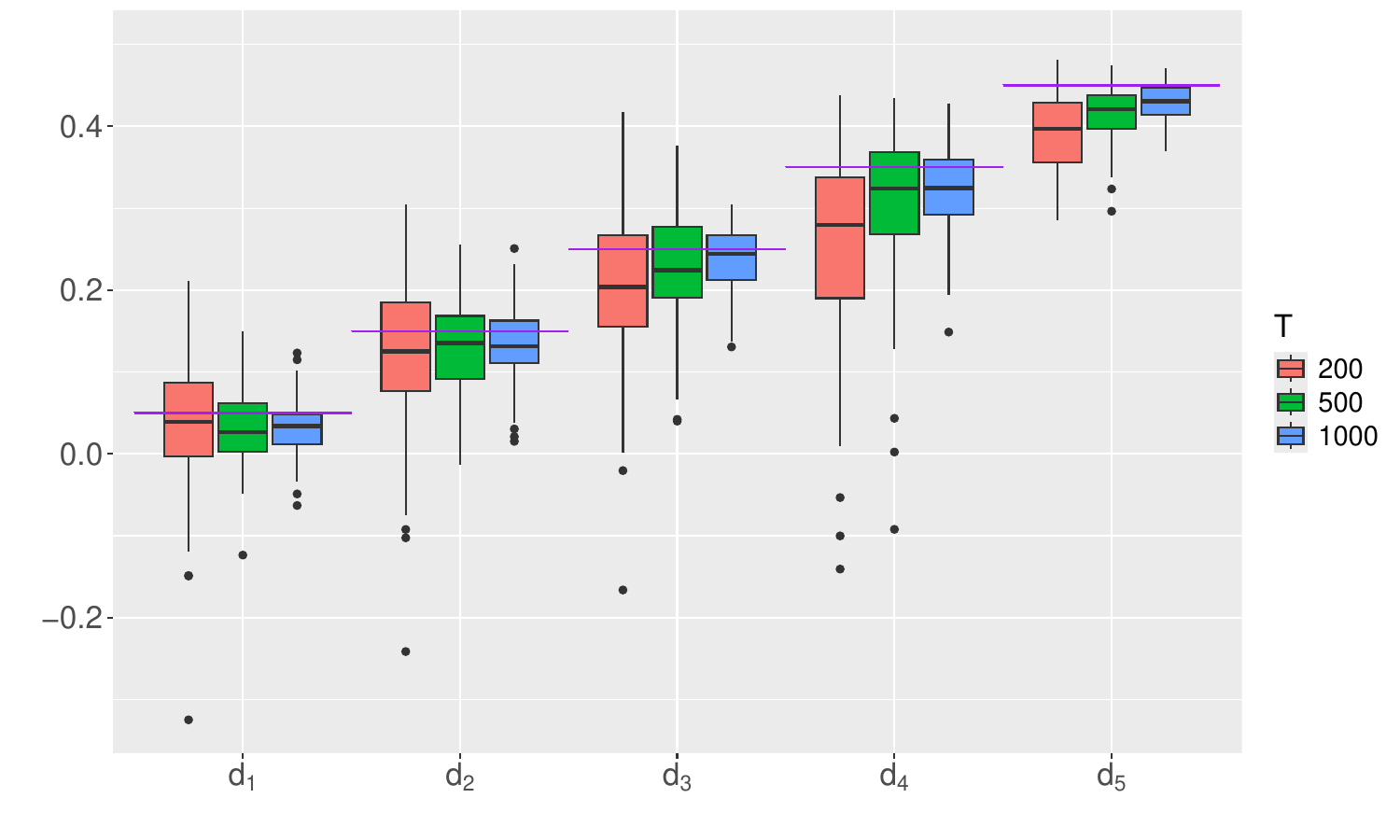}
\end{subfigure}

% Row 3
\begin{subfigure}[t]{0.48\textwidth}
    \includegraphics[width=\linewidth]{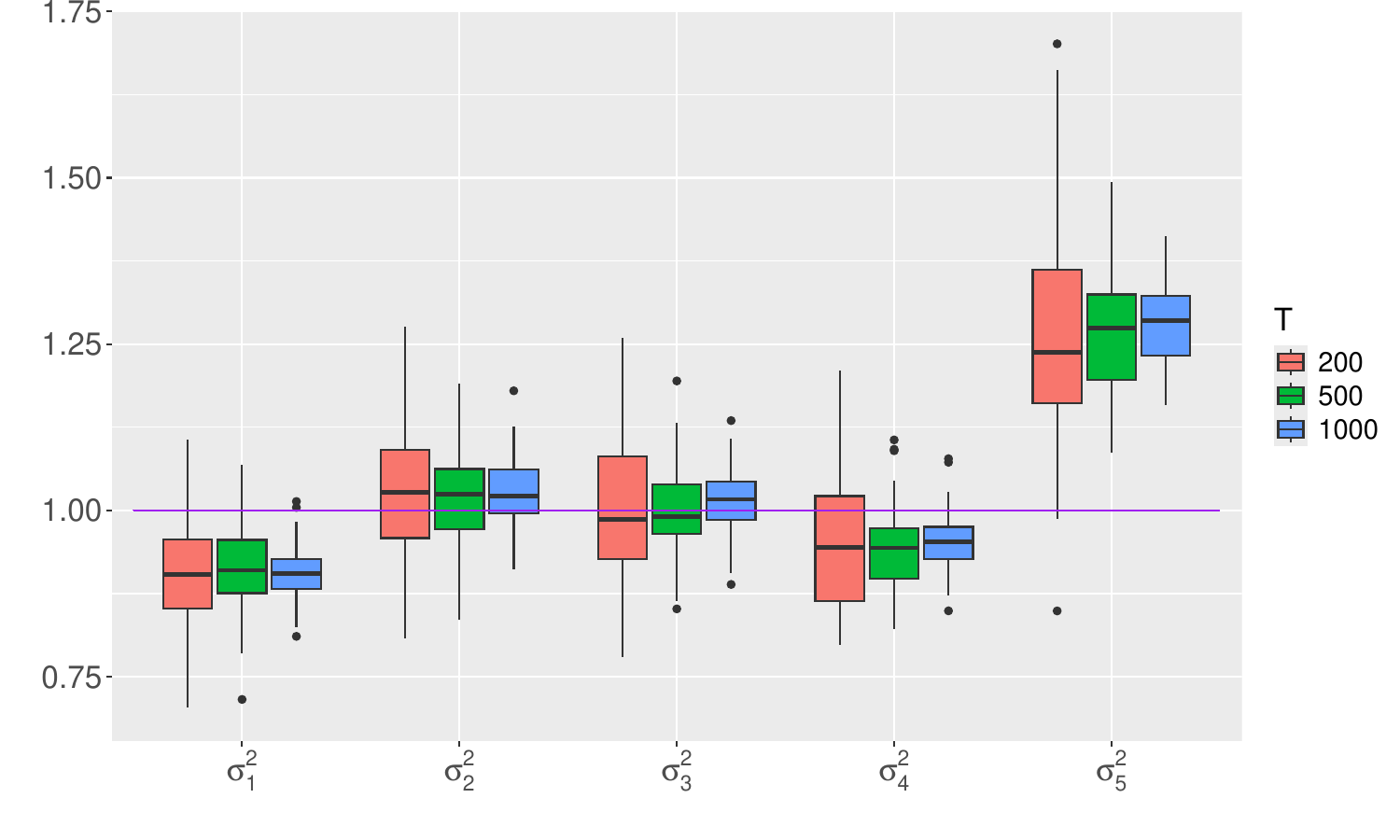}
\end{subfigure}
\hfill
\begin{subfigure}[t]{0.48\textwidth}
    \includegraphics[width=\linewidth]{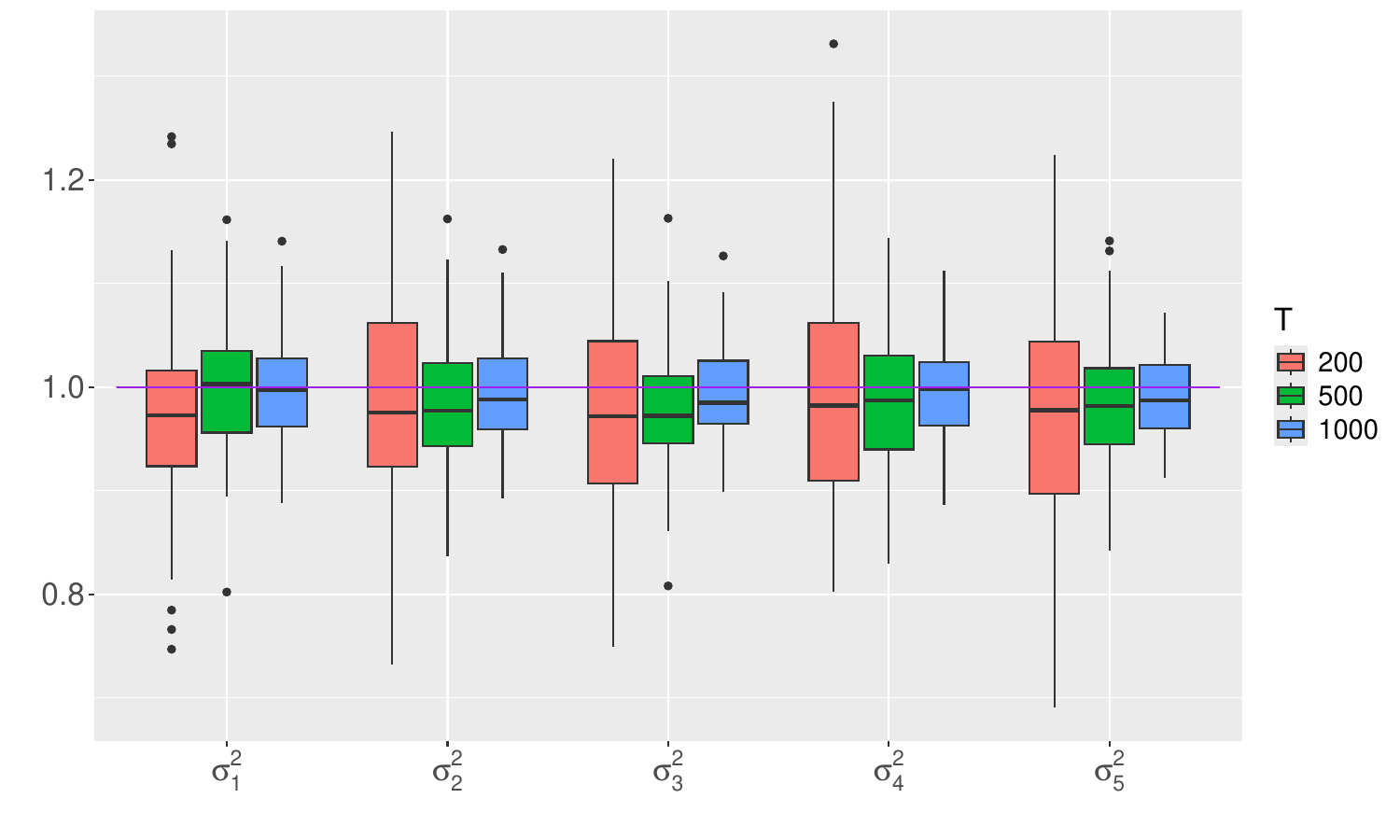}
\end{subfigure}
\caption{DGP3 - GNARFI(1,[1],$\dd$), individual-$\sigma^2$ specification: boxplots of the estimated ${\aalpha,\bbeta}$ (top), $\dd$ (middle), and $\ssigma_{\vveps}^{2}$ (bottom), shown for the standard method (left) and the conditional method (right).}\label{img:DGP3}
\end{figure}

Figures~\ref{img:DGP1sig}–\ref{img:DGP3} show boxplots of the estimated parameters for each simulation study in Section \ref{sec:simstudy}. For each parameter, three boxplots are shown for increasing sample sizes, $T=200$, $500$, and $1000$, coloured red, green, and blue, respectively. The purple horizontal lines mark the true values. GNARFI parameters are estimated with both the standard and the conditional estimation approaches. 

%% DGP1 noise covariance

% DGP2: global d FIGNAR

% DGP2: individual d FIGNAR
\begin{figure}[H]
\centering
% Row 1
\begin{subfigure}[t]{0.48\textwidth}
    \includegraphics[width=\linewidth]{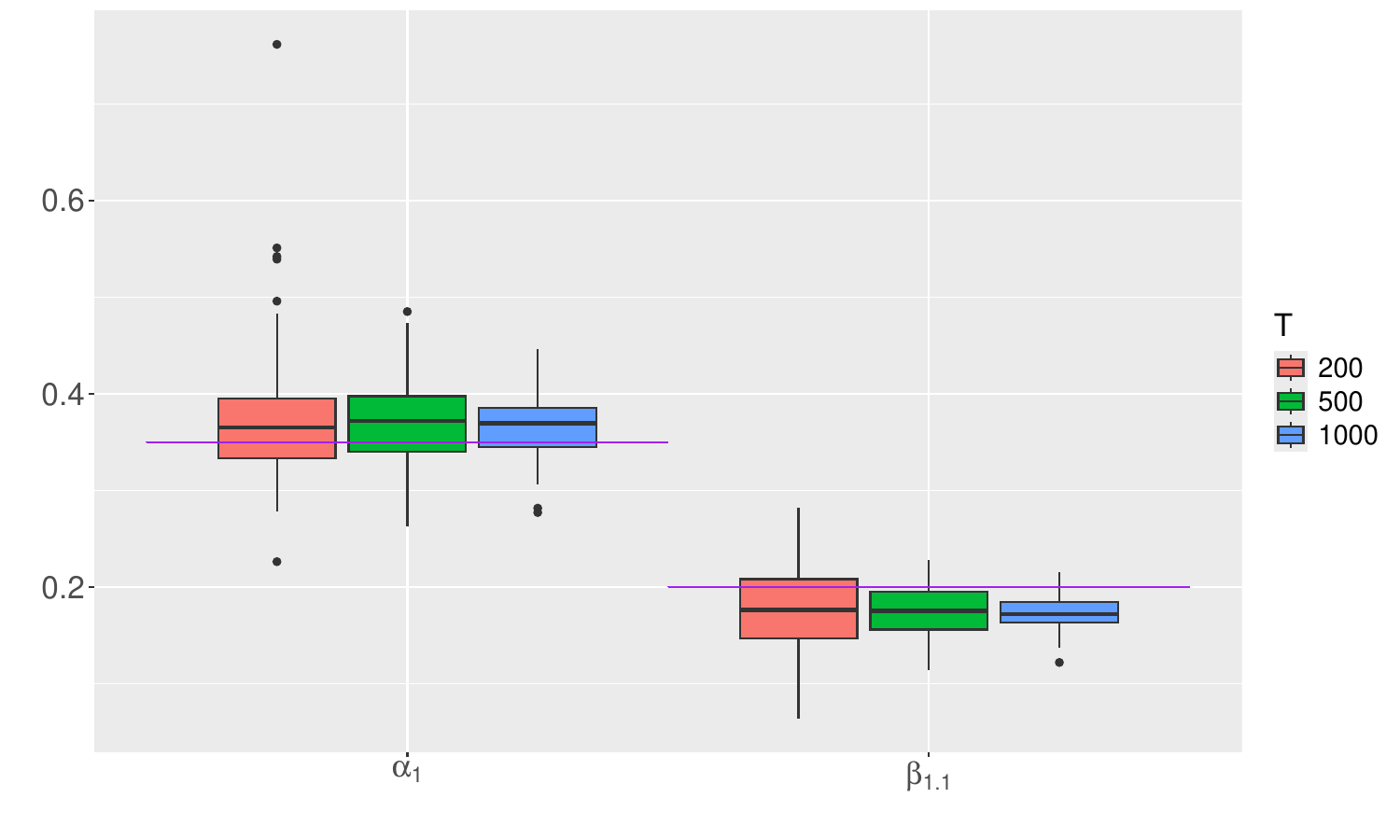}
\end{subfigure}\hfill
\begin{subfigure}[t]{0.48\textwidth}
    \includegraphics[width=\linewidth]{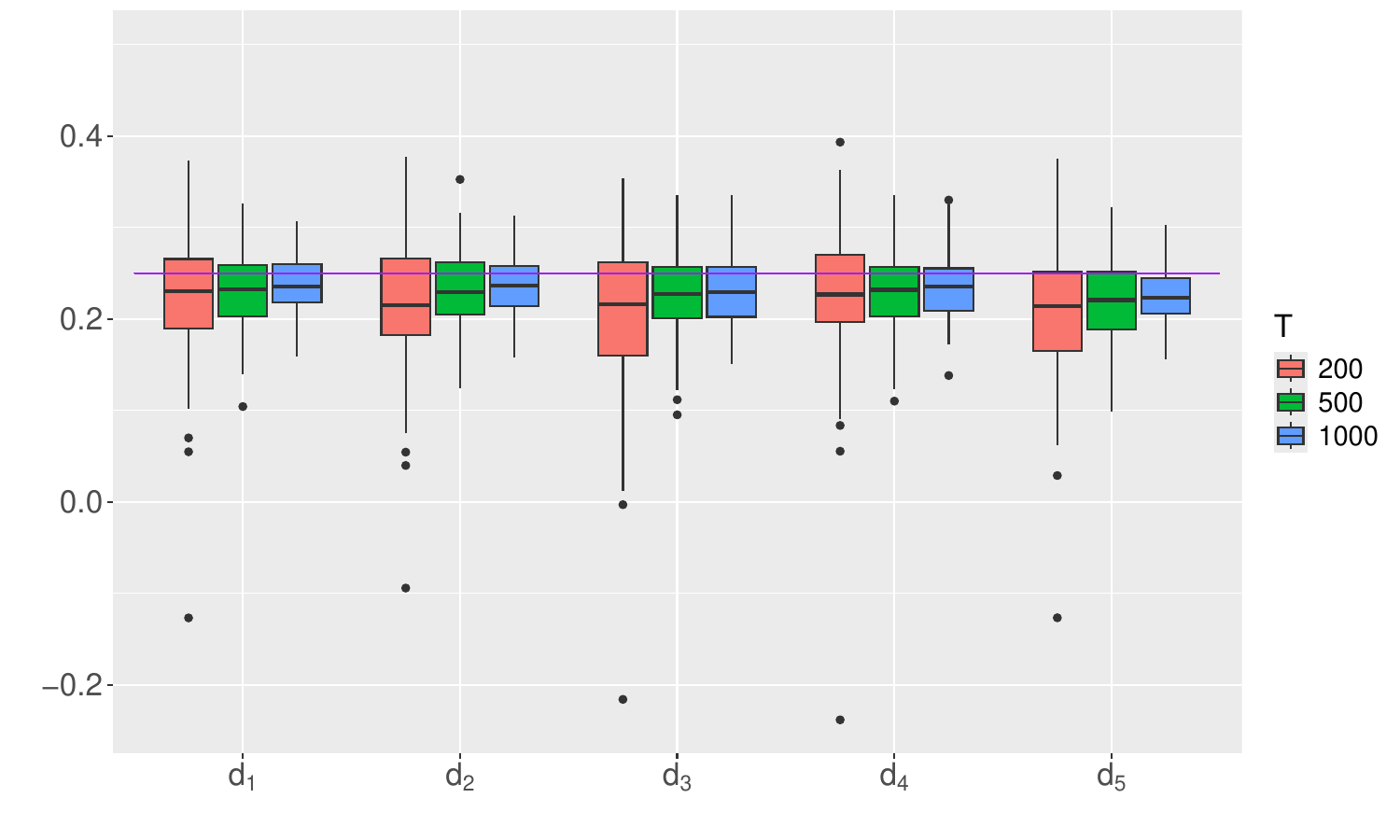}
\end{subfigure}
\begin{subfigure}[t]{0.48\textwidth}
    \includegraphics[width=\linewidth]{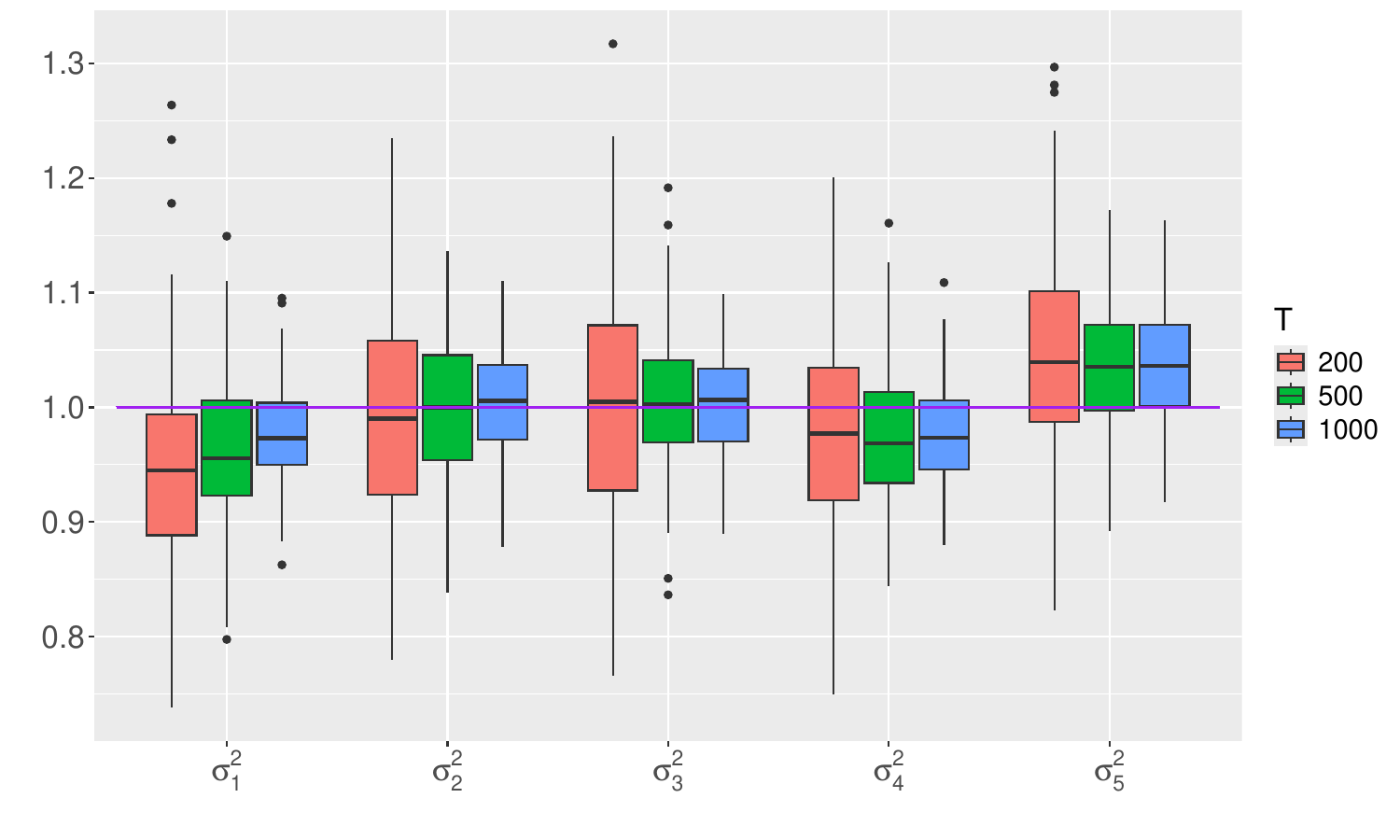}
\end{subfigure}
\caption{DGP2 - FIGNAR(1,[1],$\dd$), individual-$d$ specification: boxplots of the estimated ${\aalpha,\bbeta}$ (top left), $\dd$ (top right), and $\ssigma_{\vveps}^{2}$ (bottom).}
\end{figure}

% DGP2: global d GNARFI
\begin{figure}[H]
\centering
% Row 1
\begin{subfigure}[t]{0.48\textwidth}
    \includegraphics[width=\linewidth]{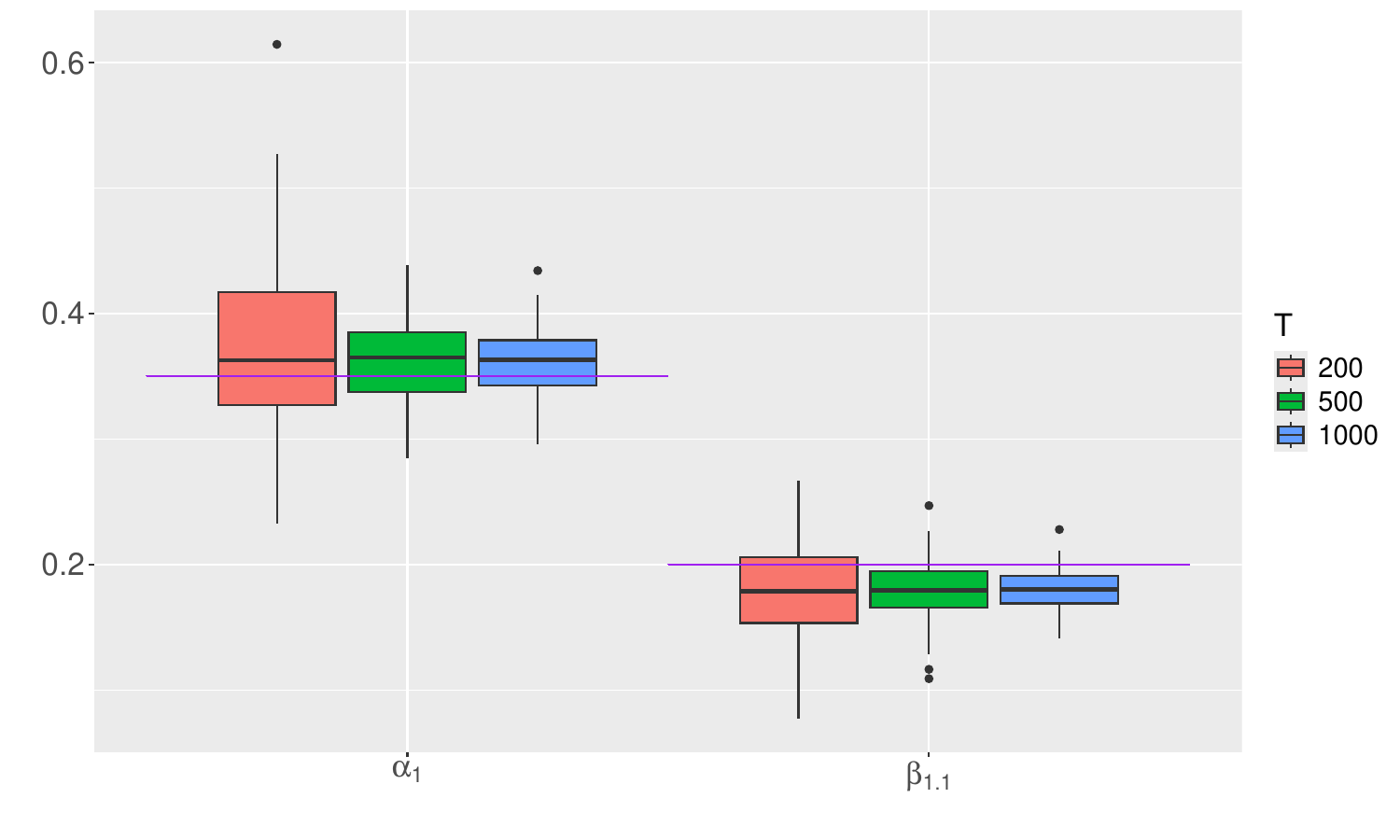}
\end{subfigure}\hfill
\begin{subfigure}[t]{0.48\textwidth}
    \includegraphics[width=\linewidth]{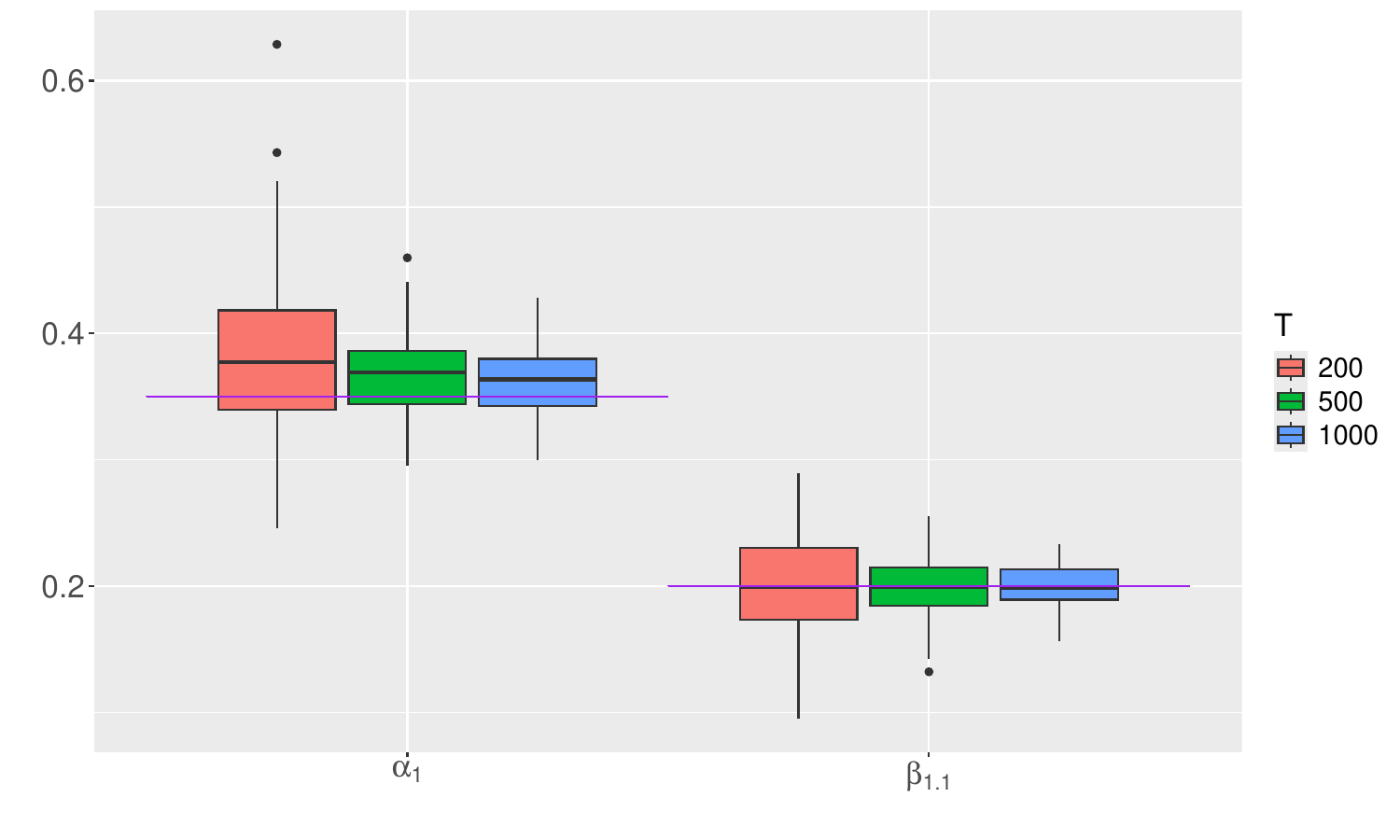}
\end{subfigure}

% Row 2
\begin{subfigure}[t]{0.48\textwidth}
    \includegraphics[width=\linewidth]{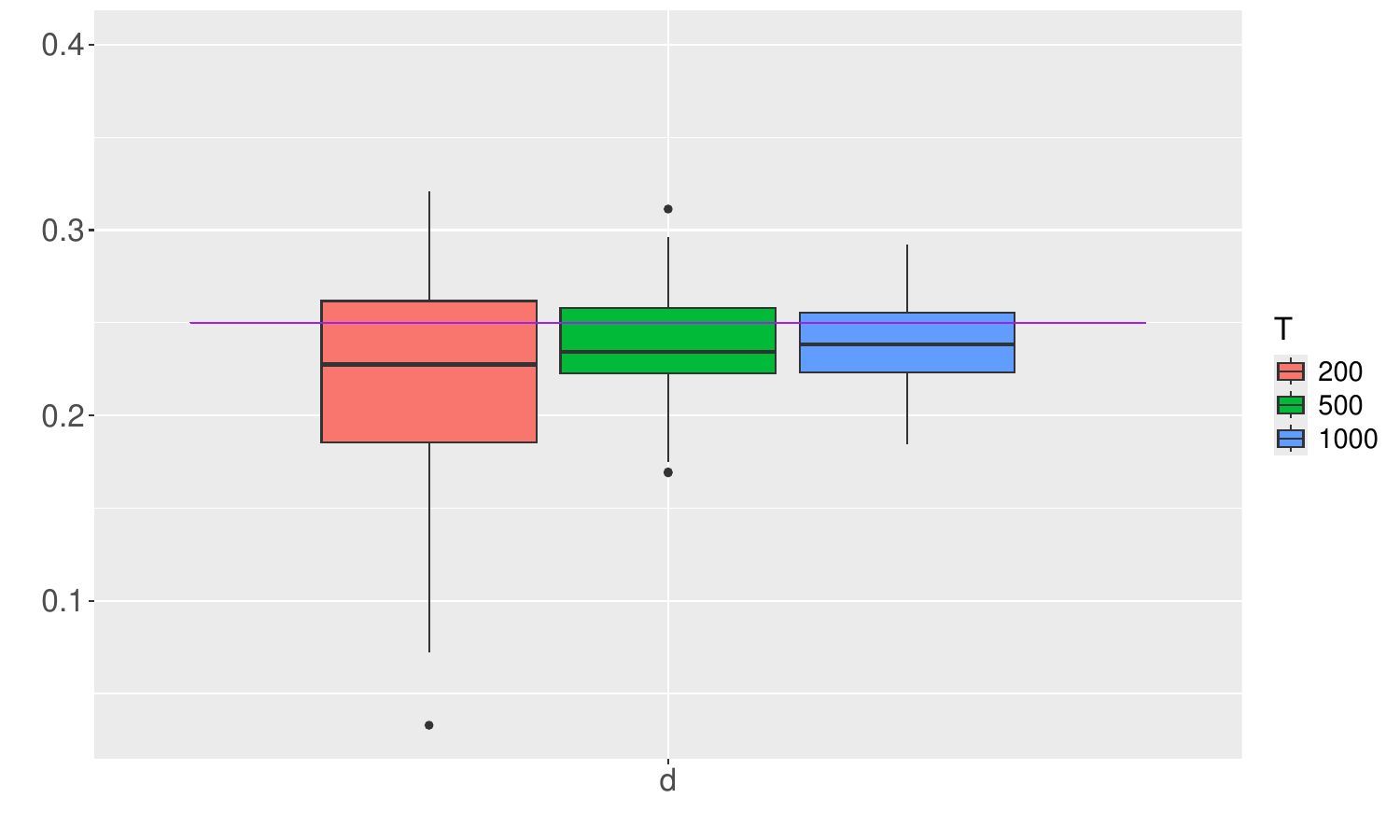}
\end{subfigure}\hfill
\begin{subfigure}[t]{0.48\textwidth}
    \includegraphics[width=\linewidth]{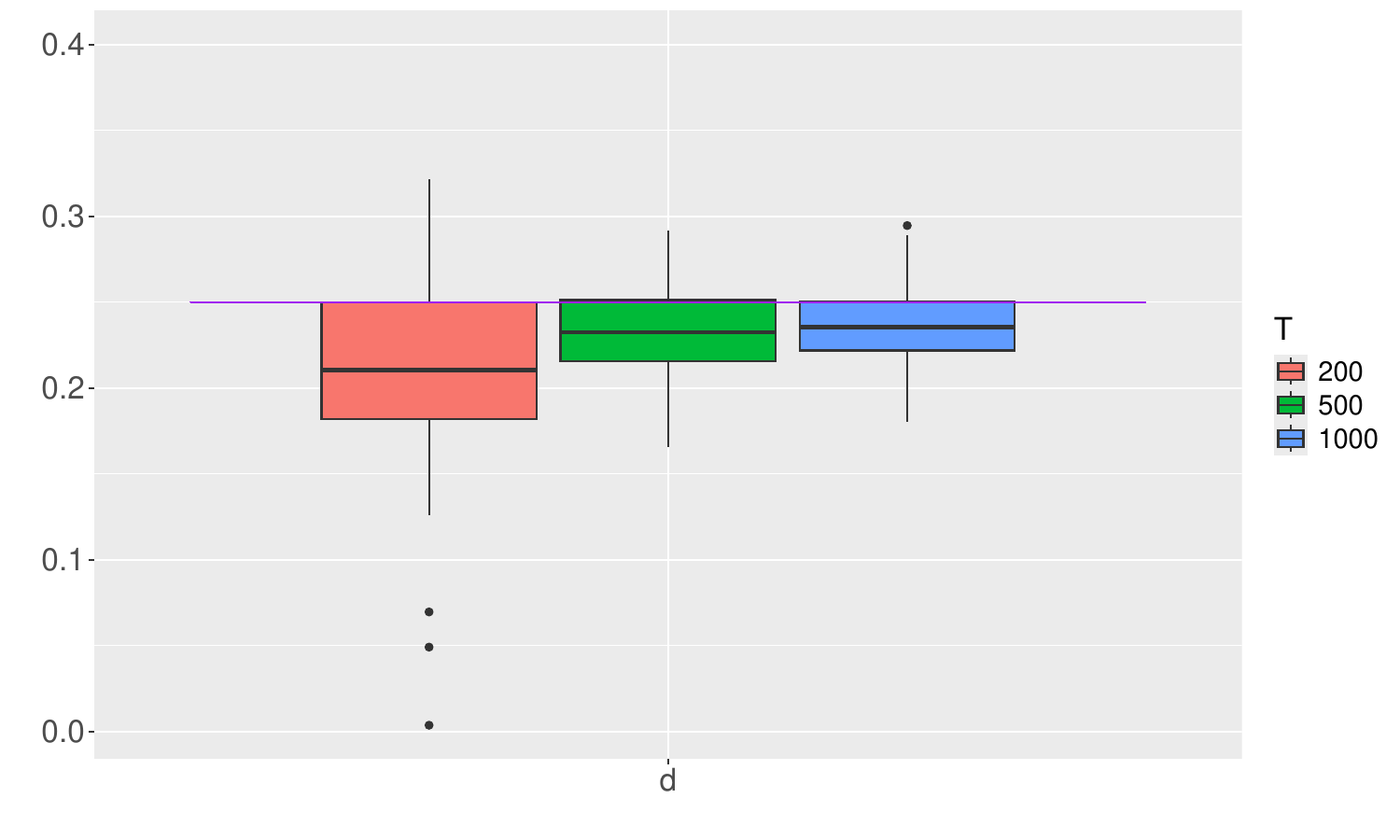}
\end{subfigure}

% Row 3
\begin{subfigure}[t]{0.48\textwidth}
    \includegraphics[width=\linewidth]{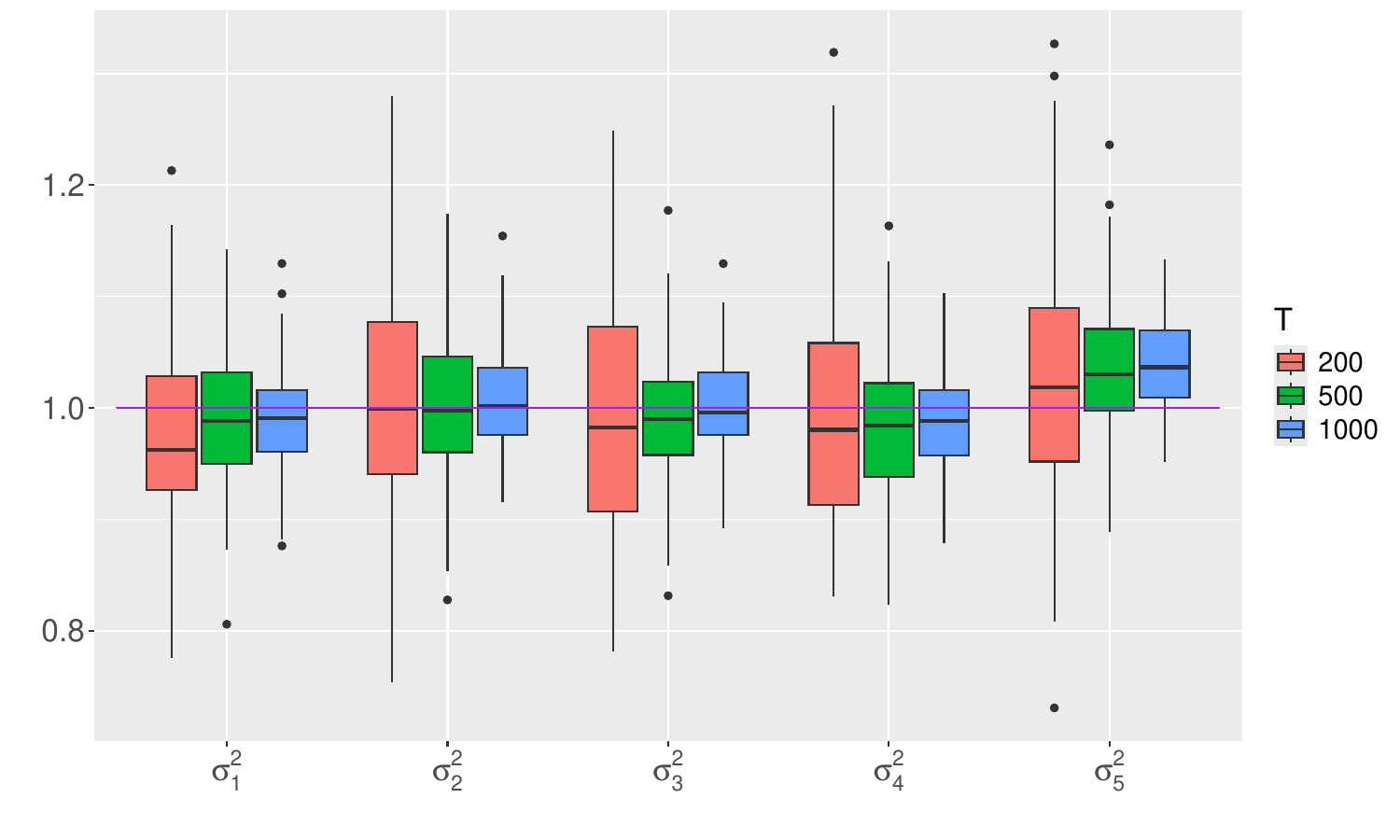}
\end{subfigure}
\begin{subfigure}[t]{0.48\textwidth}
    \includegraphics[width=\linewidth]{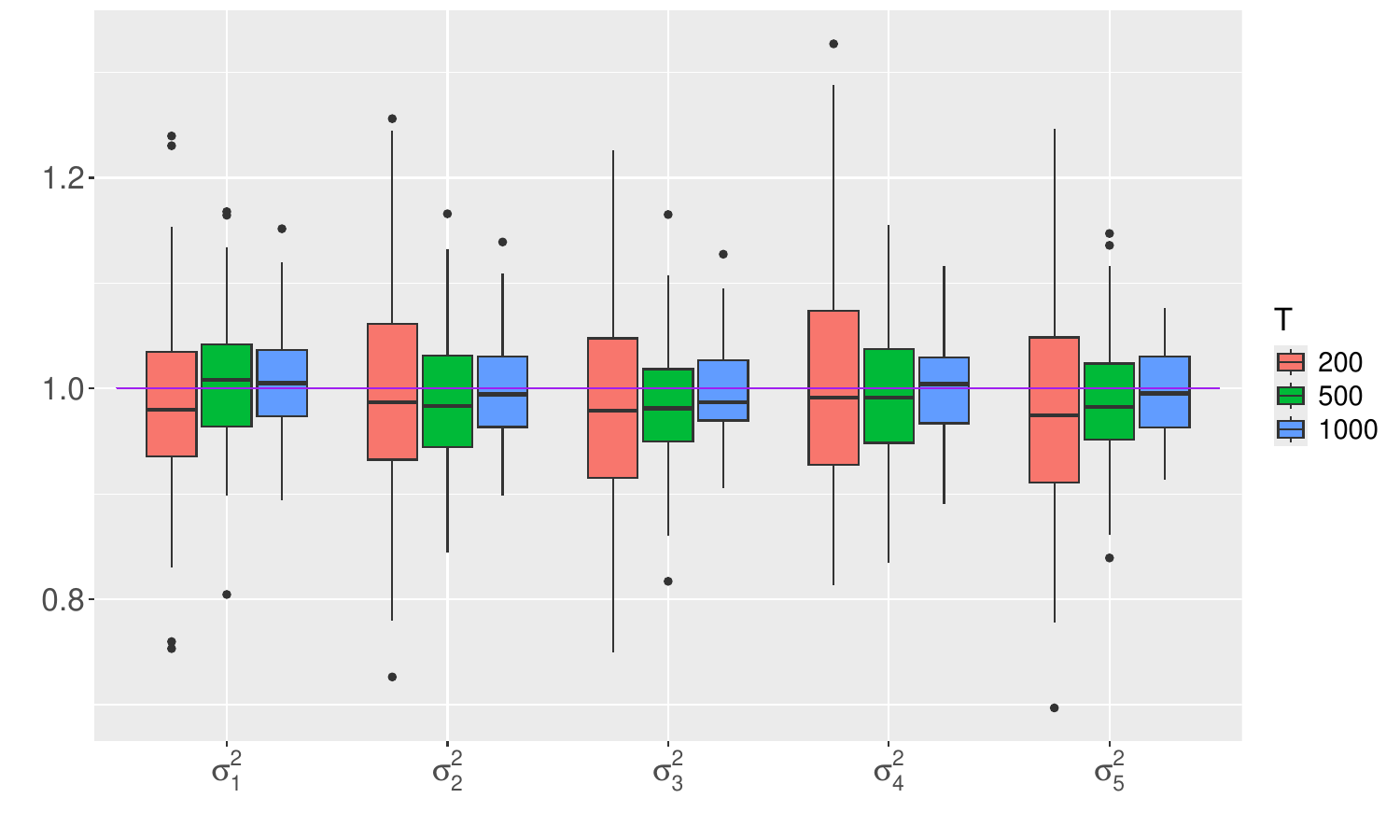}
\end{subfigure}

\caption{DGP2 - GNARFI(1,[1],$\dd$), global-$d$ specification: boxplots of the estimated ${\aalpha,\bbeta}$ (top), $\dd$ (middle), and $\ssigma_{\vveps}^{2}$ (bottom), shown for the standard method (left) and the conditional method (right).}
\end{figure}

% DGP2: global d GNARFI
\begin{figure}[H]
\centering
% Row 1
\begin{subfigure}[t]{0.48\textwidth}
    \includegraphics[width=\linewidth]{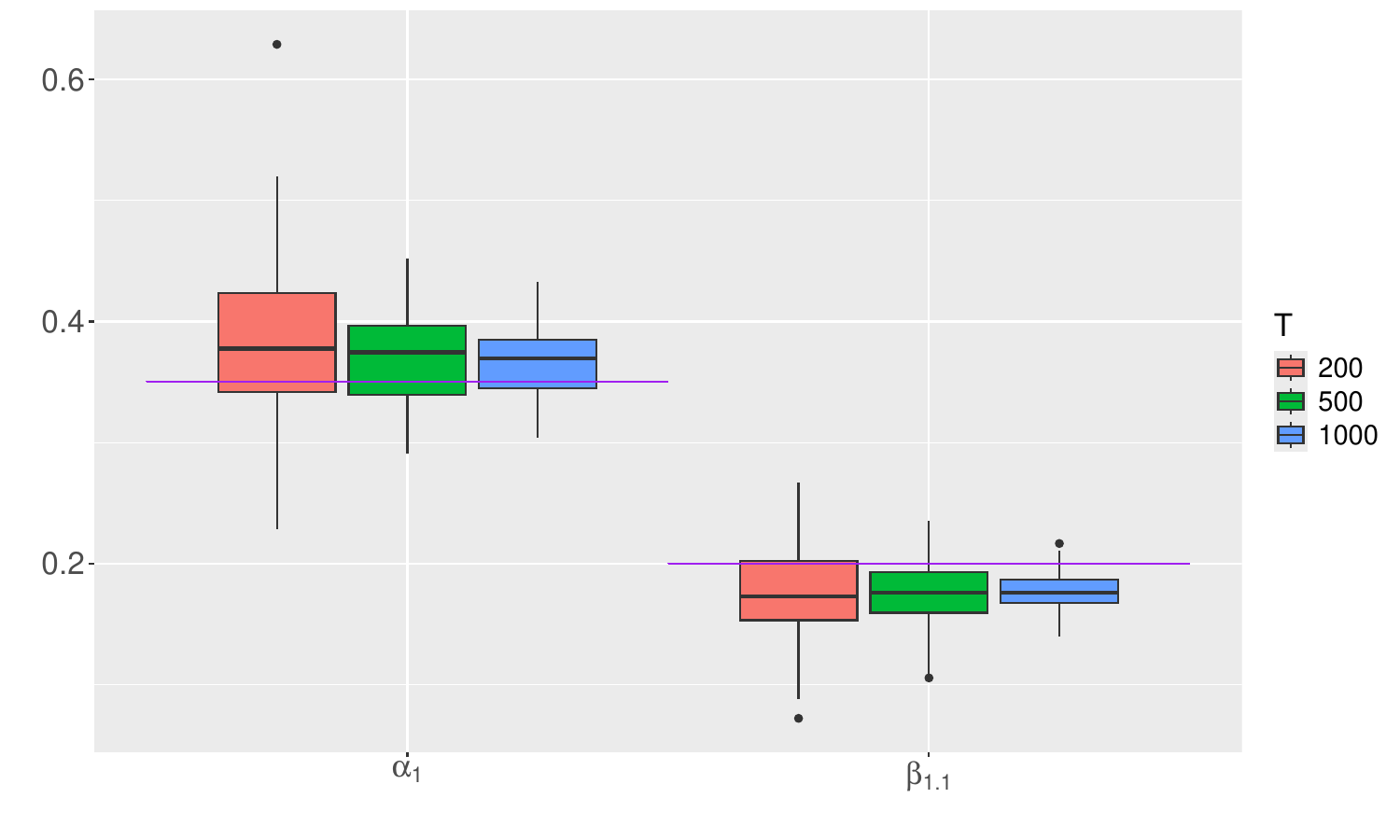}
\end{subfigure}
\hfill
\begin{subfigure}[t]{0.48\textwidth}
    \includegraphics[width=\linewidth]{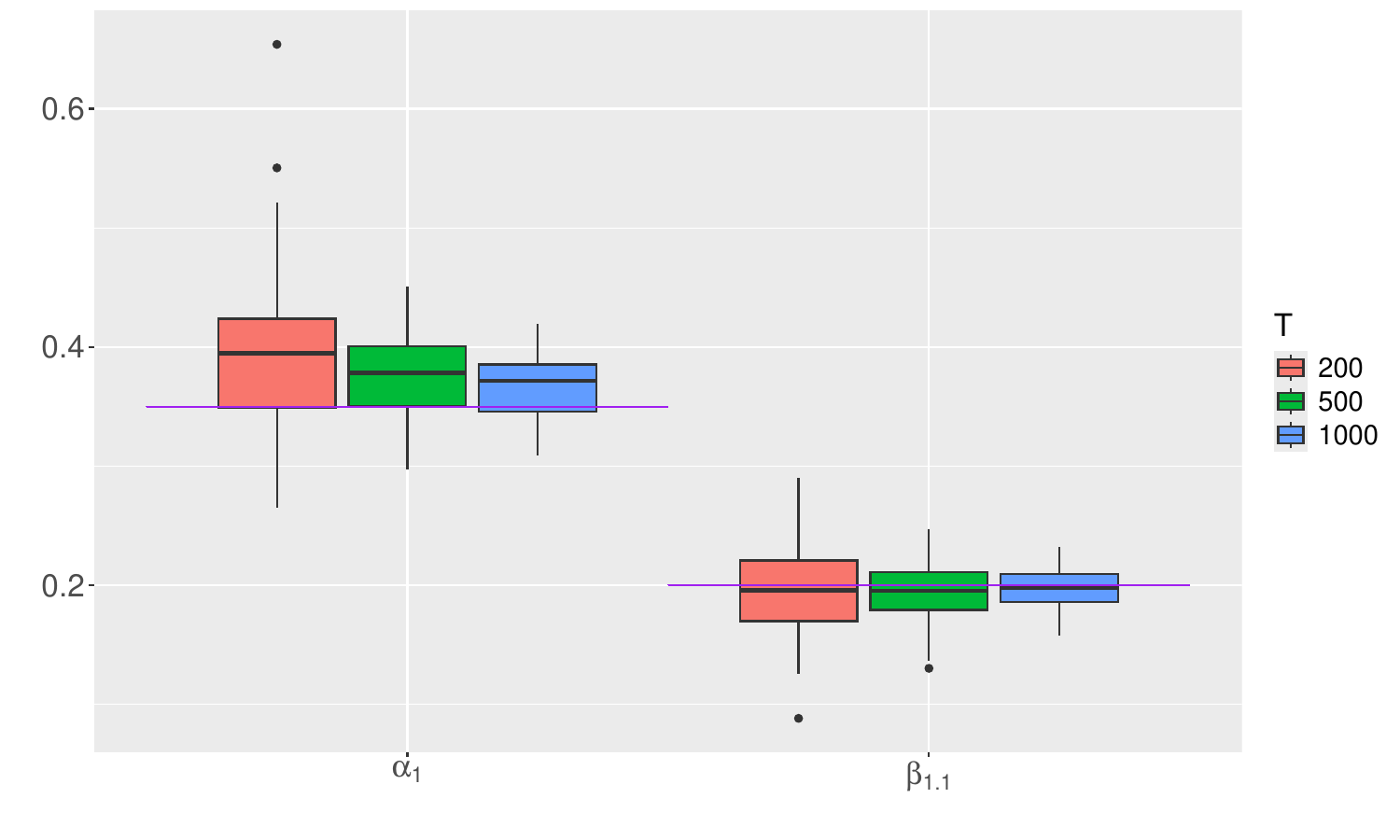}
\end{subfigure}

% Row 2
\begin{subfigure}[t]{0.48\textwidth}
    \includegraphics[width=\linewidth]{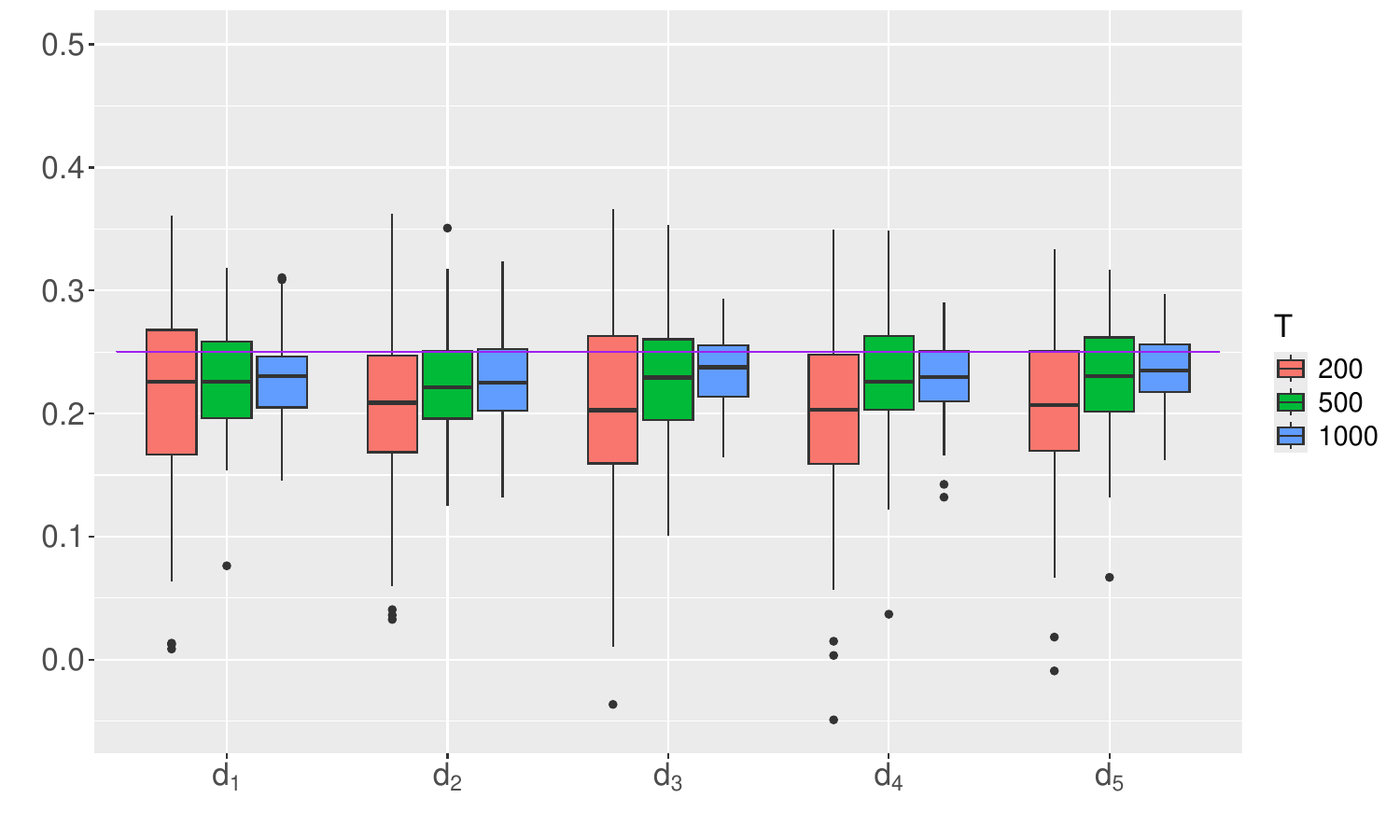}
\end{subfigure}
\hfill
\begin{subfigure}[t]{0.48\textwidth}
    \includegraphics[width=\linewidth]{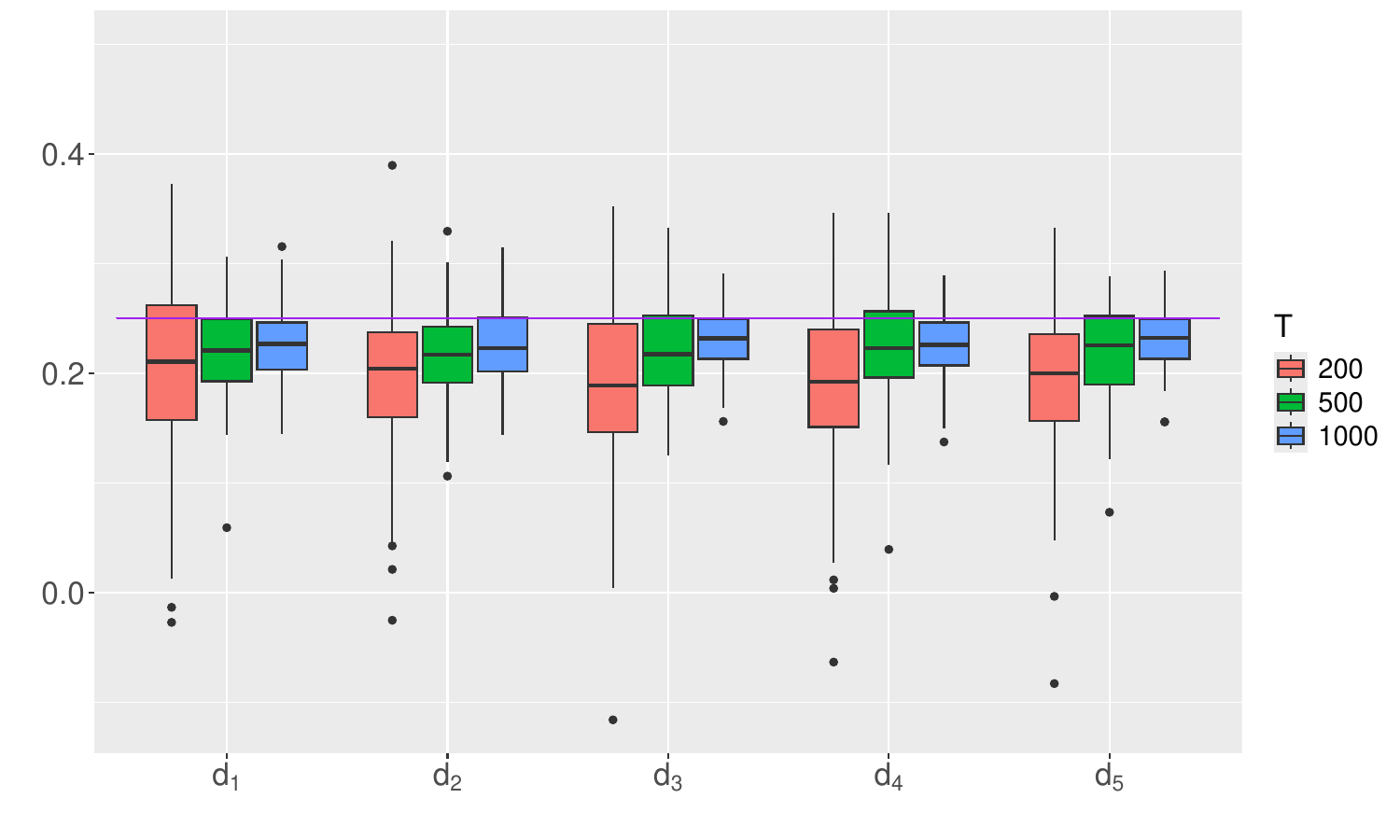}
\end{subfigure}

% Row 3
\begin{subfigure}[t]{0.48\textwidth}
    \includegraphics[width=\linewidth]{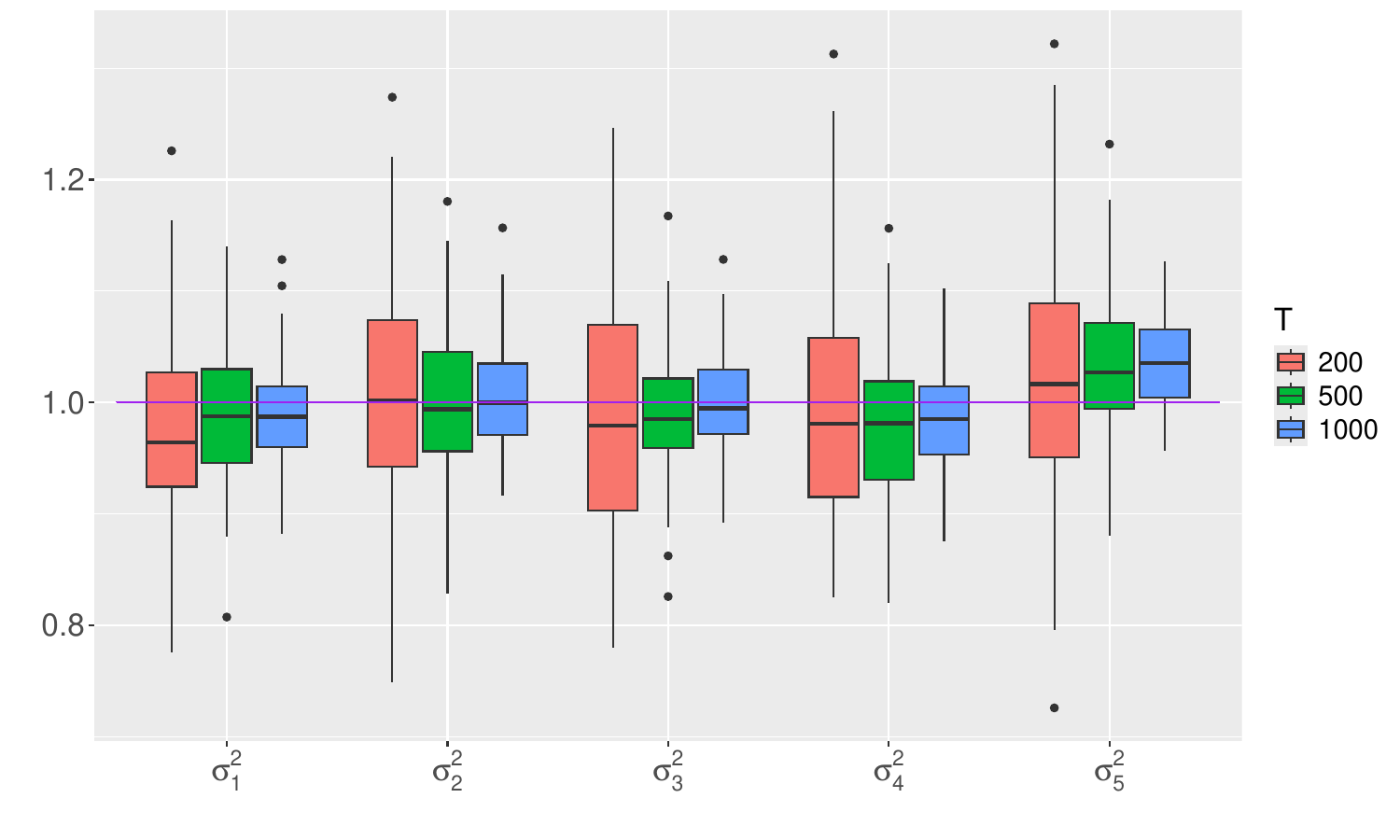}
\end{subfigure}
\hfill
\begin{subfigure}[t]{0.48\textwidth}
    \includegraphics[width=\linewidth]{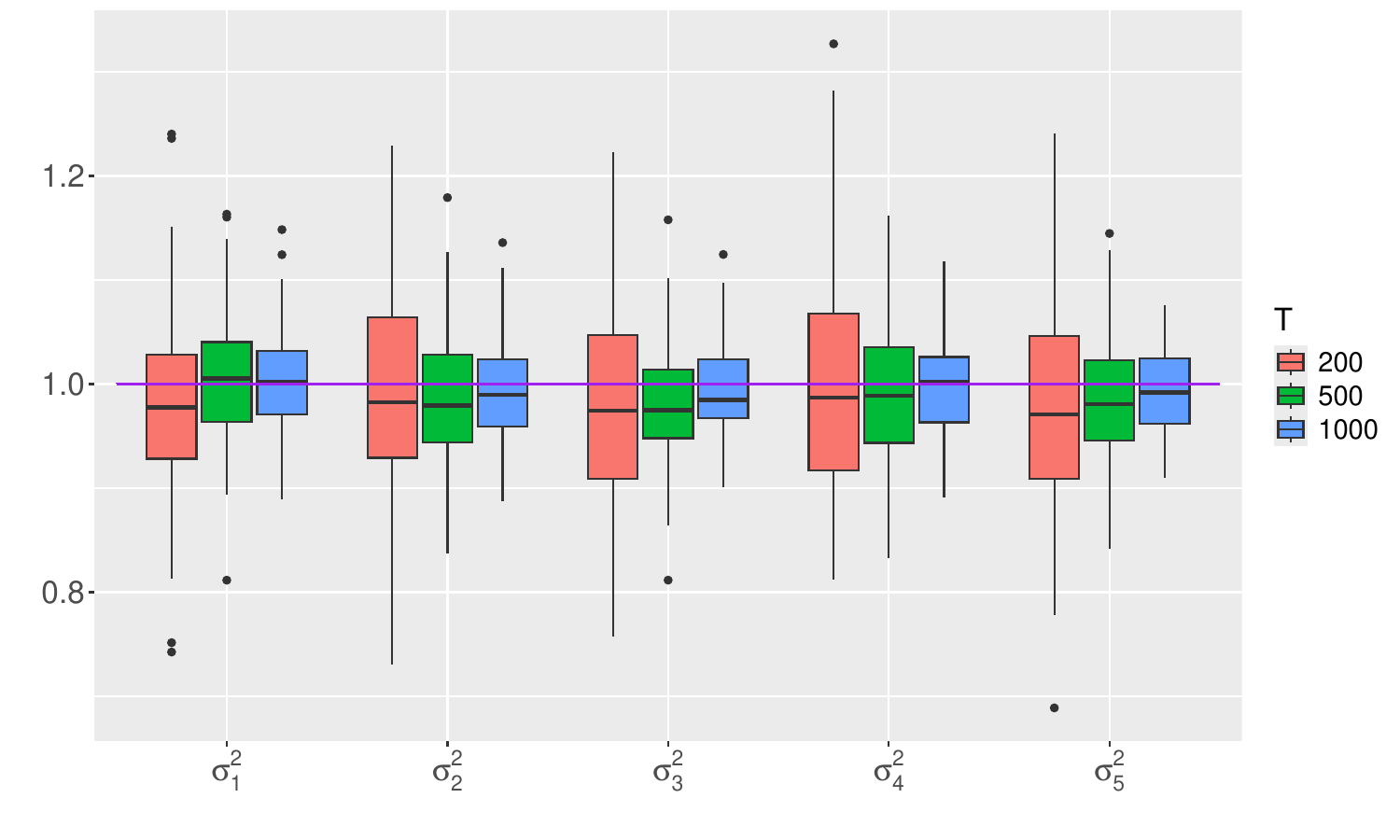}
\end{subfigure}

\caption{DGP2 - GNARFI(1,[1],$\dd$), individual-$d$ specification: boxplots of the estimated ${\aalpha,\bbeta}$ (top), $\dd$ (middle), and $\ssigma_{\vveps}^{2}$ (bottom), shown for the standard method (left) and the conditional method (right).}
\end{figure}

% DGP3: global sigma FIGNAR
\begin{figure}[H]
\centering
\begin{subfigure}[t]{0.48\textwidth}
    \includegraphics[width=\linewidth]{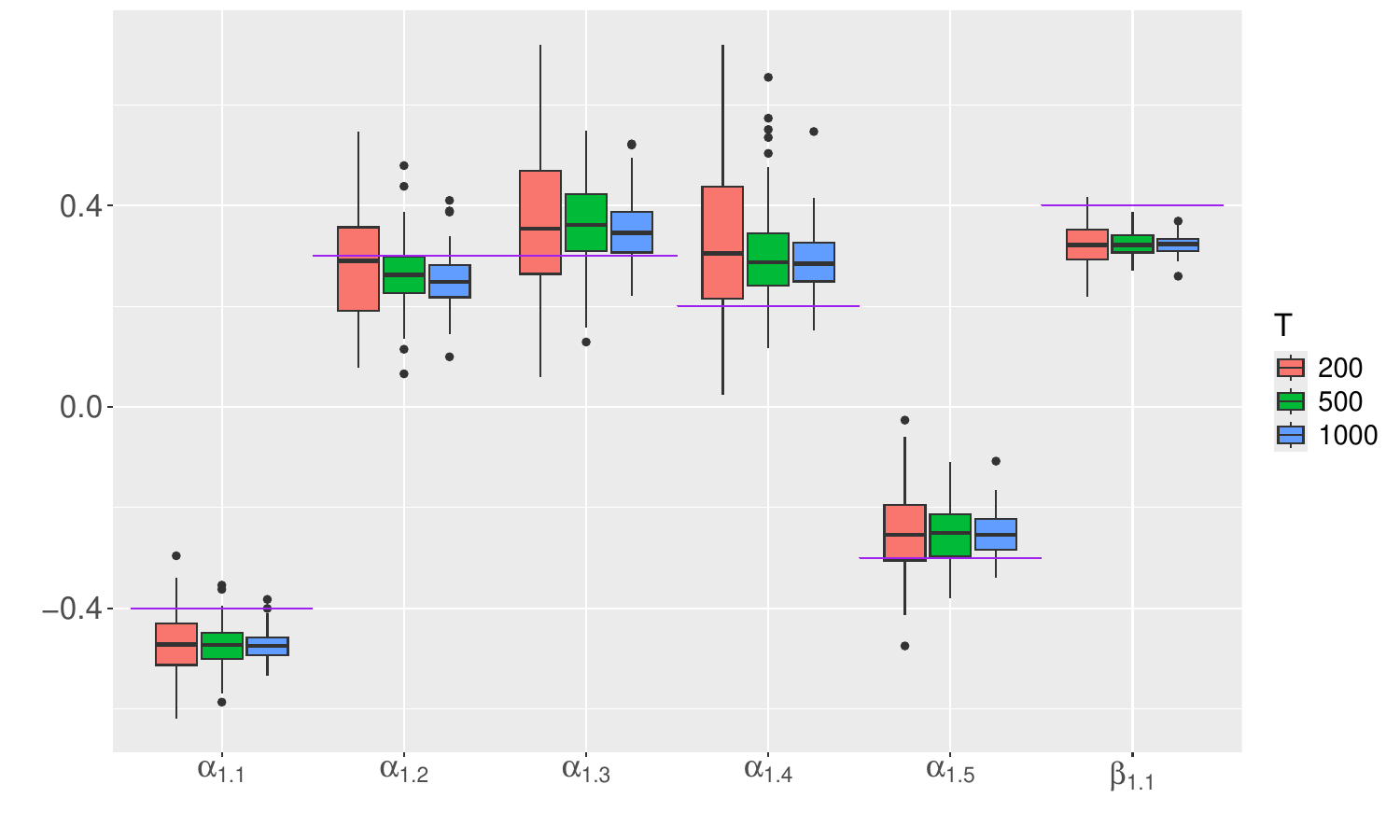}
\end{subfigure}
\hfill
\begin{subfigure}[t]{0.48\textwidth}
    \includegraphics[width=\linewidth]{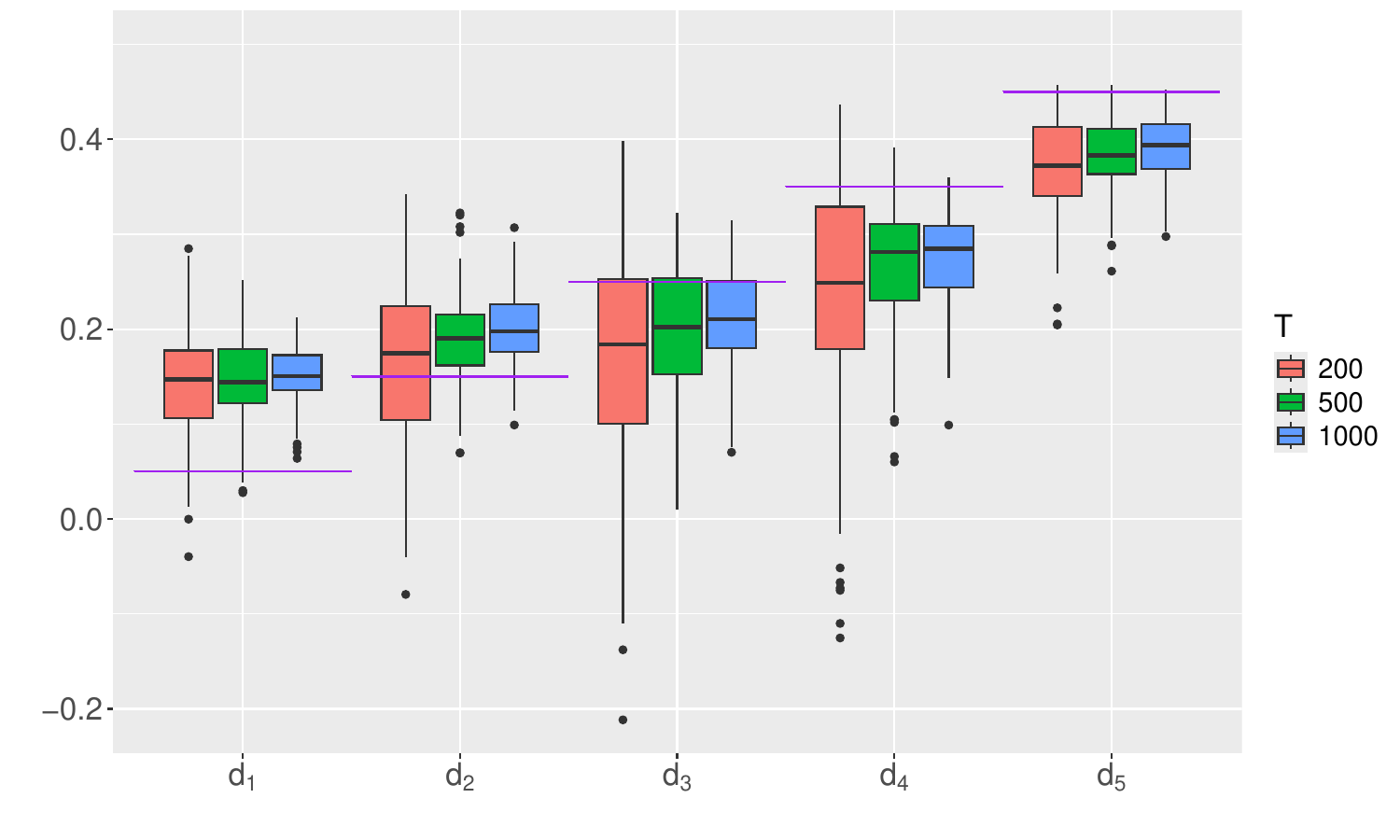}
\end{subfigure}
\begin{subfigure}[t]{0.48\textwidth}
    \includegraphics[width=\linewidth]{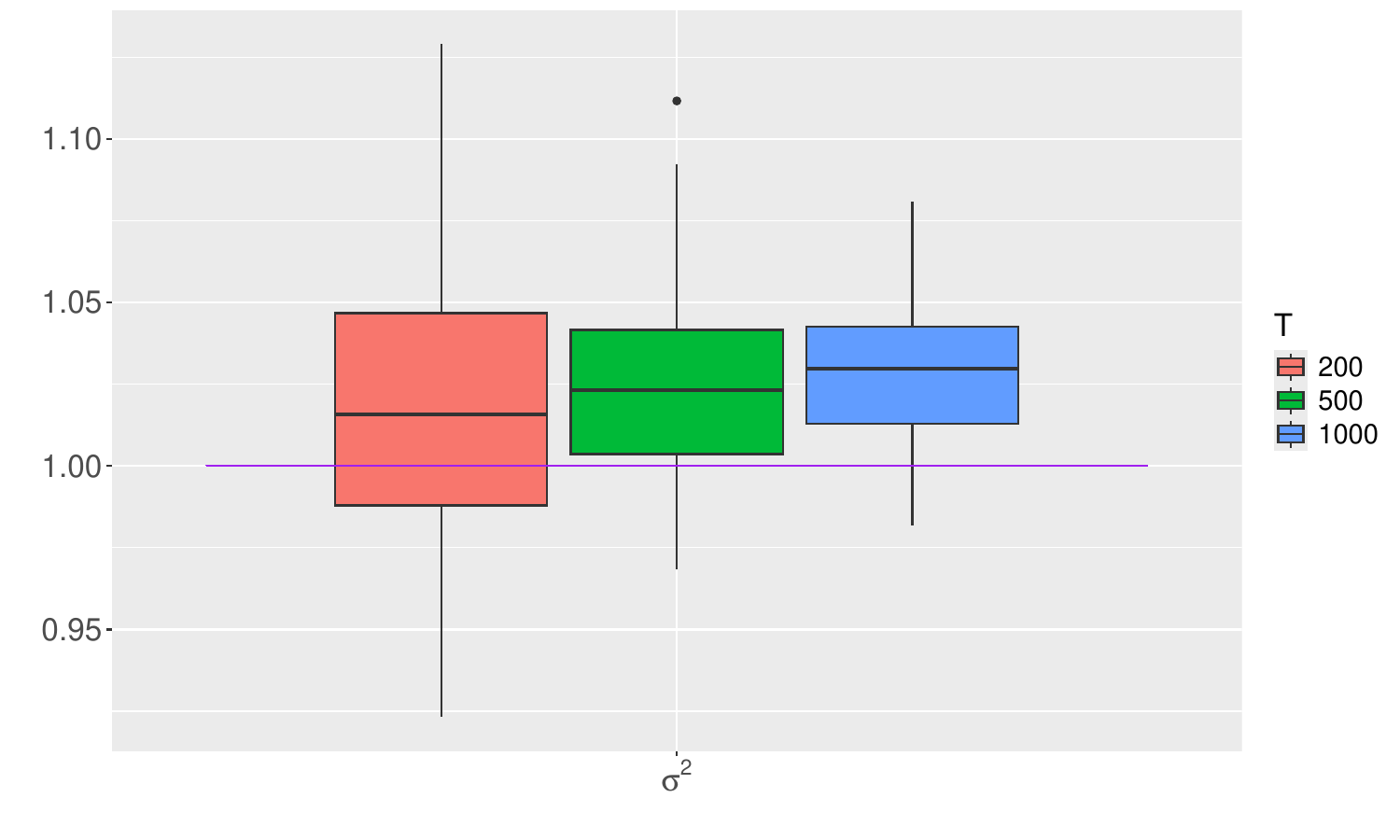}
\end{subfigure}
\caption{DGP3 - FIGNAR(1,[1],$\dd$), global-$\sigma^2$ specification: boxplots of the estimated ${\aalpha,\bbeta}$ (top left), $\dd$ (top right), and $\ssigma_{\vveps}^{2}$ (bottom).}
\end{figure}

% DGP3: individual sigma FIGNAR
\begin{figure}[H]
\centering
\begin{subfigure}[t]{0.48\textwidth}
    \includegraphics[width=\linewidth]{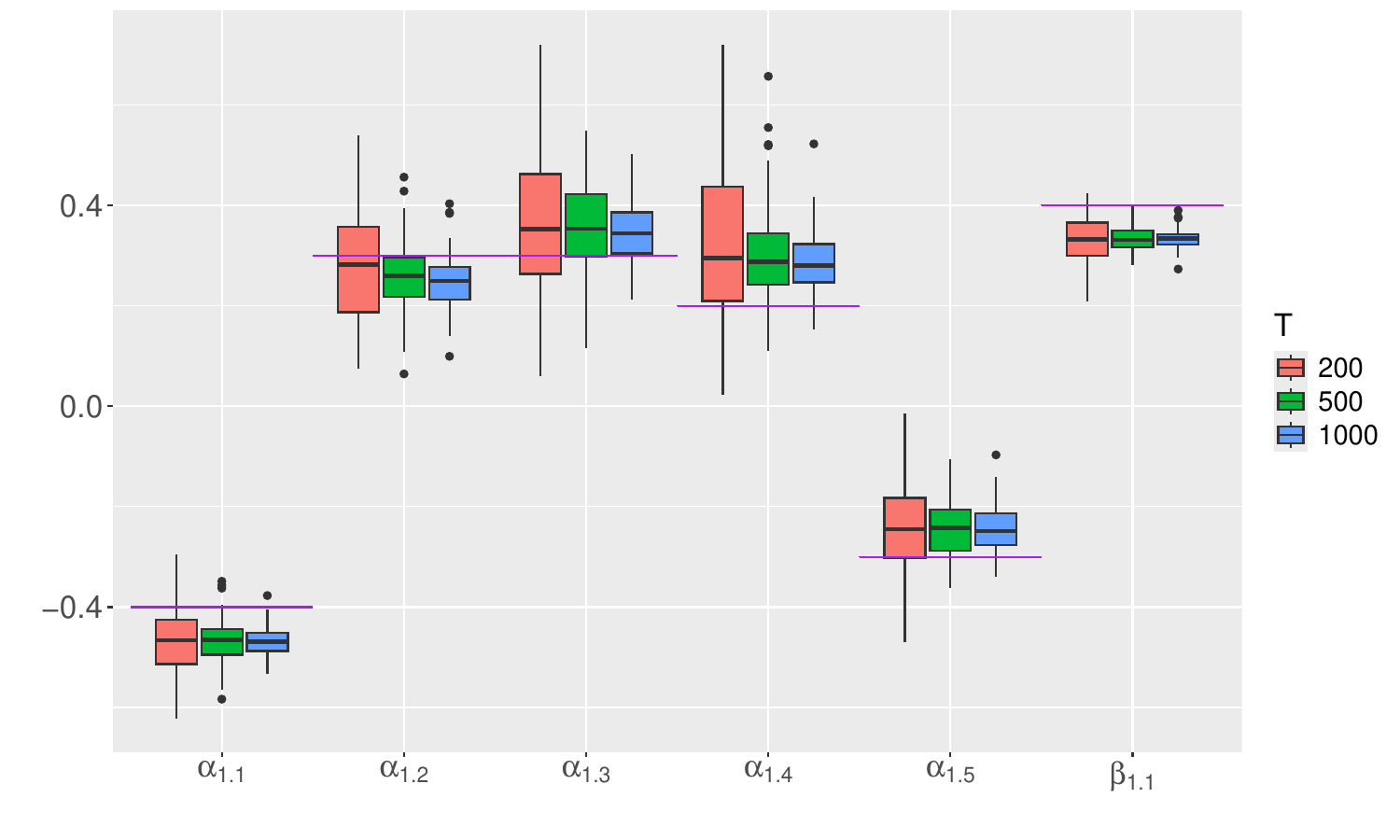}
\end{subfigure}
\hfill
\begin{subfigure}[t]{0.48\textwidth}
    \includegraphics[width=\linewidth]{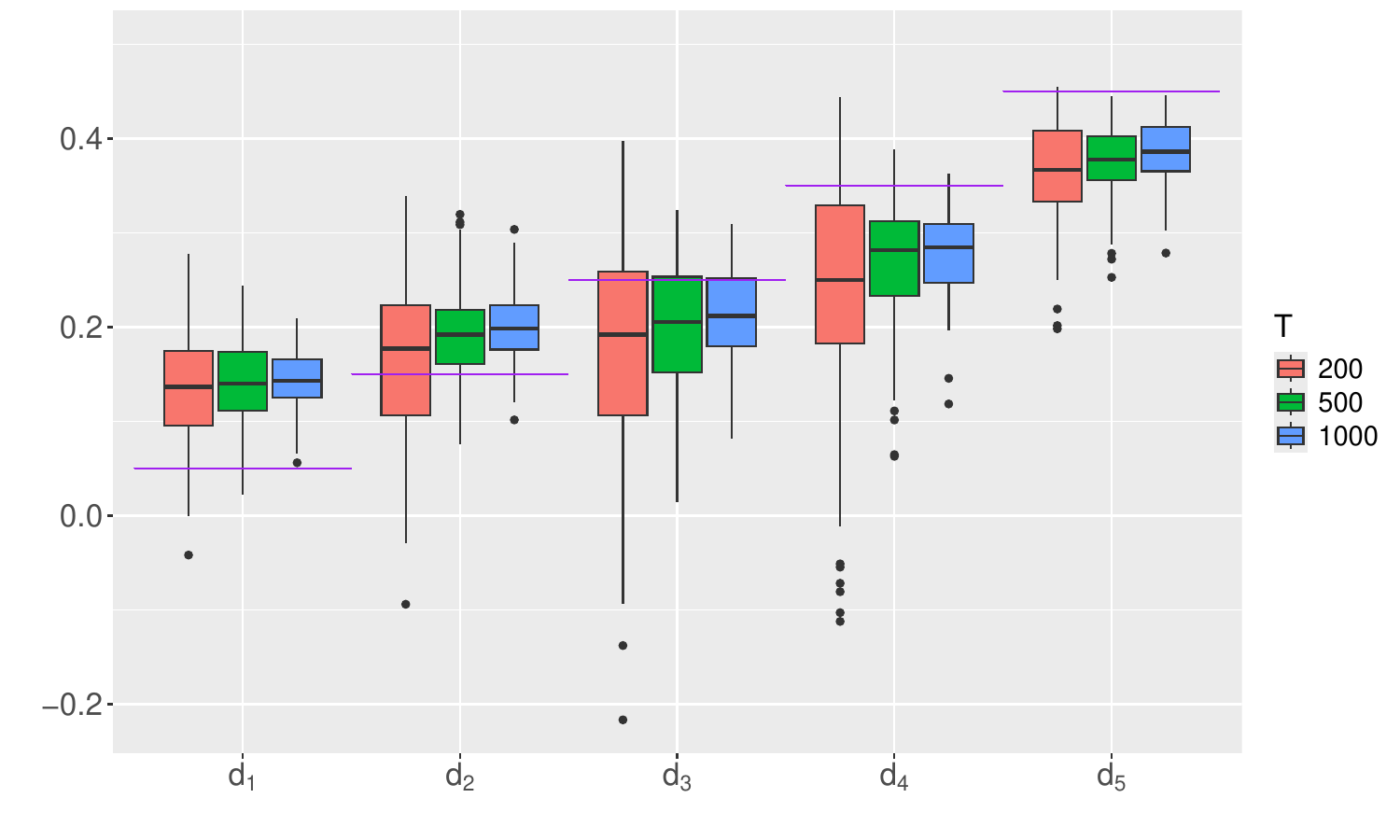}
\end{subfigure}

\begin{subfigure}[t]{0.48\textwidth}
    \includegraphics[width=\linewidth]{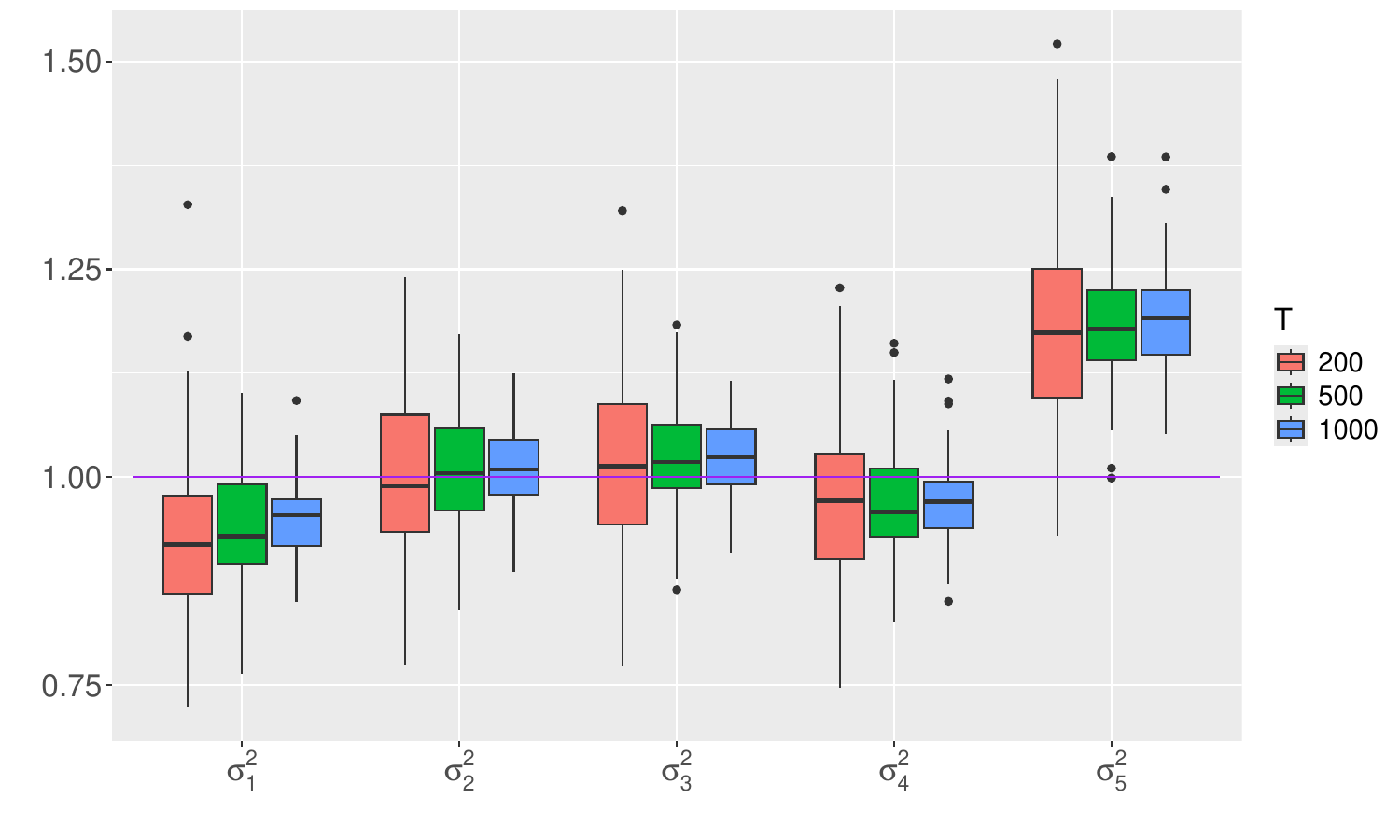}
\end{subfigure}
\caption{DGP3 - FIGNAR(1,[1],$\dd$), individual-$\sigma^2$ specification: boxplots of the estimated ${\aalpha,\bbeta}$ (top left), $\dd$ (top right), and $\ssigma_{\vveps}^{2}$ (bottom).}
\end{figure}

\begin{figure}[H]
\centering
% Row 1
\begin{subfigure}[t]{0.48\textwidth}
    \includegraphics[width=\linewidth]{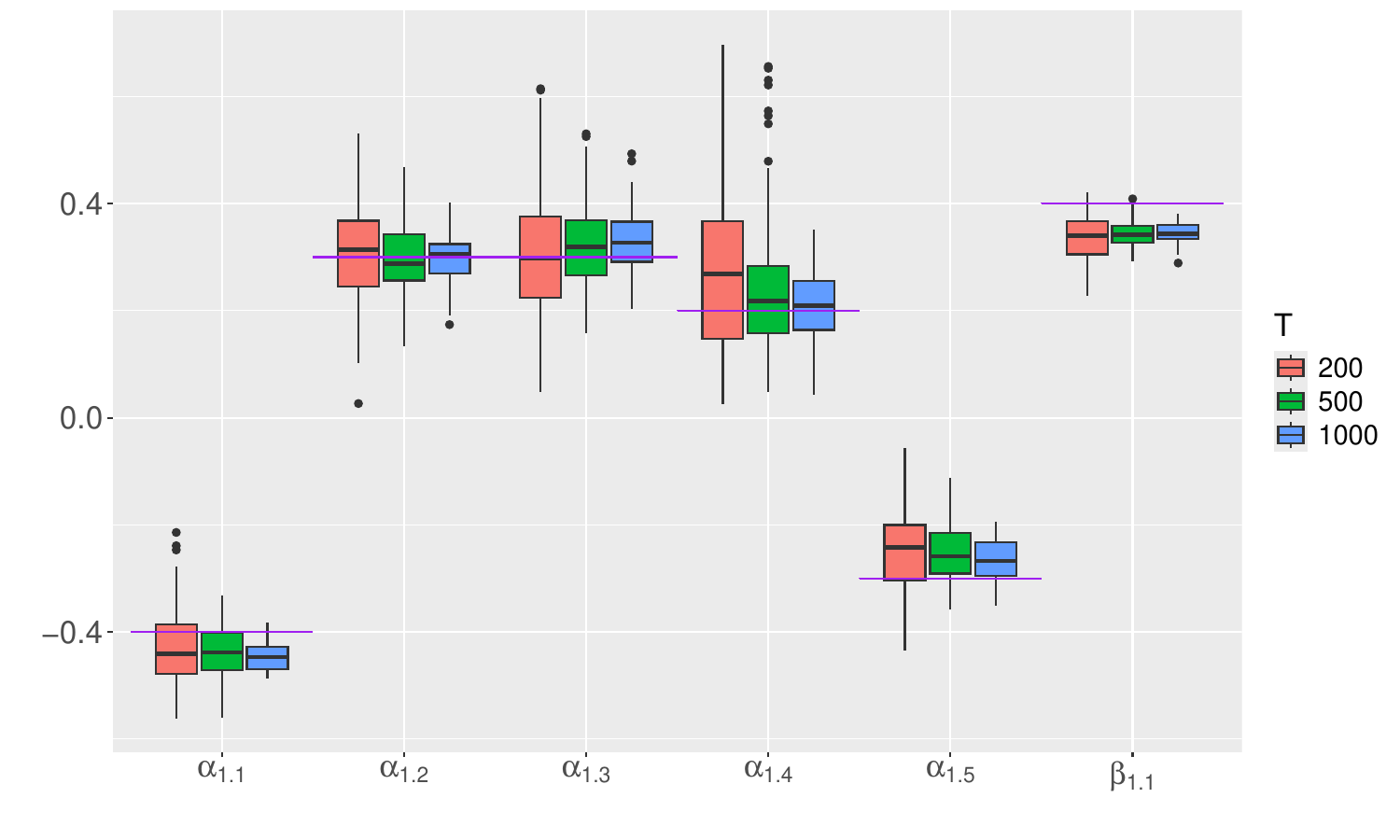}
\end{subfigure}
\hfill
\begin{subfigure}[t]{0.48\textwidth}
    \includegraphics[width=\linewidth]{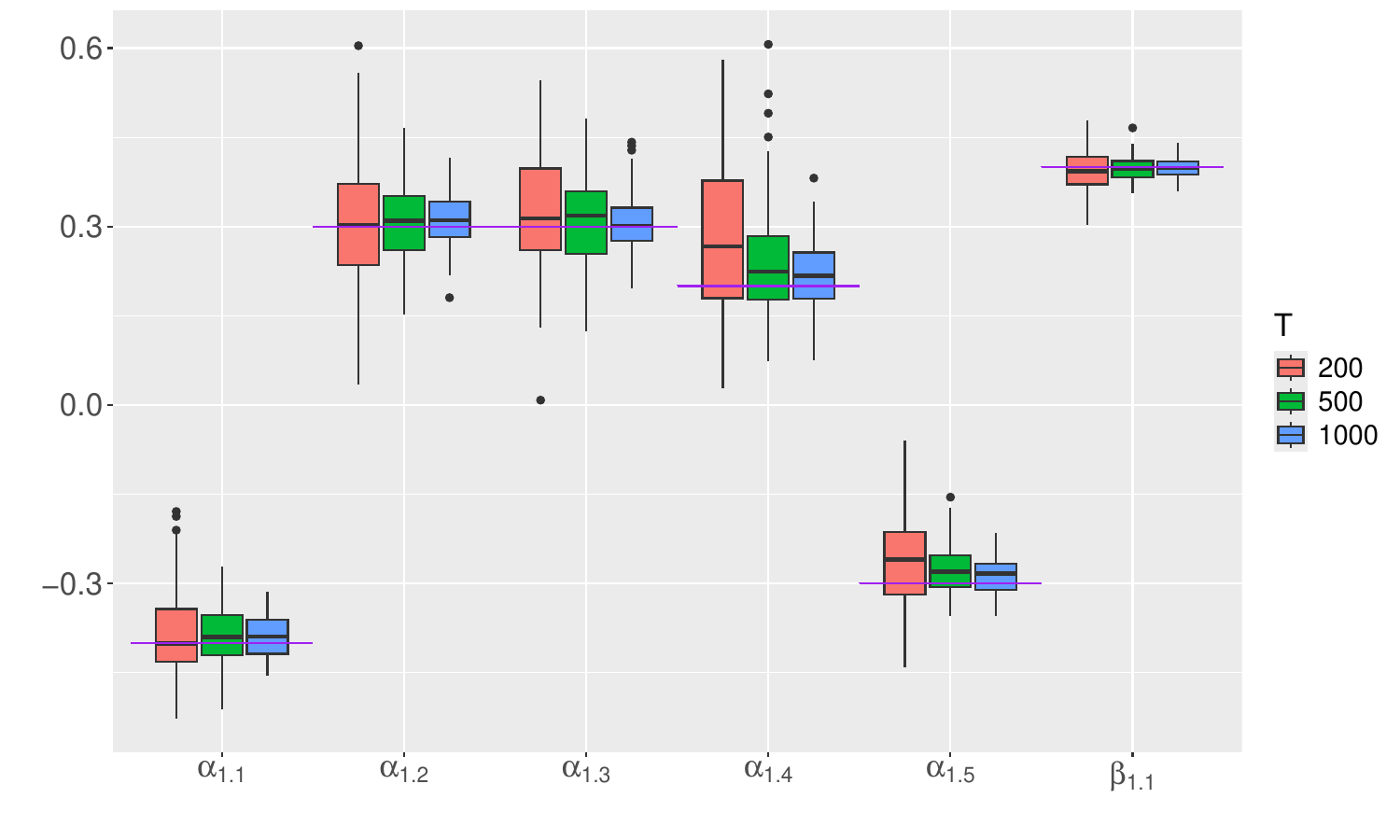}
\end{subfigure}

% Row 2
\begin{subfigure}[t]{0.48\textwidth}
    \includegraphics[width=\linewidth]{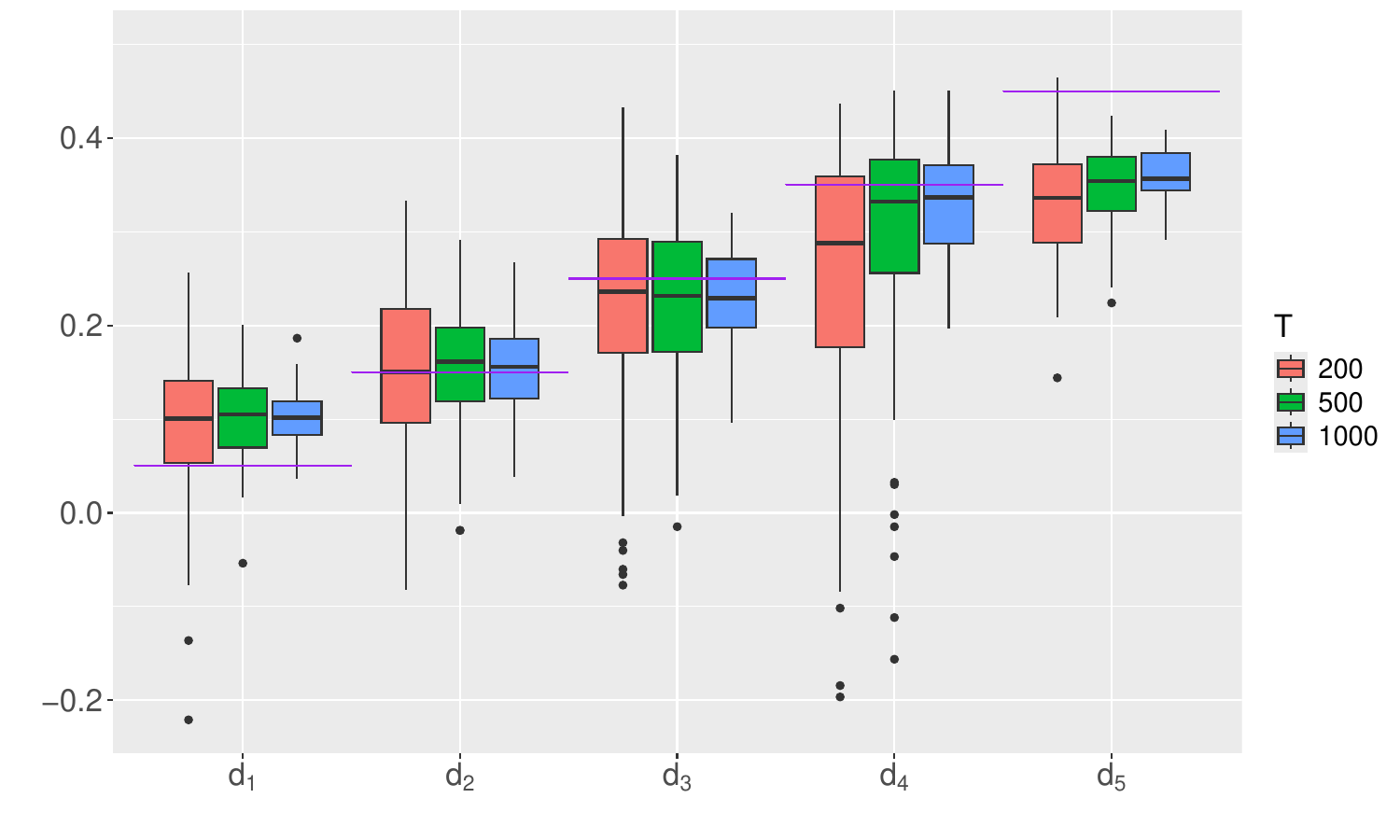}
\end{subfigure}
\hfill
\begin{subfigure}[t]{0.48\textwidth}
    \includegraphics[width=\linewidth]{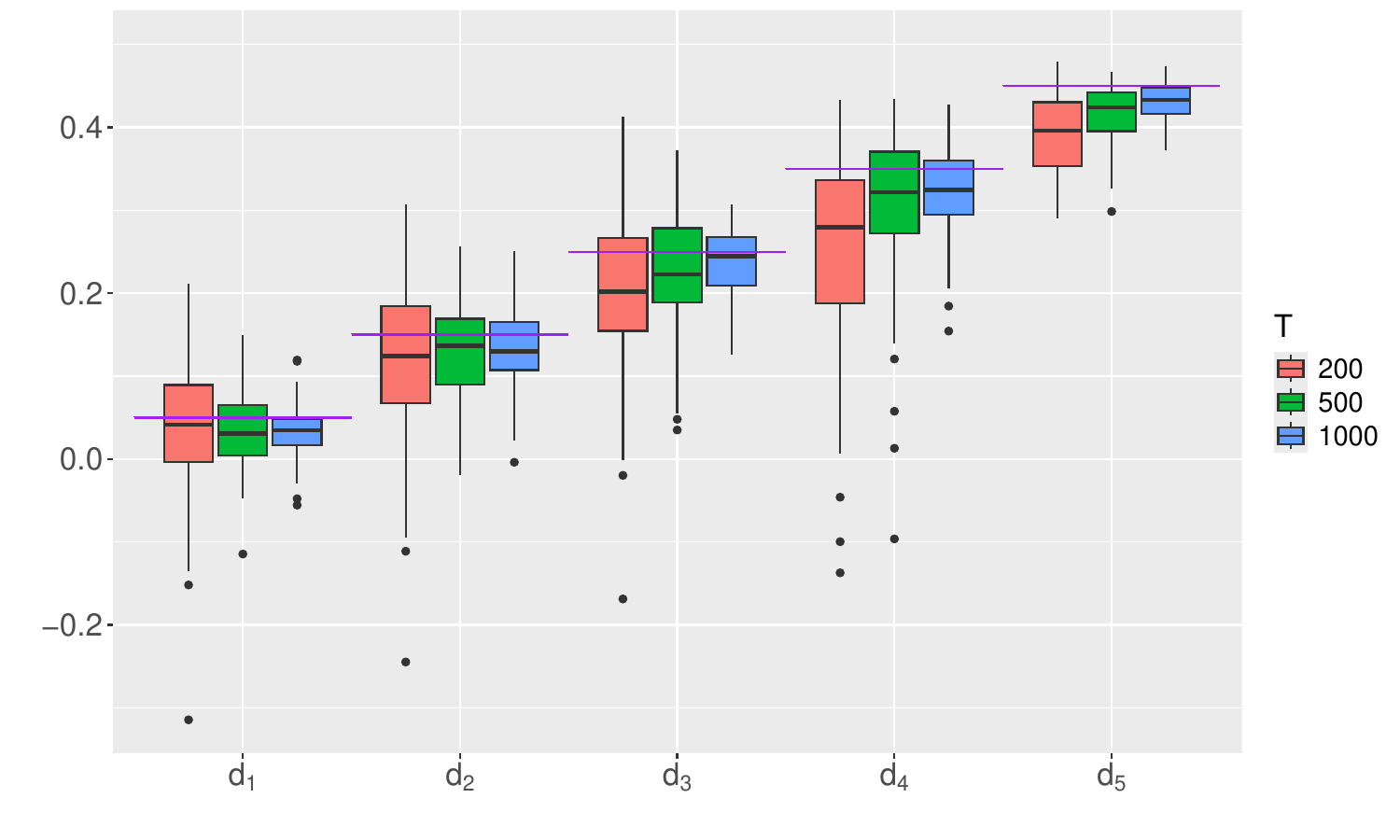}
\end{subfigure}

% Row 3
\begin{subfigure}[t]{0.48\textwidth}
    \includegraphics[width=\linewidth]{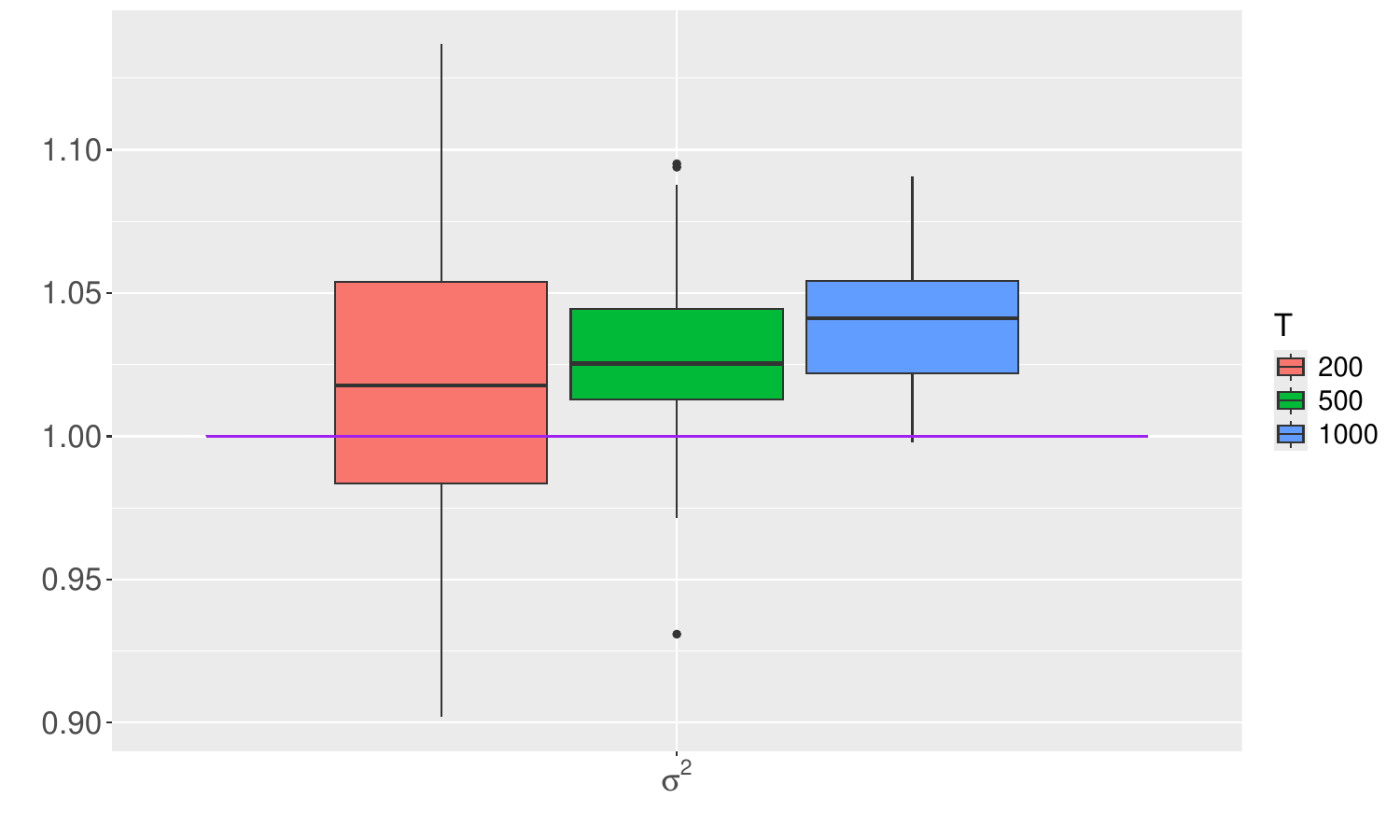}
\end{subfigure}
\hfill
\begin{subfigure}[t]{0.48\textwidth}
    \includegraphics[width=\linewidth]{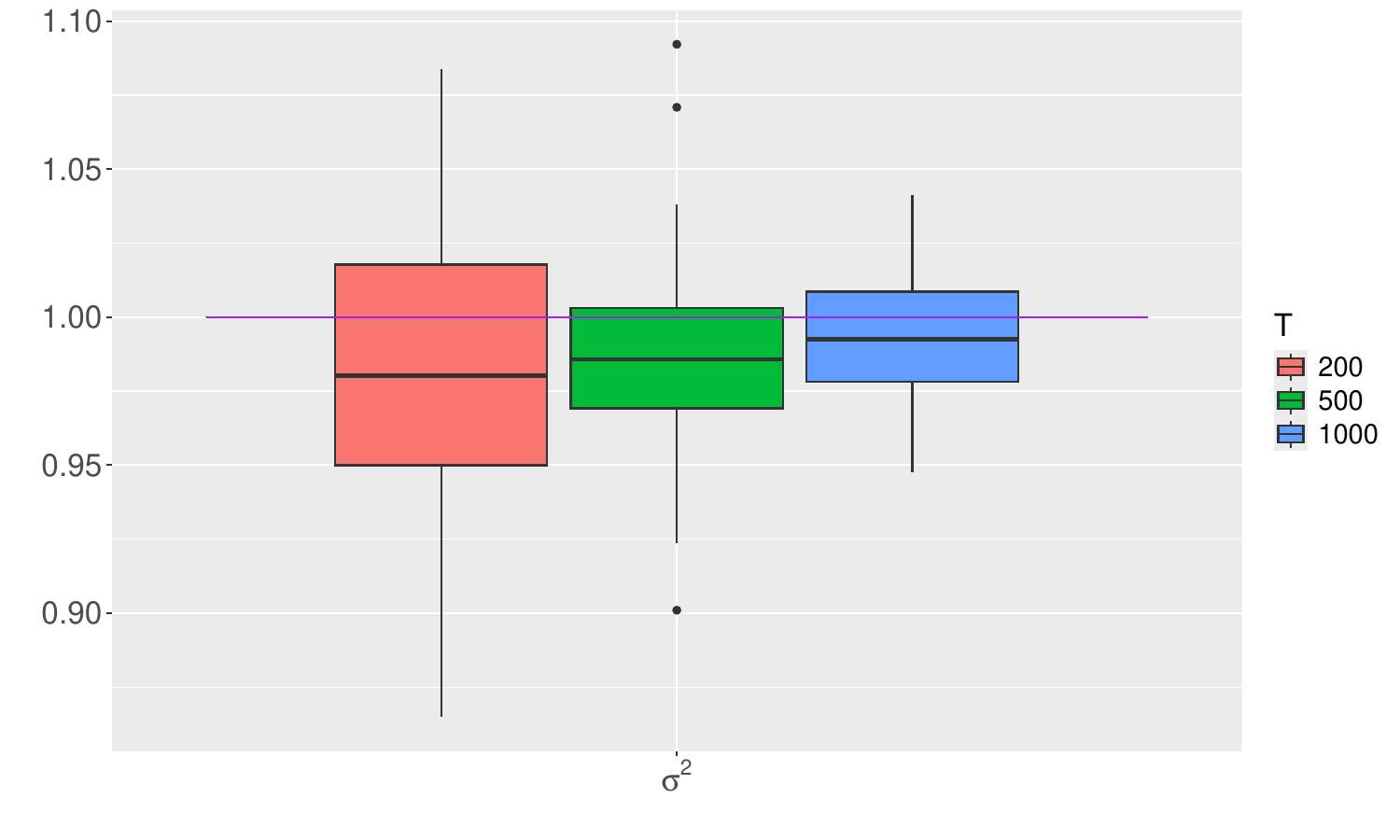}
\end{subfigure}

\caption{DGP3 - GNARFI(1,[1],$\dd$), global-$\sigma^2$ specification: boxplots of the estimated ${\aalpha,\bbeta}$ (top), $\dd$ (middle), and $\ssigma_{\vveps}^{2}$ (bottom), shown for the standard method (left) and the conditional method (right).}
\end{figure}

% DGP3: individual sigma

%-------------------------------------------------------------------
\newpage
\section{Additional Results for Forecasting}\label{app:extraforecast}
\begin{figure}[H]
\centering
\begin{subfigure}[t]{0.7\textwidth}
    \includegraphics[width=\linewidth]{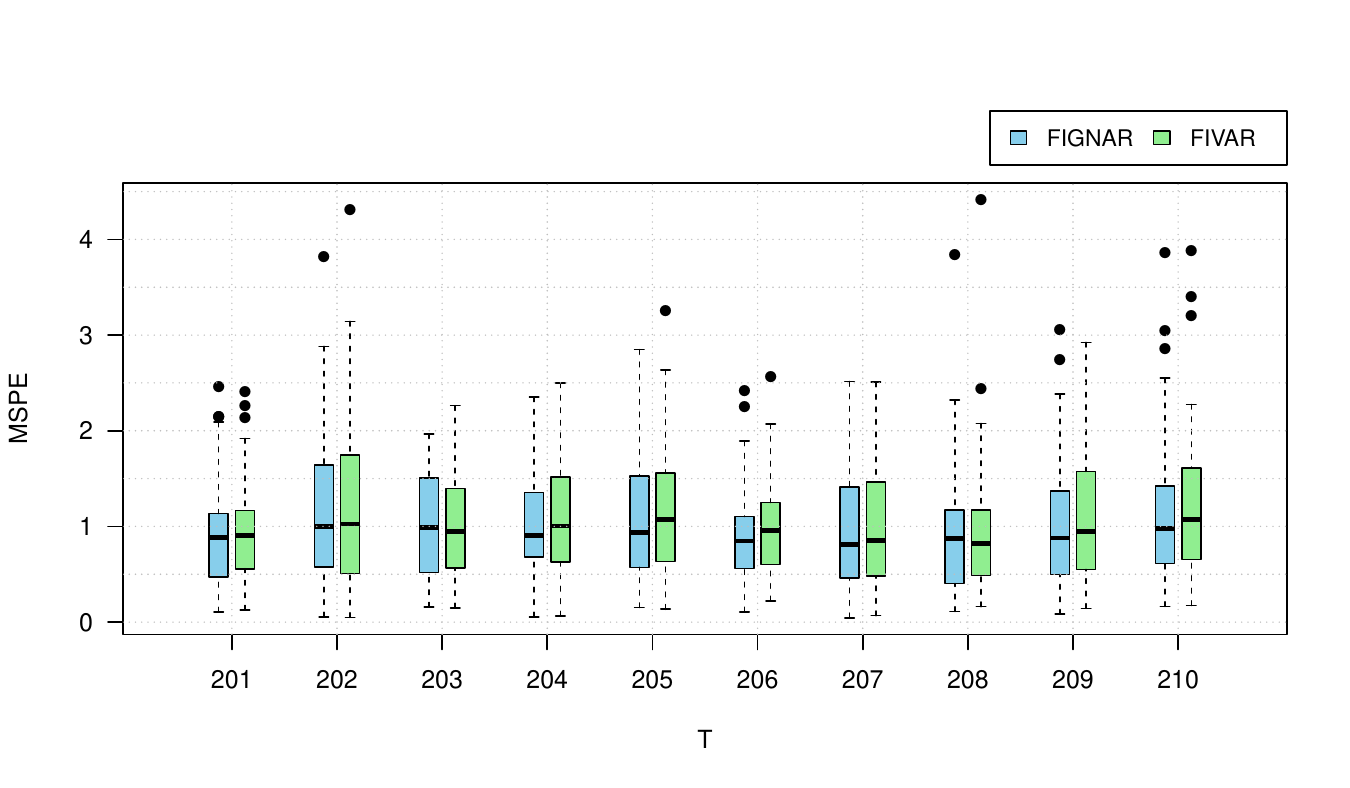}
\end{subfigure}
\begin{subfigure}[t]{0.7\textwidth}
    \includegraphics[width=\linewidth]{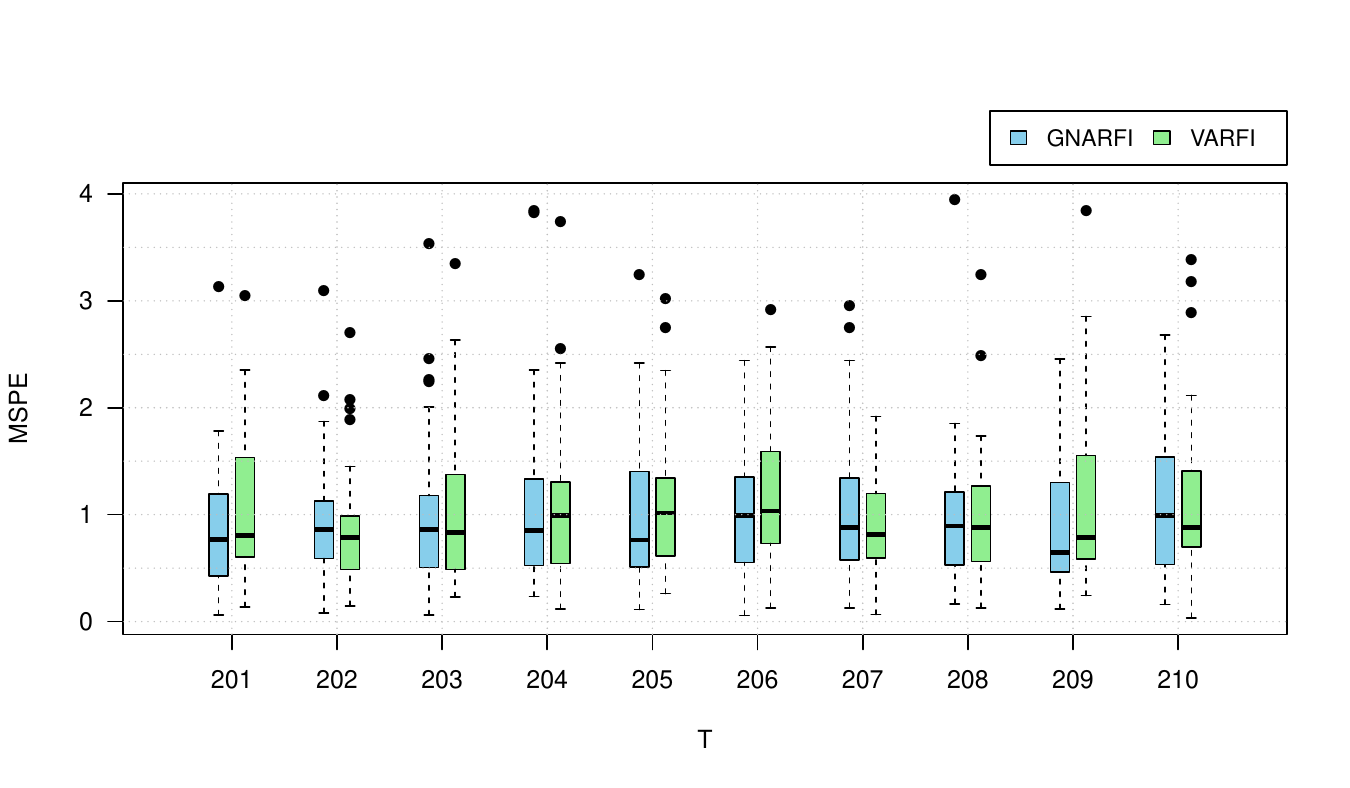}
\end{subfigure}
\begin{subfigure}[t]{0.7\textwidth}
    \includegraphics[width=\linewidth]{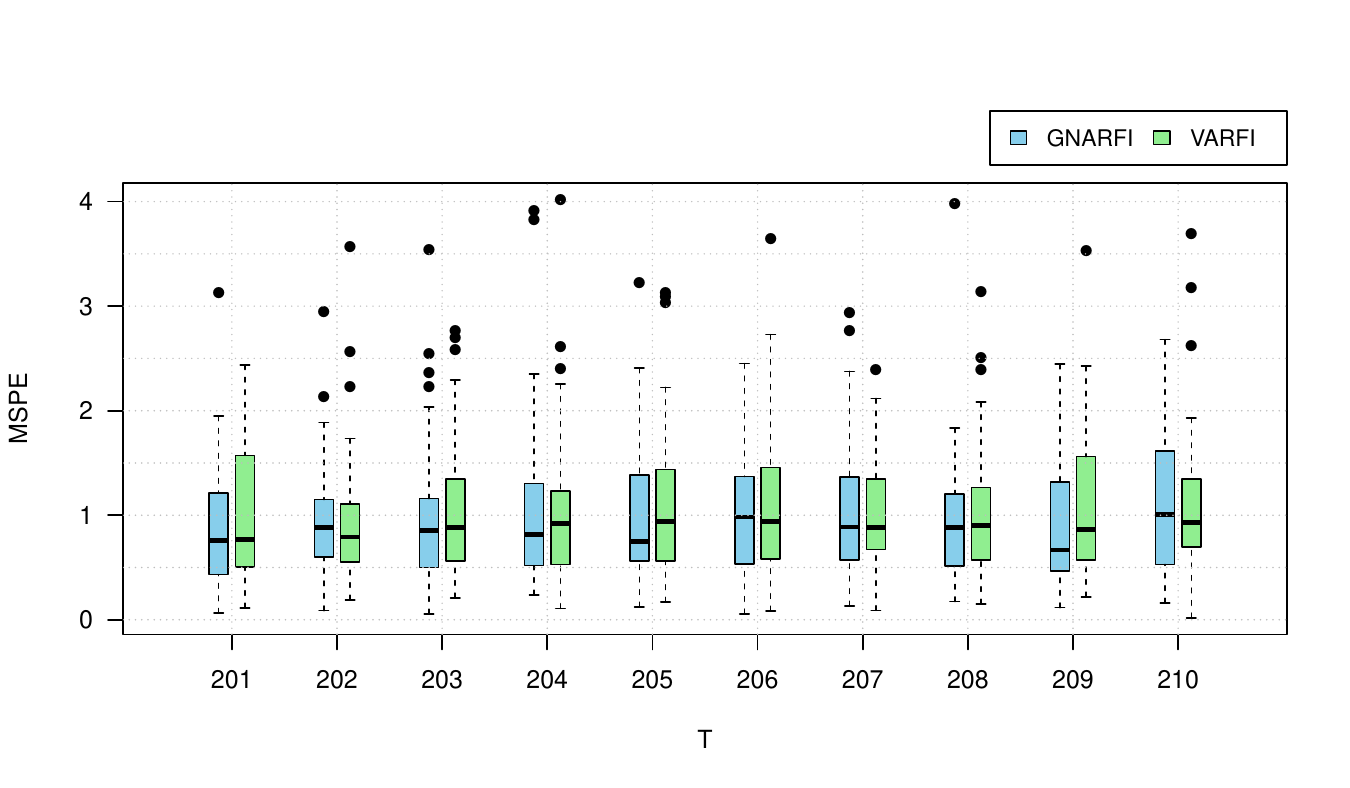}
\end{subfigure}
\caption{Boxplots of MSPEs for the rolling window predictions on data of length $T=200$ generated under DGP1. Top: FIGNAR/FIVAR models. Bottom: GNARFI/VARFI models fitted with the standard (left) and conditional (right) parameter estimation methods.}\label{img:forecast-rw}
\end{figure}

\begin{figure}[H]
\centering
\begin{subfigure}[t]{0.7\textwidth}
    \includegraphics[width=\linewidth]{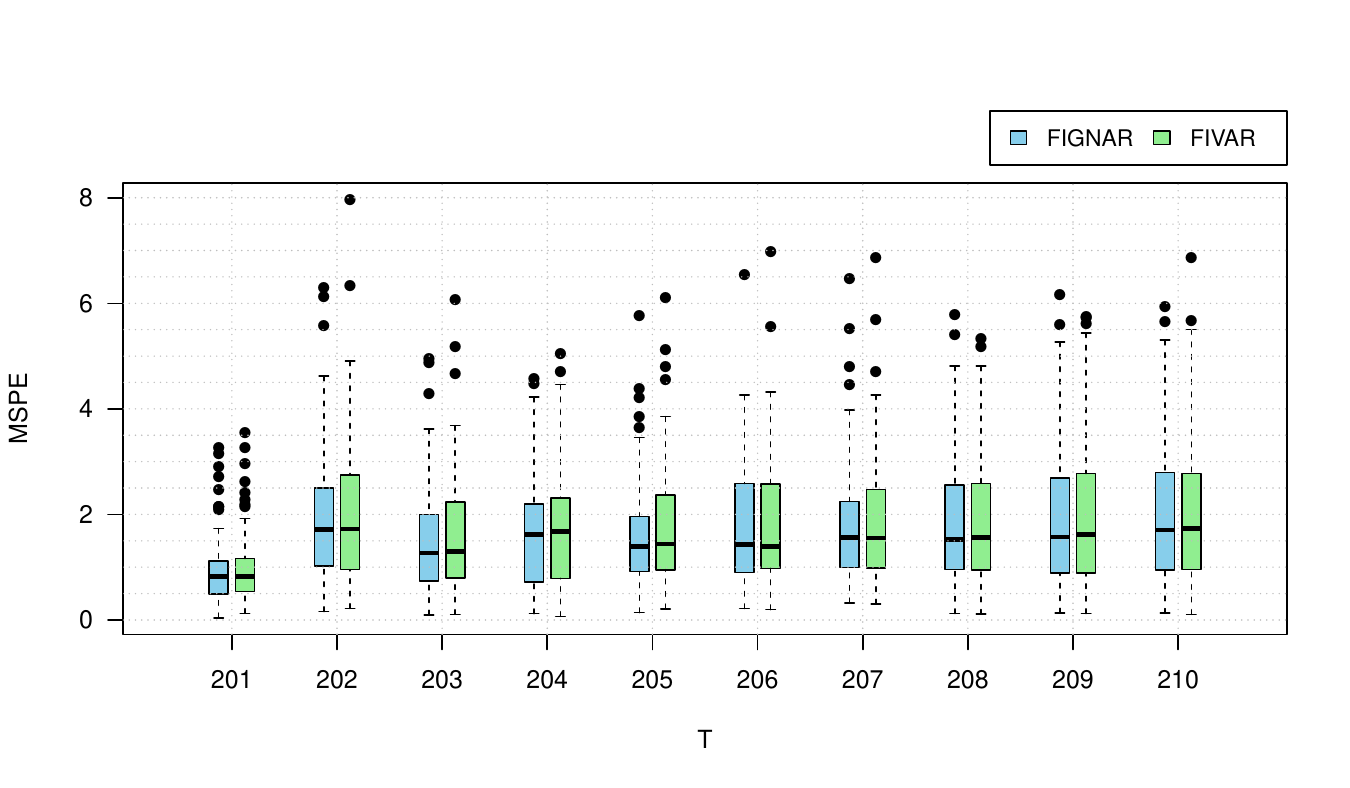}
\end{subfigure}
\begin{subfigure}[t]{0.7\textwidth}
    \includegraphics[width=\linewidth]{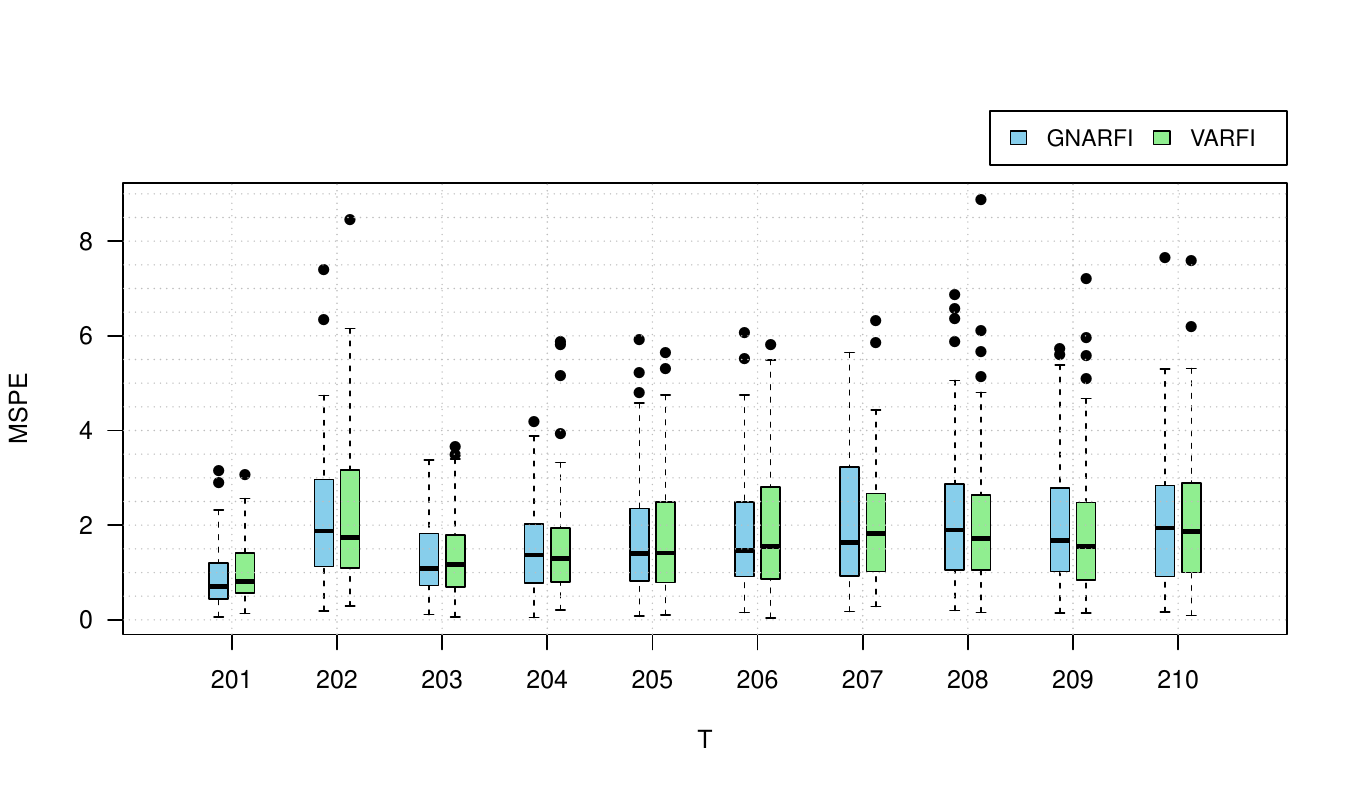}
\end{subfigure}
\begin{subfigure}[t]{0.7\textwidth}
    \includegraphics[width=\linewidth]{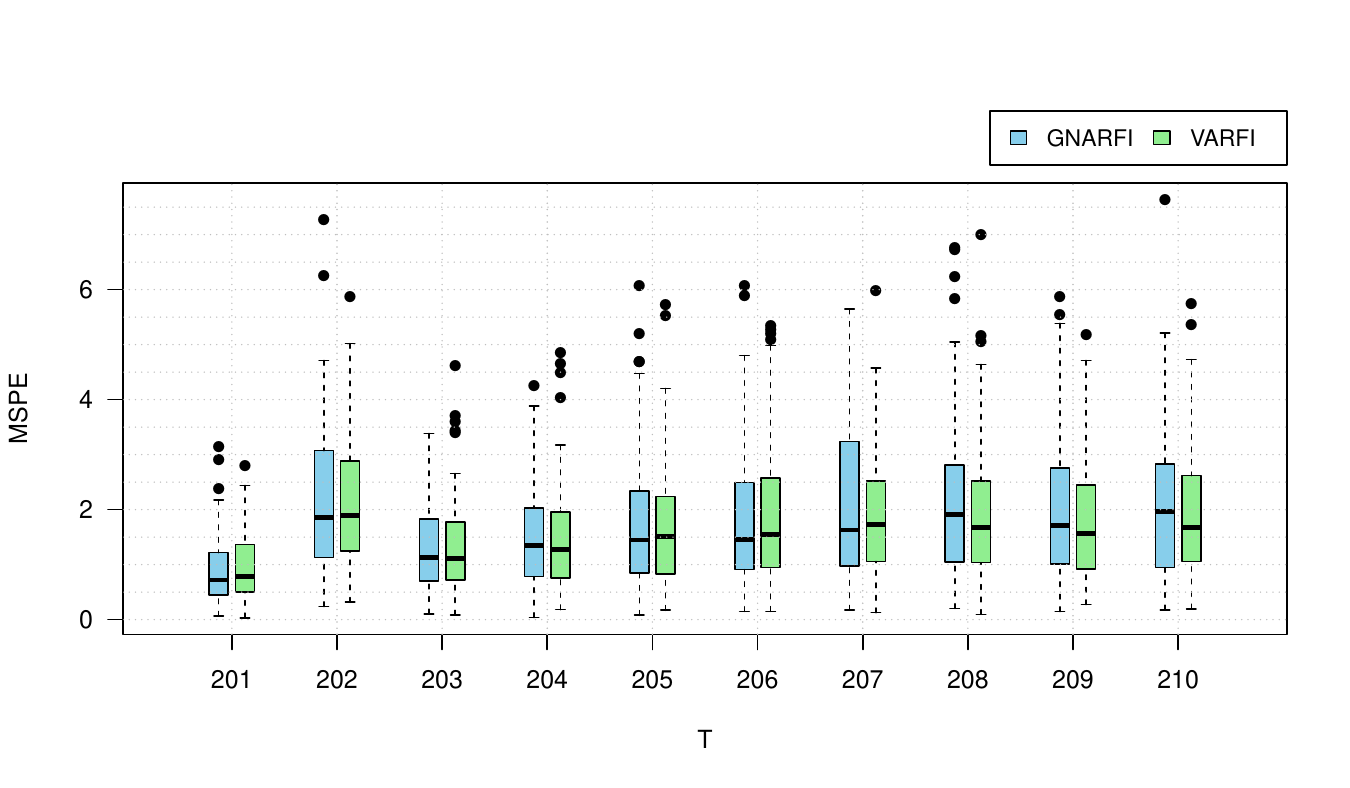}
\end{subfigure}
\caption{Boxplots of MSPEs for $h$-step-ahead predictions ($h=1,\ldots,10$) on data of length $T=200$ generated under DGP1. Top: FIGNAR/FIVAR models. Bottom: GNARFI/VARFI models fitted with the standard (left) and conditional (right) parameter estimation methods.}\label{img:forecast-long}
\end{figure}

In this appendix, we provide additional results on the forecasting experiments performed in Section \ref{sec:forecasting}. In particular, Figure \ref{img:forecast-rw} shows the boxplots of the rolling-window forecast analysis, whose mean numbers are reported in Table \ref{tab:forecast-rw}. Similarly, Figure \ref{img:forecast-long} correspond to the $h$-step ahead forecast analysis of Table \ref{tab:forecast-long}.

% Rolling window forecasts - Network model and 200 obs

% h>0 forecasts - Network model and 200 obs

%-------------------------------------------------------------------
\newpage
\section{Additional Results of Application to Real Data}\label{app:extrarealdata}

In this section, we provide additional details related to the real data analysis performed in Section \ref{sec:application}.

\subsection{Irish Wind Data}

\begin{table}[!ht]
\centering
\begin{NiceTabular}{l|l||ccccc}
Code & Station & Mean & SD & Skew & Kurt & AC1 \\
\hline\hline
RPT & Roche's Point & 12.483 & 5.898 & 0.743 & 3.522 & 0.484 \\
VAL & Valentia & 10.466 & 4.957 & 0.625 & 3.331 & 0.479 \\
ROS & Rosslare & 11.888 & 5.492 & 0.797 & 3.428 & 0.478 \\
KIL & Kilkenny & 7.106 & 4.006 & 0.979 & 4.396 & 0.451 \\
SHA & Shannon & 11.107 & 4.969 & 0.679 & 3.564 & 0.507 \\
BIR & Birr & 7.860 & 4.112 & 0.538 & 3.105 & 0.520 \\
DUB & Dublin & 10.602 & 5.220 & 0.713 & 3.318 & 0.556 \\
CLA & Claremorris & 9.339 & 4.616 & 0.454 & 3.127 & 0.477 \\
MUL & Mullingar & 8.635 & 4.295 & 0.563 & 3.240 & 0.503 \\
CLO & Clones & 9.921 & 4.502 & 0.513 & 3.230 & 0.503 \\
BEL & Belmullet & 13.393 & 5.946 & 0.504 & 2.998 & 0.540 \\
MAL & Malin Head & 14.343 & 6.477 & 0.566 & 3.185 & 0.548 \\
\end{NiceTabular}
\caption{Summary statistics for Irish wind time series: mean, standard deviation, skewness, kurtosis, and first-order autocorrelation.}\label{tab:irish-wind-data}
\end{table}

\begin{figure}[!ht]
    \centering
    \includegraphics[width=0.9\textwidth]{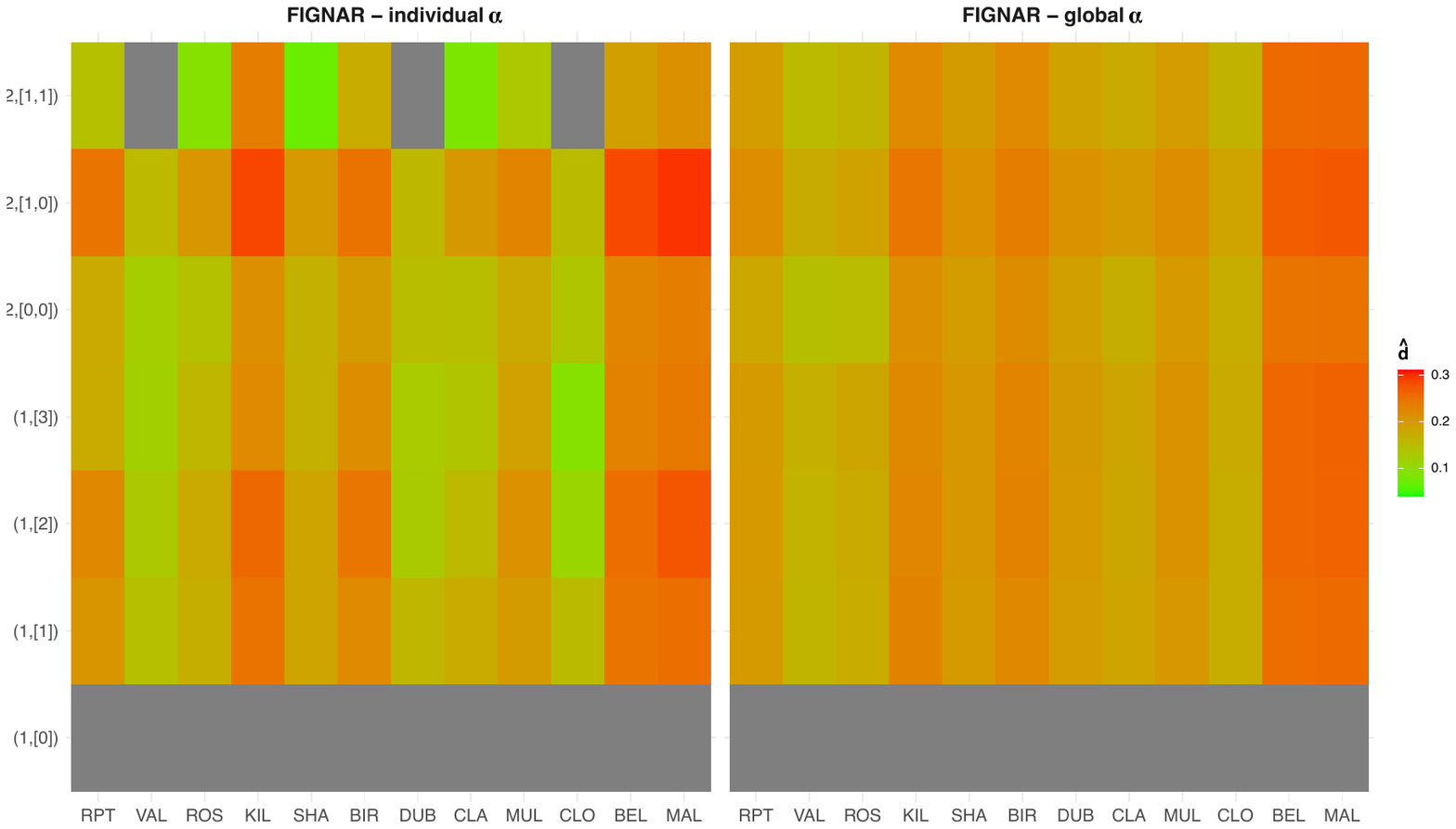}
    \centering
    \includegraphics[width=0.9\textwidth]{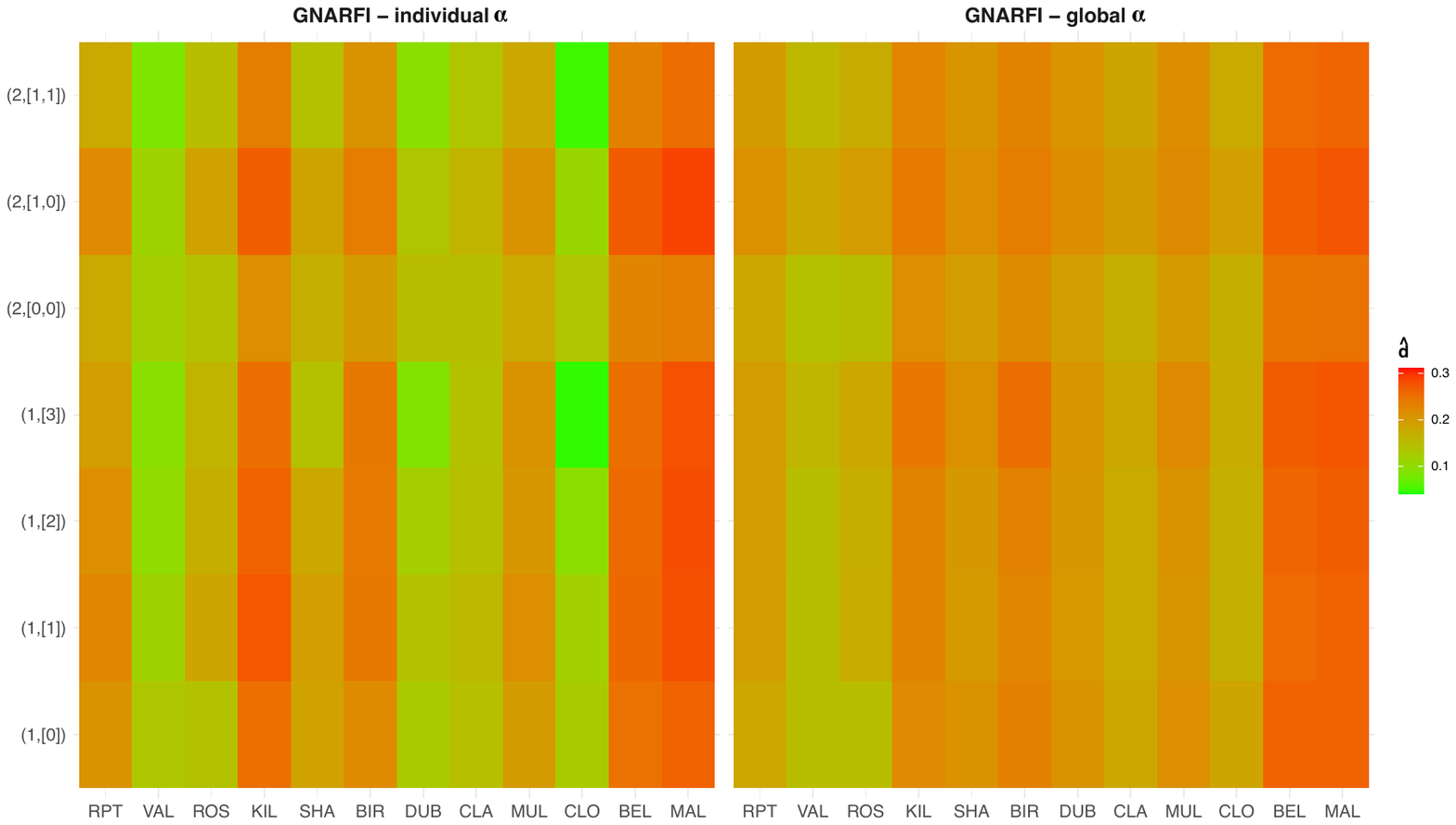}
    \captionof{figure}{Estimated long memory parameters for the 12 Irish wind stations.}\label{img:dhat-irishwind}
\end{figure}

Figure \ref{img:dhat-irishwind} reports the average values of the long memory estimates $\hat\dd$ across the various model orders. Recall that the estimation method of models with order (1,[0]) did not converge, and method converged to unreliable solution for all individual-$\alpha$ FIGNAR(2,[1,1],$\dd$) models (e.g. $\hat{d}_{i} \notin (0,\frac{1}{2})$ for some $i$). 
In general, we see no substantial differences in the estimated long memory component across the various model orders. The plots highlight a clear geographical pattern, with long memory increasing from the southern stations to the northern ones. They also illustrate the two ends of the spectrum—lower versus higher long-range dependence—especially in the individual-$\alpha$ models, which are able to capture behaviour at both extremes.

% ------------------------------------------------------------------
\subsection{Realised Volatility Data}
\begin{table}[H]
\centering
{\fontsize{10pt}{12pt}\selectfont  % 10pt text with 12pt baselineskip
\begin{NiceTabular}{l|l||cccccc}
Tick & Stock & Mean & SD & Skew & Kurt & AC1 & \% NA \\
\hline\hline
AEX & AEX Index & -10.263 & 0.814 & 0.517 & 3.619 & 0.726 & 2.2 \\
BFX & Bell 20 Index  & -10.155 & 0.723 & 0.527 & 3.622 & 0.679 & 2.3 \\
BVLG & PSI All Shares Gross Return Index & -10.636 & 0.644 & 1.050 & 4.770 & 0.736 & 2.3 \\
FCHI & CAC 40 & -10.071 & 0.776 & 0.450 & 3.409 & 0.668 & 2.2 \\
FTMIB & FTSE MIB & -9.875 & 0.807 & 0.354 & 4.061 & 0.724 & 2.8 \\
FTSE & FTSE 100 & -10.121 & 0.827 & 0.613 & 4.717 & 0.577 & 3.1 \\
GDAXI & DAX  & -10.019 & 0.788 & 0.257 & 3.433 & 0.673 & 3.2 \\
IBEX & IBEX 35 Index & -9.721 & 0.735 & 0.853 & 4.918 & 0.699 & 2.3 \\
OMXC20 & OMX Copenhagen 20 Index & -9.755 & 0.816 & 1.086 & 5.439 & 0.559 & 4.4 \\
OMXHPI & OMX Helsinki All Share Index & -10.353 & 0.701 & 0.674 & 3.807 & 0.647 & 3.8 \\
OMXSPI & OMX Stockholm All Share Index & -10.496 & 0.727 & 0.600 & 4.101 & 0.632 & 3.7 \\
OSEAX & Oslo Exchange All-share Index & -9.998 & 0.870 & 0.860 & 4.358 & 0.572 & 5.3 \\
SMSI & Madrid General Index & -9.863 & 0.723 & 0.958 & 5.186 & 0.727 & 2.4 \\
SSMI & Swiss Stock Market Index  & -10.342 & 0.631 & 0.707 & 4.117 & 0.722 & 4.2 \\
\end{NiceTabular}}
\caption{Summary statistics for log-realized variance time series: mean, standard deviation, skewness, kurtosis, first-order autocorrelation, and percentage of missing values.}\label{tab:rel-vol-data}
\end{table}

\begin{figure}[H]
    \centering
    \includegraphics[width=0.9\textwidth]{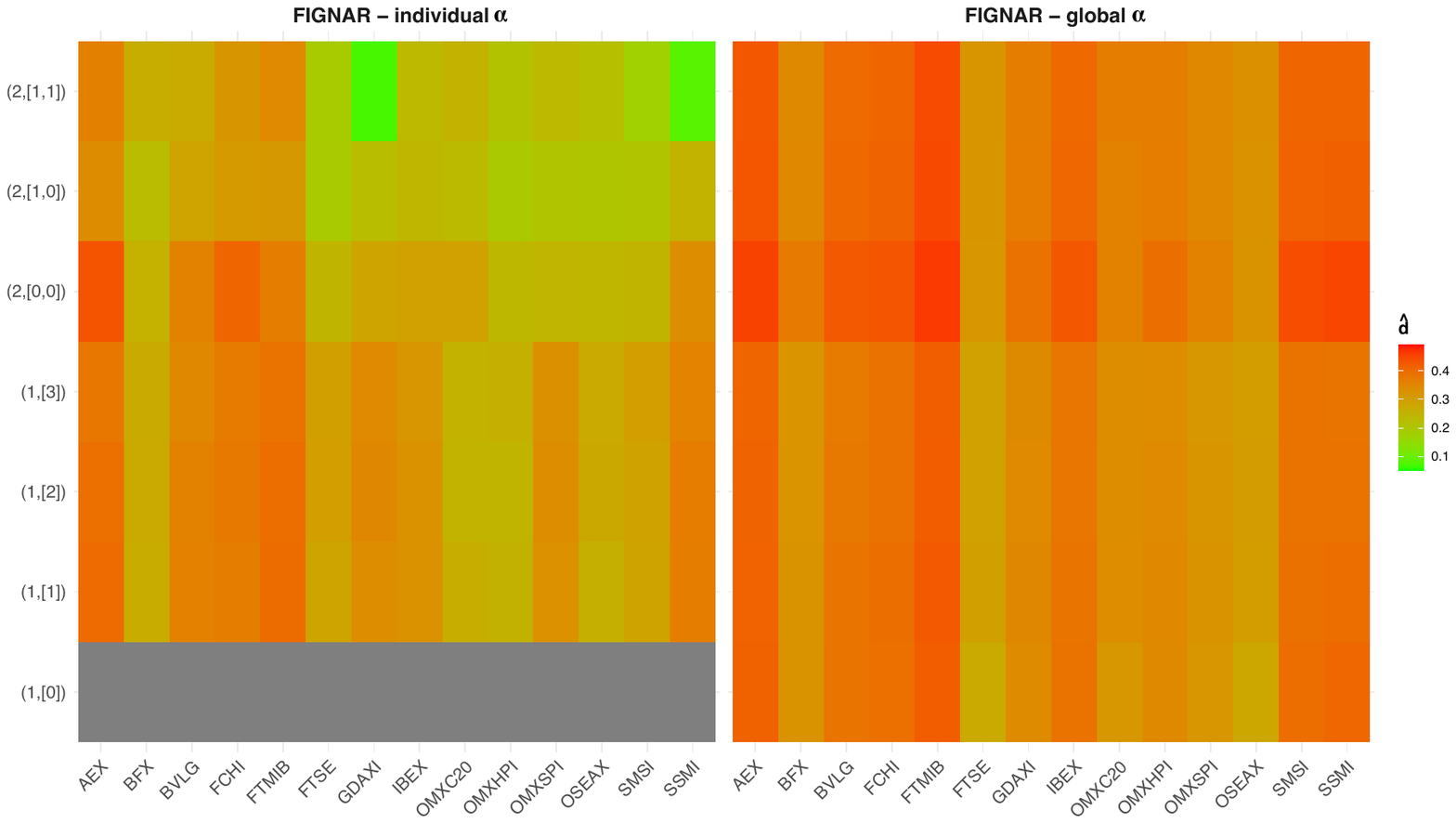}
    \centering
    \includegraphics[width=0.9\textwidth]{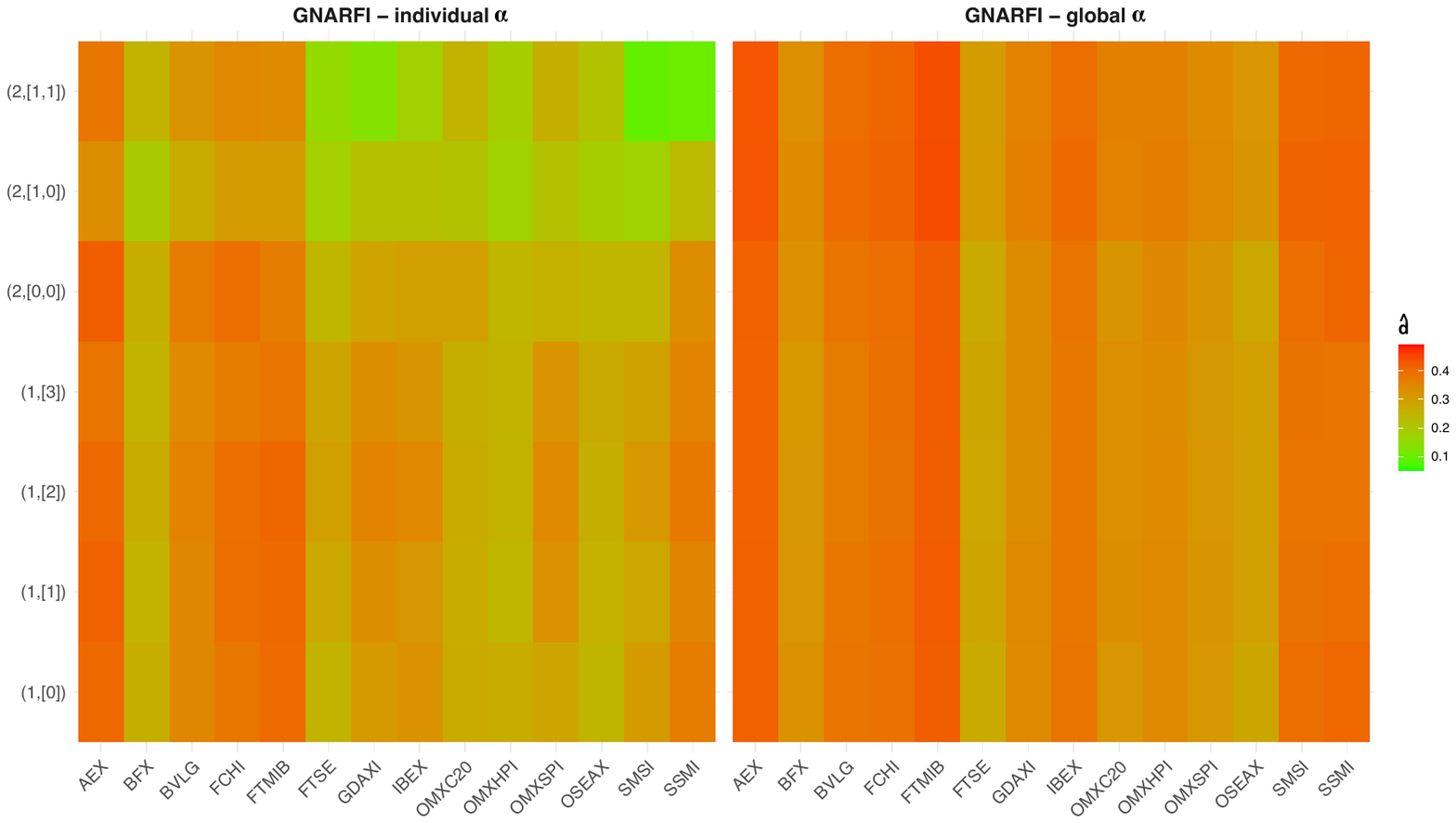}
    \captionof{figure}{Estimated long memory parameters for the 14 European stock data.}\label{img:dhat-rv}
\end{figure}

Figure \ref{img:dhat-rv} shows the average long memory estimates obtained by fitting the network models across four graphs. The estimates display similar patterns, with global-$\alpha$ values slightly larger than those under individual-$\alpha$. As expected, the strength of long memory varies across stocks, while remaining of similar magnitude across the different fitted models.

\end{document}